\documentclass[preprint,aps,12pt,notitlepage,nofootinbib,tightenlines]{revtex4}
\usepackage[utf8]{inputenc}
\usepackage{amsmath}
\usepackage{bm}
\usepackage{times}
\usepackage{braket}
\usepackage{color}
\usepackage{epsfig}
\usepackage{slashed}
\usepackage{multirow}
\usepackage{booktabs}
\usepackage{array}
\usepackage{graphicx}
\usepackage{subcaption}

\newcommand{\gev}{\text{GeV}}

\usepackage{makecell}

\usepackage{svg}
\usepackage{epstopdf}



\usepackage{doi}
\hypersetup{
  colorlinks=true,
  linkcolor=red,
  citecolor=green,
  urlcolor=magenta,
  pdfborder={0 0 0},
  breaklinks=true,
}

\begin{document}
%%%%%%%%%%%%%%%%%%%%%%%%%%%%%%%%%%%%%%%%%%%%%%%%%%%%%%%%%%%%%%%%%%%%%%%%%%%
\title{Probing Anomalous $t{\bar q}Z$ Interactions at Muon Colliders}

\author{Liangliang Shang$^{1}$\footnote{Email: shangliangliang@htu.edu.cn; liangliang.shang@physics.uu.se}, Lu Zhang$^1$\footnote{Email: zhanglu@stu.htu.edu.cn}, Bingfang Yang$^1$\footnote{E-mail: yangbingfang@htu.edu.cn}, Stefano Moretti$^{2,3}$\footnote{Email: s.moretti@soton.ac.uk; stefano.moretti@physics.uu.se},}

\affiliation{\footnotesize
$^1$School of Physics, Henan Normal University, Xinxiang 453007,
PR China\\
$^2$Department of Physics and Astronomy, University of Southampton, Southampton, SO17 1BJ, United Kingdom\\
$^3$Department of Physics and Astronomy, Uppsala University, Box 516, SE-751 20 Uppsala, Sweden
}

\begin{abstract}
\vspace*{1cm}

In the framework of effective field theory, we study the anomalous $t{\bar q}Z$ interaction through the process $\mu^+\mu^- \to t{\bar q}Z$ at future muon colliders with
$\sqrt s=$ 3, 10, 14\,\text{TeV}. Based on the top quark decay modes involving $W$ and $Z$ bosons, we first divide the signal into six cases. Then, in order to obtain the limits on the corresponding branching ratios, we perform a detector simulation for both signals and Standard Model backgrounds. To enhance the signal significance, we exploit the polarization of the muon beams and employ the fat jet method to reconstruct signals in hadronic final states. For $\sqrt s=$ 14\,\text{TeV} with $20\,\text{ab}^{-1}$, we find that the upper limit on the branching ratio for $t\to qZ$ can reach the order of $\mathcal{O}(10^{-8})$, which exceeds the limits provided by the CMS and ATLAS collaborations by 2 to 3 orders of magnitude. Our study thus demonstrates that TeV-scale muon colliders can provide an efficient and complementary platform for probing  rare top quark interactions. 

\end{abstract}

\maketitle

\newpage

\section{Introduction}

Research on rare top quark physics processes represents an important avenue for exploring physics Beyond the Standard Model (BSM). Being the most massive elementary particle in the SM and possessing direct coupling to the Electro-Weak Symmetry Breaking (EWSB) sector, the top quark occupies a unique position among SM particles, making it an ideal probe for BSM physics~\cite{probe_for_BSM_physics}. Among these processes, Flavor Changing Neutral Current (FCNC) interactions involving the top quark have attracted particular attention over the years: in fact, such processes are strictly forbidden at tree level in the SM and are also strongly suppressed at the loop level due to the Glashow-Iliopoulos-Maiani (GIM) mechanism~\cite{GIM_mechanism_1,GIM_mechanism_2}. The predicted Branching Ratio (BR) for the top quark decay channels $t \to qZ$ ($q = u, c$) in the SM is only of the order of $10^{-14}$~\cite{only_of_the_order_of_10^(-14)}, which is far below  the current detection capability of the Large Hadron Collider (LHC). This extreme suppression effect implies that any experimental observation of a signal deviating from SM predictions would necessarily indicate the existence of new physical degrees of freedom or modifications to the interaction structure of the top quark. To this end, various BSM scenarios have investigated the detectability of top quark FCNCs providing deviations from the SM predictions.

According to different physics mechanisms, these models can be roughly categorized into four classes. {The first class} uses new particle mixing to break the GIM mechanism, i.e., by introducing new fermions (such as vector-like top quarks) that mix with SM quarks, the diagonal form of the top quark mass  and coupling matrices is destroyed, thereby generating FCNC couplings at the tree or one loop level: examples include the quark singlet model~\cite{the_quark_singlet_model} and the warped extra dimension model~\cite{the_warped_extra_dimension_model}. {The second class} extends the Higgs sector and introduces new couplings, i.e., increasing the number of Higgs doublets leads to direct coupling between neutral Higgs bosons and quarks of different generations (hence, tree level FCNCs) or to contributions from new particles (such as charged Higgs at the loop level): for instance, the 2-Higgs Doublet Model (2HDM) without without a $Z_2$ symmetry~\cite{2HDM_without_flavor_changing_neutral_currents}. {The third class} introduces Supersymmetry (SUSY) and its breaking, so that SUSY partner particles contribute at the loop level by disrupting the GIM cancellation (if R-parity violation is further allowed, entirely new tree level FCNCs can be introduced): examples include the Minimal Supersymmetric Standard Model (MSSM)~\cite{MSSM} (and the R-parity violating MSSM~\cite{R-parity_violating_MSSM}). {The fourth class} involves new loop contributions from so-called  mirror fermions that mediate FCNC interactions at the loop level: for example, like in the extended mirror fermion model~\cite{mirror_fermion_model}.
Therefore, searches for such FCNC processes involving the top quark are of great theoretical significance, to the extent that, once this process is observed, it will be a clear signal of the existence of BSM physics~\cite{new_physics_beyond_the_Standard_Model}.

On the experimental side, 
the ATLAS and CMS collaborations at the LHC continue to conduct related studies~\cite{ATLAS_CMS_conduct_related_studies_1,ATLAS_CMS_conduct_related_studies_2,ATLAS_CMS_conduct_related_studies_3,ATLAS_CMS_conduct_related_studies_4,ATLAS_CMS_conduct_related_studies_5,ATLAS_CMS_conduct_related_studies_6,ATLAS_CMS_conduct_related_studies_7}. From existing literature, it is clear that the LHC provides the most stringent experimental upper limits on the $t\to u(c)Z$ decay BRs at the 95\% Confidence Level (CL). In fact, the CMS Collaboration concluded  the following: $\text{BR}(t\to uZ)<2.4\times10^{-4}$ and $\text{BR}(t\to cZ)<4.5\times10^{-4}$~\cite{ATLAS_CMS_conduct_related_studies_5} at $\sqrt{s}=13$ TeV with $L = 35.9 \ \text{fb}^{-1}$. Furthermore, the ATLAS Collaboration set the upper limits as $\text{BR}(t\to uZ)<6.2\times10^{-5}$ and $\text{BR}(t\to cZ)<1.2\times10^{-4}$~\cite{ATLAS_CMS_conduct_related_studies_6} at $\sqrt{s}=13$ TeV with $L = 139 \ \text{fb}^{-1}$.

However, noting that the $u$ and $c$ quarks generates jets and that the $Z$ boson decay predominantly into hadrons, it is clear that the LHC environment is not the
ideal one to pursue searches for FCNC interactions of top quarks: in addition to QCD backgrounds emerging from the final state, one also has 
initial state radiation, in turn inducing numerous background events unrelated to the hard scattering and dictating complex event topologies for the signals, both of which inherently limits the achievable experimental sensitivity. As a complement to hadron collider experiments, lepton collider projects have then been extensively studied at the phenomenological level, as their key advantage lies in the cleaner experimental environment, crucially, without interference from initial state strong interactions. Against this backdrop, a large number of works have explored the discovery sensitivity of future $e^+e^-$ colliders to anomalous FCNC couplings of the top quark within the framework of Effective Field Theory (EFT)~\cite{effective_field_theory_1,effective_field_theory_2,effective_field_theory_3,effective_field_theory_5,effective_field_theory_6,effective_field_theory_7}. However, there is an intrinsic limitation in such colliders too, i.e., the large emission of photon radiation from the electro-positron beams (especially when accelerated circularly), which limits the energy reach of such colliders, in turn affecting the cross section for top quark production and decay via a reduced 
phase space (and, potentially, cross section), and, more importantly, which reduces the collider luminosity.
Among various planned (more generally) `lepton' collider schemes, though, the muon collider occupies a unique position. Firstly, such a kind of facilities were prominently proposed in the ``2020 European Strategy for Particle Physics''~\cite{2020_European_Strategy_1,2020_European_Strategy_2,2020_European_Strategy_3}. Secondly, due to the large mass of the muon (in comparison to the electron one), the described synchrotron radiation effect is significantly suppressed, thus allowing even a circular collider to achieve the multi-TeV scale while maintaining high luminosity.

Therefore, benefiting from three major advantages including ultra-high collision energy, large statistical event samples and a clean experimental environment, a muon collider provides a powerful and complementary platform for advancing the search for top quark FCNC processes, specifically, of the aforementioned decay processes, with its detection sensitivity expected to largely surpass the limits achievable in existing hadron collider experiments.

This paper is organized as follows. In Section II, we calculate the production cross section of the signal processes and discuss the effect of muon polarization. In Section III, we perform a Monte Carlo (MC) simulation for signal and background events, eventually presenting the 95\% CL exclusion limits in correspondence to the six decay channels we mentioned. Finally, in Section IV, we summarize the main results and draw the conclusions.

\section{Production and decay processes with top quark FCNC interactions}

In this section, we describe the structure of the EFT embedding $t{\bar q}Z$ couplings and quantify the cross sections of the corresponding production and decay processes at a muon collider. Unless otherwise specified, all results in this work include the contributions from charge conjugation.

\subsection{The $t{\bar q}Z$ Anomalous Couplings}

To study the $t\to qZ$ FCNC, we adopt the effective Lagrangian approach of Ref.~\cite{Lagrangian_approach}, so as to investigate top quark FCNC interactions in a model-independent manner. The relevant effective Lagrangian is given as follows~\cite{Lagrangian_approach,dif_chan_0} :
\begin{align}
\mathcal{L}_{\rm eff}  &=\sum_{q=u,c} [\frac{g}{4c_{W}m_{Z}}\kappa_{t{\bar q}Z} \bar{q}\sigma^{\mu\nu}(\kappa_{L}P_{L}+\kappa_{R}P_{R})tZ_{\mu\nu}  \nonumber \\
&+ \frac{g}{2c_{W}}\lambda_{t{\bar q}Z} \bar{q}\gamma^{\mu}(\lambda_{L}P_{L}+\lambda_{R}P_{R})tZ_{\mu}] + h.c.,
\label{lag}
\end{align}
where $c_{W}=\cos\theta_{W}$, with $\theta_{W}$ being the Weinberg angle, $P_{L,R}$ are the Left-Handed (LH) and Right-Handed (RH) chirality projector operators while  $\kappa_{t{\bar q}Z}$ and $\lambda_{t{\bar q}Z}$ are effective couplings for the corresponding vertices. Thus, the BSM interactions are parameterized by SM parameters (i.e., the coupling constant $g$ and the mixing angle $\theta_{W}$) as well as the complex chiral parameters $\kappa_{L,R}$ and $\lambda_{L,R}$, which are normalized as $|\kappa_{L}|^2 + |\kappa_{R}|^2=|\lambda_{L}|^2 + |\lambda_{R}|^2=1$ (in addition to the $t{\bar q}Z$ effective couplings). Finally, $Z^\mu$ and $Z^{\mu\nu}$ are the $Z$ boson field and its stress (or energy-momentum) tensor, respectively, contracting with the usual Dirac $\gamma$-matrix structures.

The partial widths for the discussed FCNC decays of the top quark, wherein we separate the contributions of the two tensor structures entering the above Lagrangian equation,  are given by,
\begin{align}
\Gamma(t\to qZ)~(\sigma^{\mu\nu})  &=\frac{\alpha}{128s_{W}^{2}c_{W}^{2}}|\kappa_{t{\bar q}Z}|^2\frac{m_{t}^{3}}{m_{Z}^{2}}\left[1-\frac{m_{Z}^{2}}{m_{t}^{2}}\right]^{2}\left[2+\frac{m_{Z}^{2}}{m_{t}^{2}}\right], \nonumber \\
\Gamma(t\to qZ)~(\gamma^{\mu})  &=\frac{\alpha}{32s_{W}^{2}c_{W}^{2}}|\lambda_{t{\bar q}Z}|^2\frac{m_{t}^{3}}{m_{Z}^{2}}\left[1-\frac{m_{Z}^{2}}{m_{t}^{2}}\right]^{2}\left[1+2\frac{m_{Z}^{2}}{m_{t}^{2}}\right]. 
\end{align}
After neglecting all the light quark masses and assuming the dominant top decay partial width to be that of $t \to bW$~\cite{t_to_wb},
\begin{align}
\Gamma(t\to bW^{+})=\frac{\alpha}{16s_{W}^{2}}|V_{tb}|^2\frac{m_{t}^{3}}{m_{W}^{2}}\left[1-3\frac{m_{W}^{4}}{m_{t}^{4}}+2\frac{m_{W}^{6}}{m_{t}^{6}}\right], 
\end{align}
 the BR($t \to qZ$) can be approximated  by,
\begin{align}
{\rm BR}(t \to qZ)~(\sigma^{\mu\nu})  &= 0.172|~\kappa_{t{\bar q}Z}|^2,  \nonumber  \\
{\rm BR}(t \to qZ)~(\gamma^{\mu})     &= 0.471|~\lambda_{t{\bar q}Z}|^2.
\end{align}

\begin{figure}[htbp]
  \centering
  \begin{subfigure}{0.22\textwidth}
    \centering
    \includegraphics[width=\linewidth]{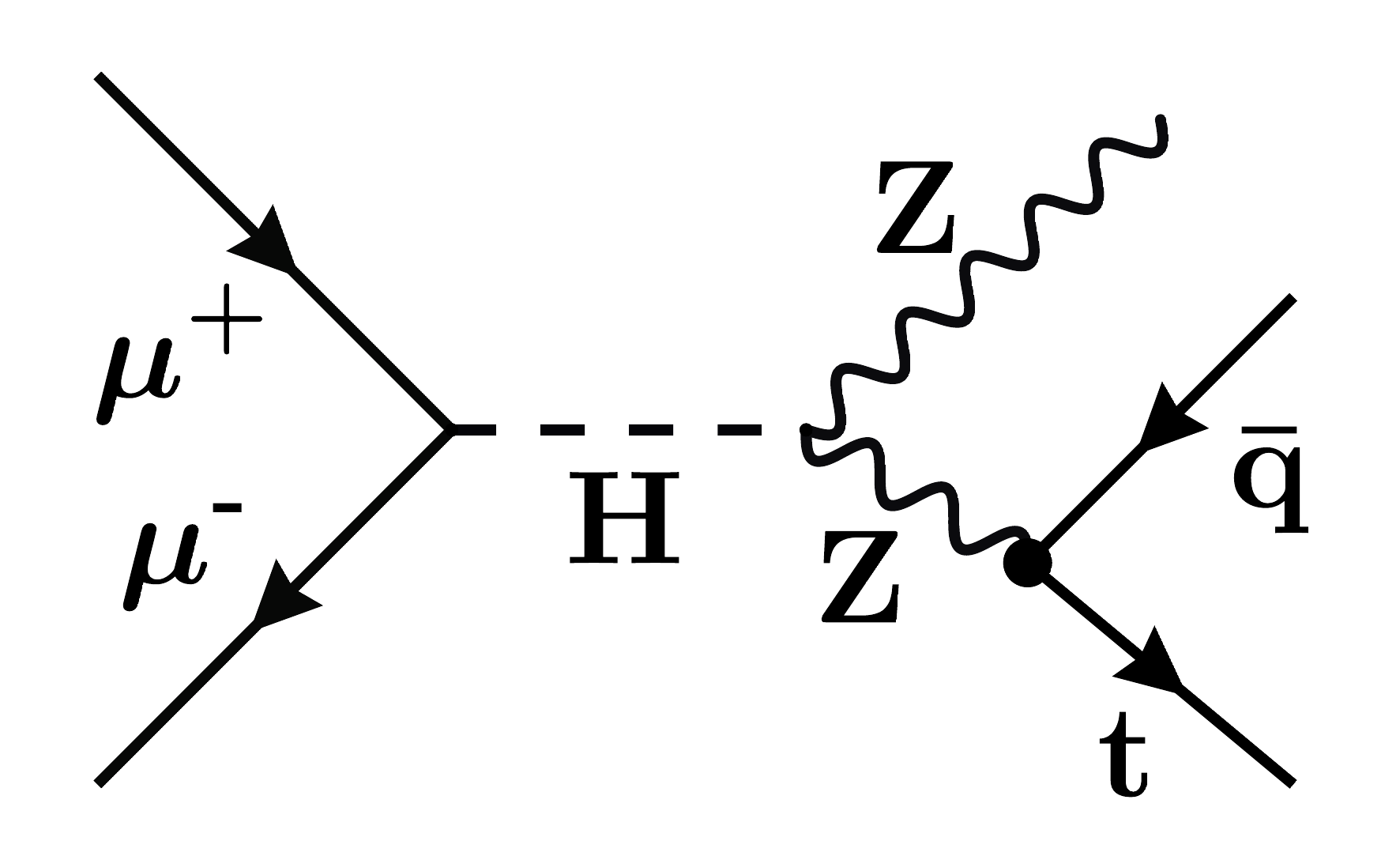}
    \raisebox{-0.5\height}{\makebox[\linewidth][c]{(a)}}
  \end{subfigure}
  \begin{subfigure}{0.22\textwidth}
    \centering
    \includegraphics[width=\linewidth]{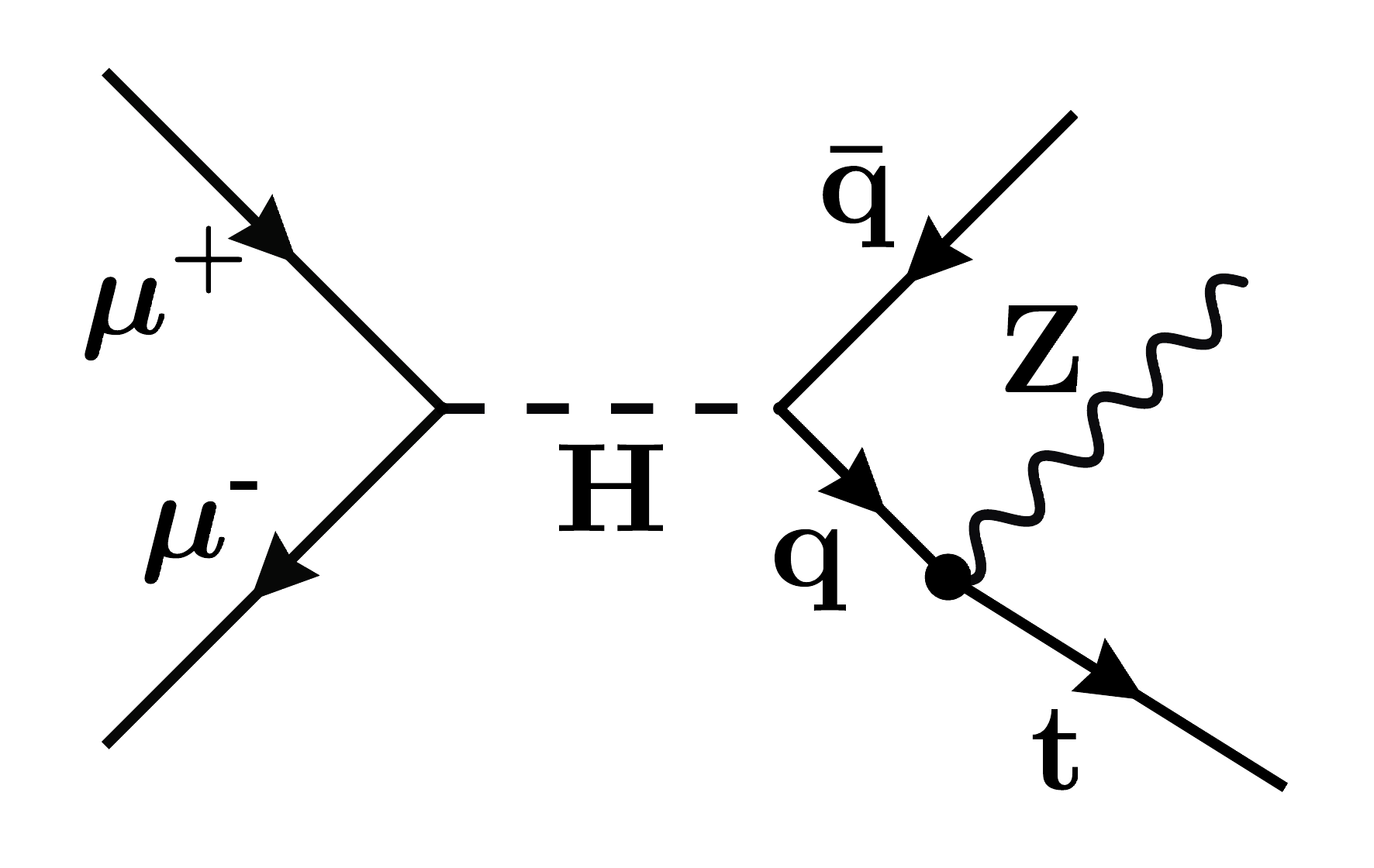}
    \raisebox{-0.5\height}{\makebox[\linewidth][c]{(b)}}
  \end{subfigure}
  \begin{subfigure}{0.22\textwidth}
    \centering
    \includegraphics[width=\linewidth]{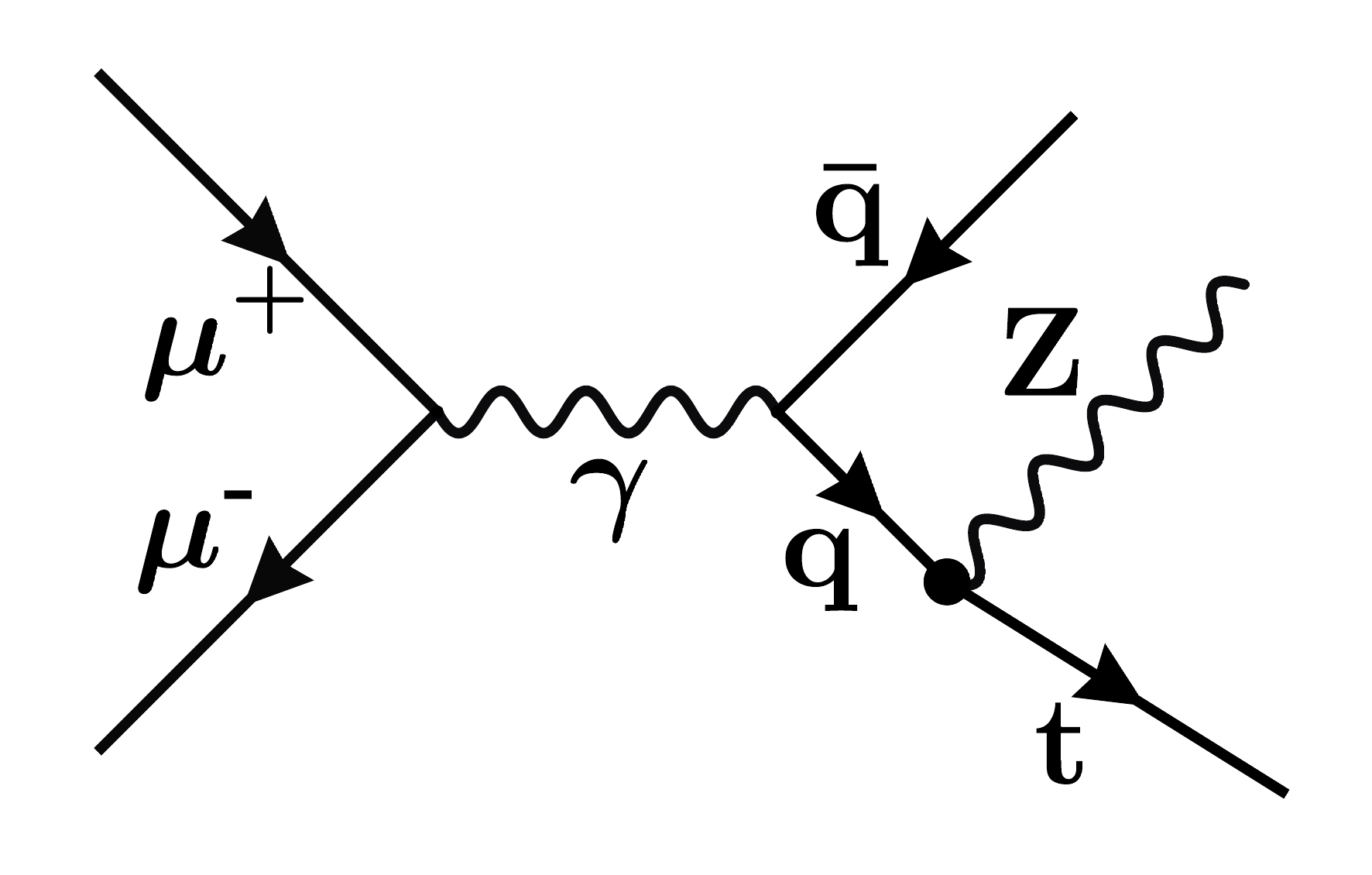}
    \raisebox{-0.5\height}{\makebox[\linewidth][c]{(c)}}
  \end{subfigure}
  \begin{subfigure}{0.22\textwidth}
    \centering
    \includegraphics[width=\linewidth]{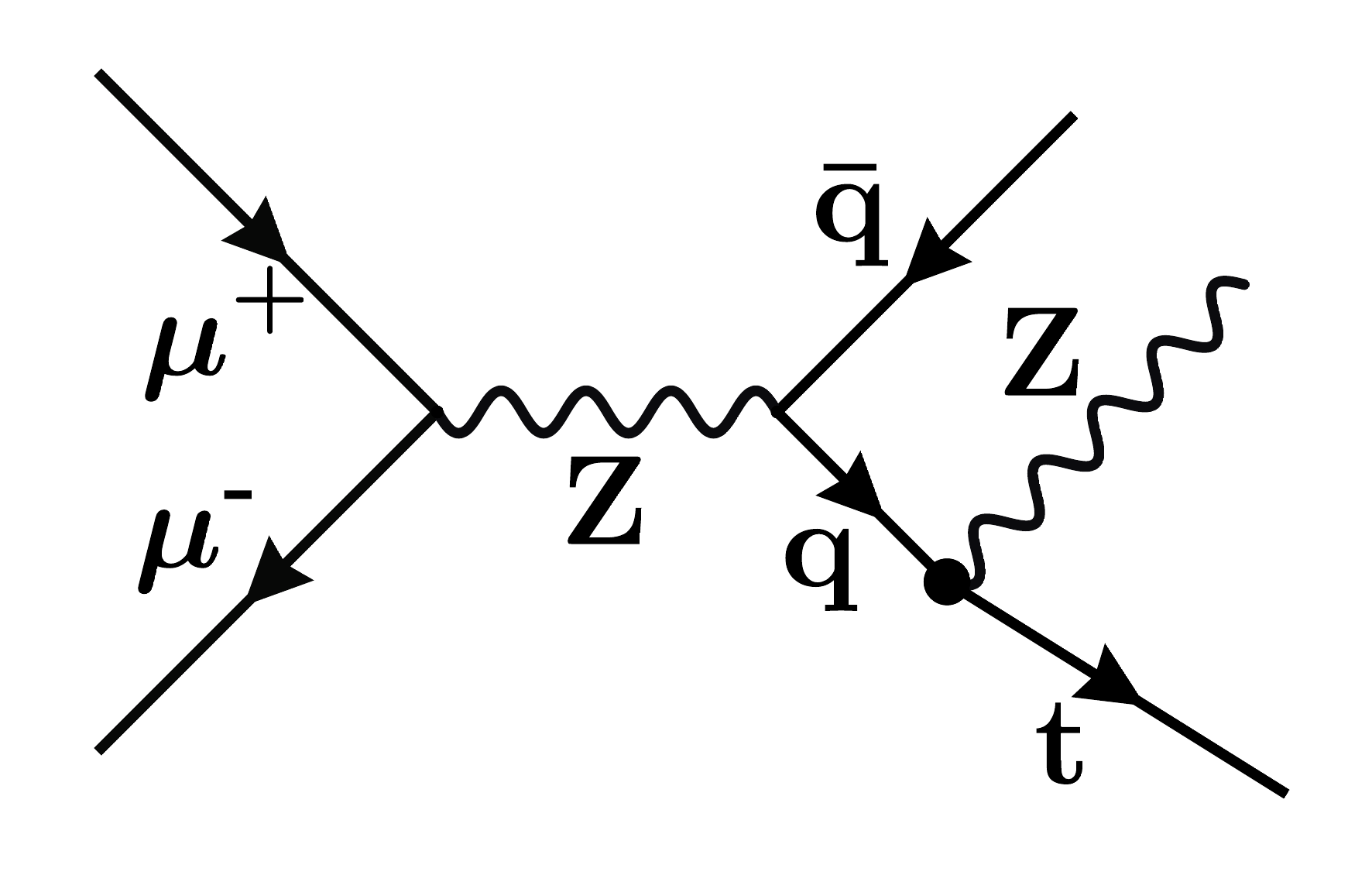}
    \raisebox{-0.5\height}{\makebox[\linewidth][c]{(d)}}
  \end{subfigure}
  
  \vspace{0.5em}
  \begin{subfigure}{0.22\textwidth}
    \centering
    \includegraphics[width=\linewidth]{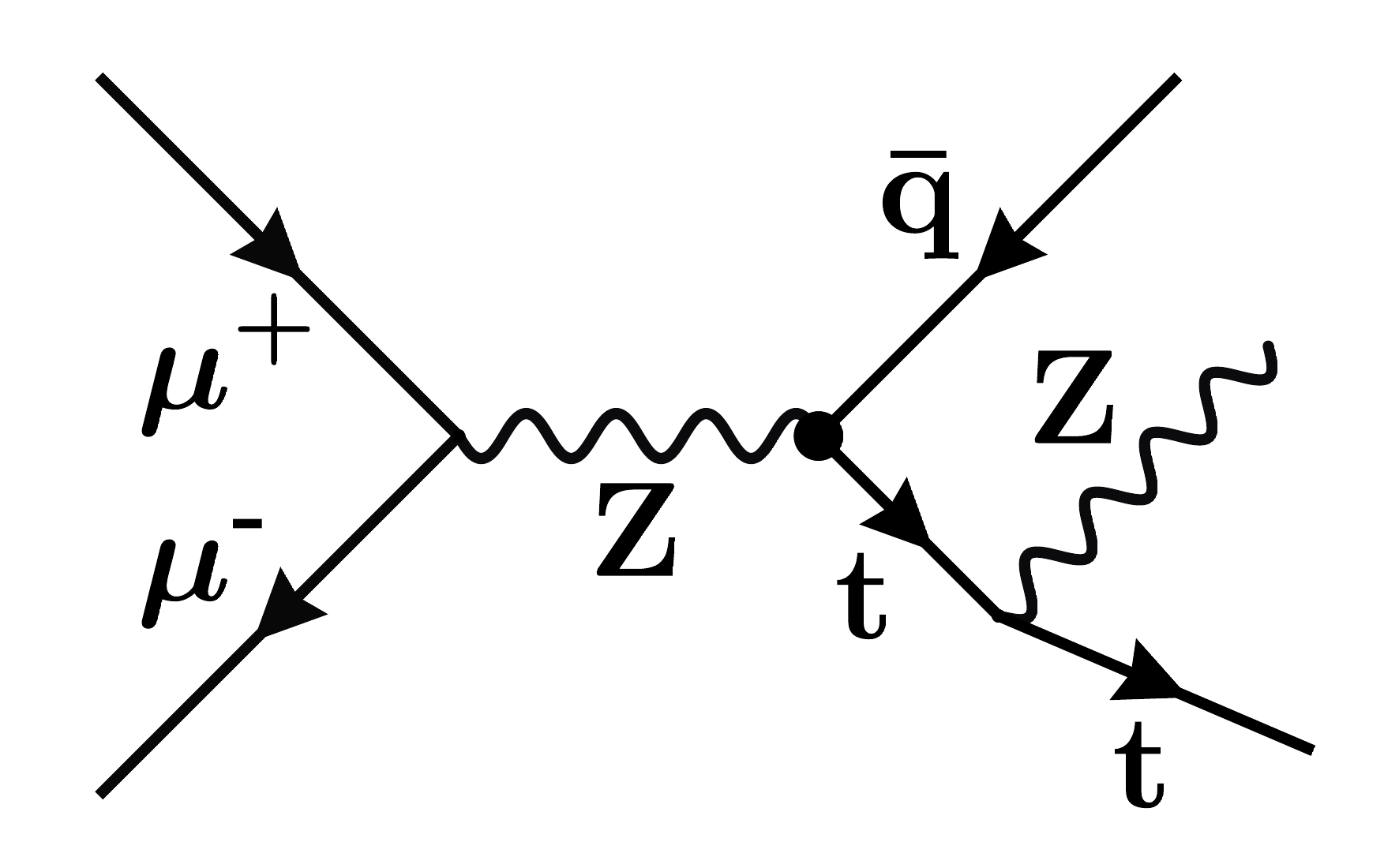}
    \raisebox{-0.5\height}{\makebox[\linewidth][c]{(e)}}
  \end{subfigure}
  \begin{subfigure}{0.22\textwidth}
    \centering
    \includegraphics[width=\linewidth]{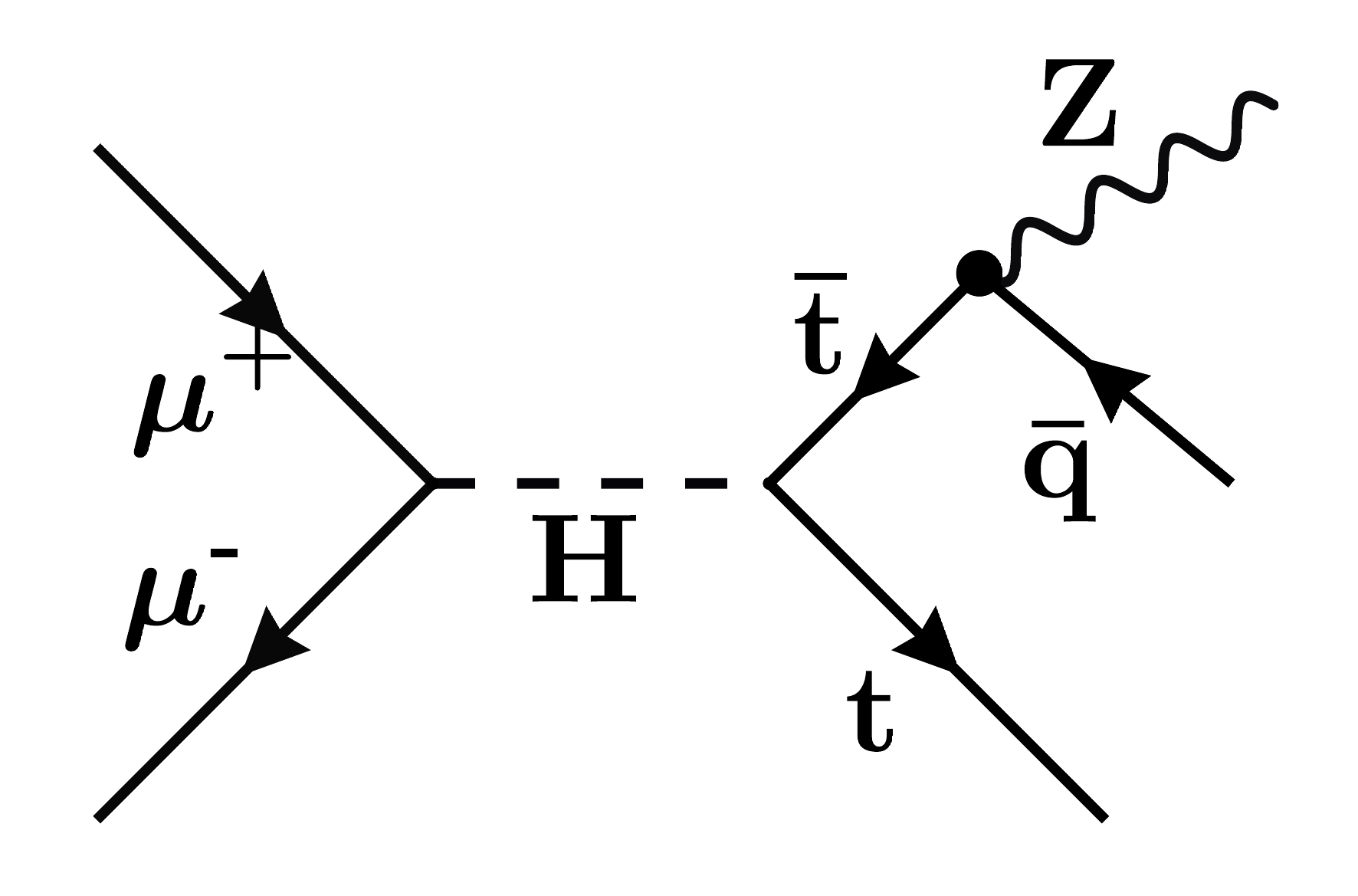}
    \raisebox{-0.5\height}{\makebox[\linewidth][c]{(f)}}
  \end{subfigure}
  \begin{subfigure}{0.22\textwidth}
    \centering

    \includegraphics[width=\linewidth]{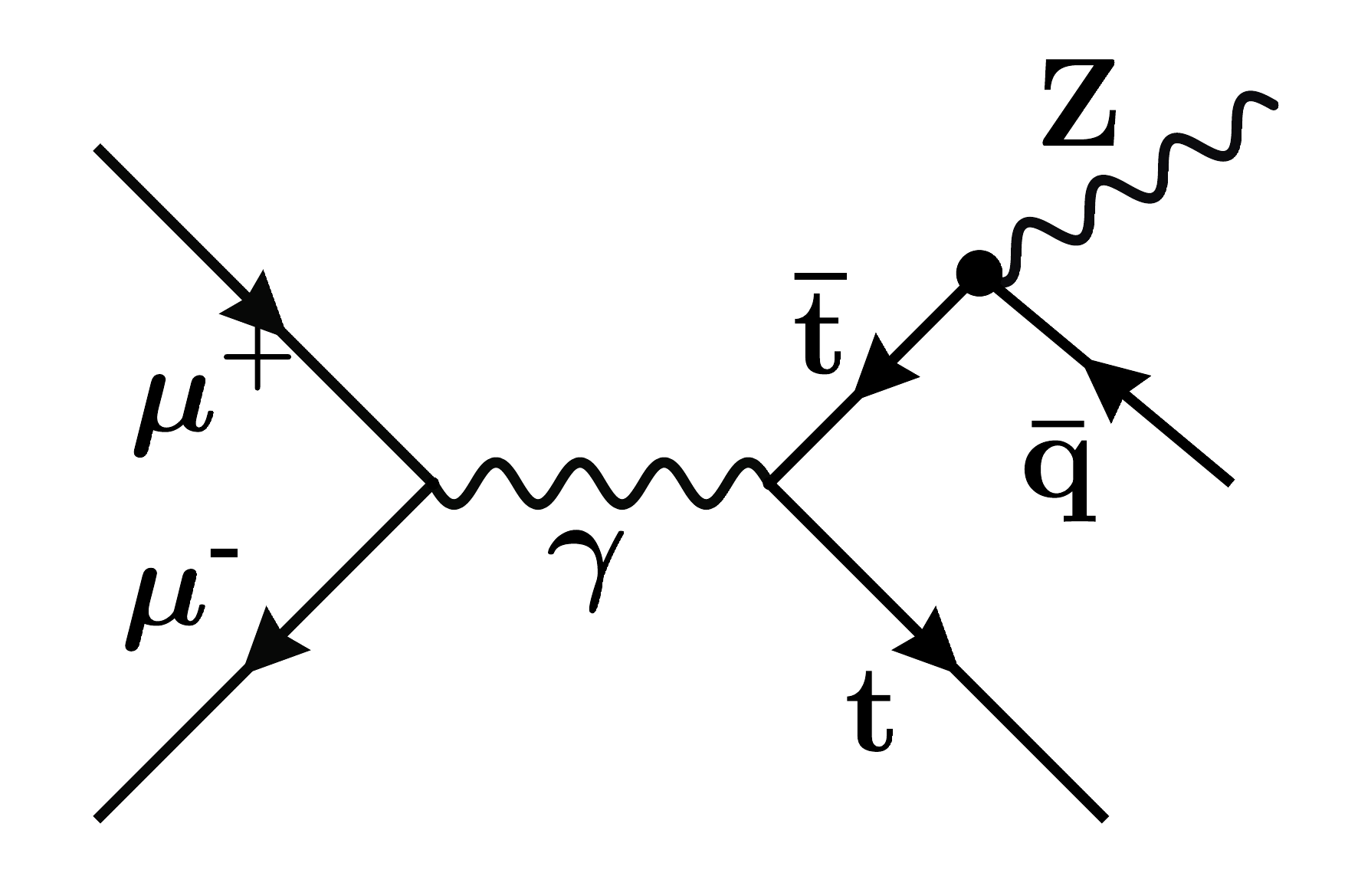}
    \raisebox{-0.5\height}{\makebox[\linewidth][c]{(g)}}
  \end{subfigure}
  \begin{subfigure}{0.22\textwidth}
    \centering
    \includegraphics[width=\linewidth]{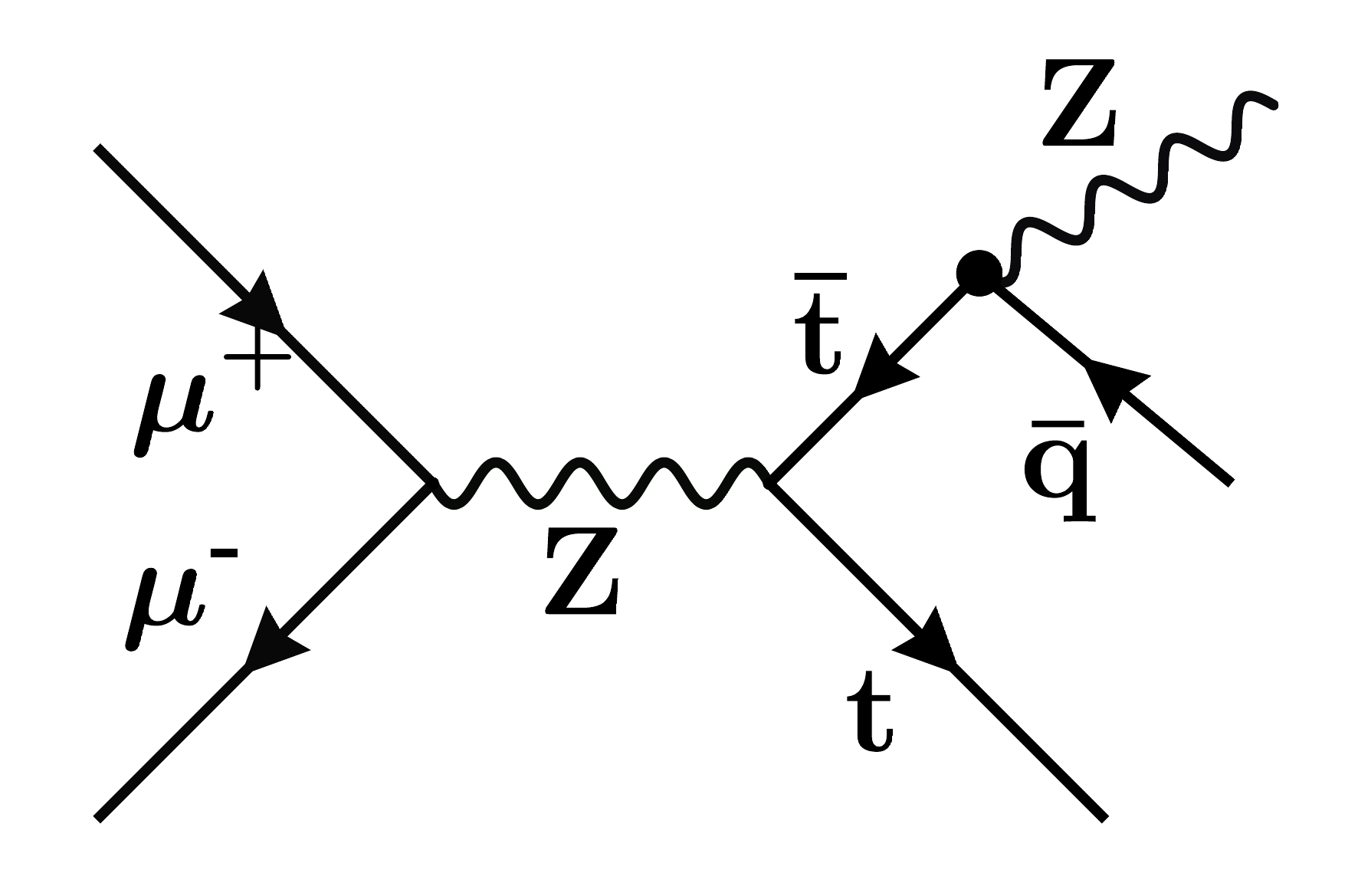}
    \raisebox{-0.5\height}{\makebox[\linewidth][c]{(h)}}
  \end{subfigure}
  
  \vspace{0.5em}
  \begin{subfigure}{0.22\textwidth}
    \centering
    \includegraphics[width=\linewidth]{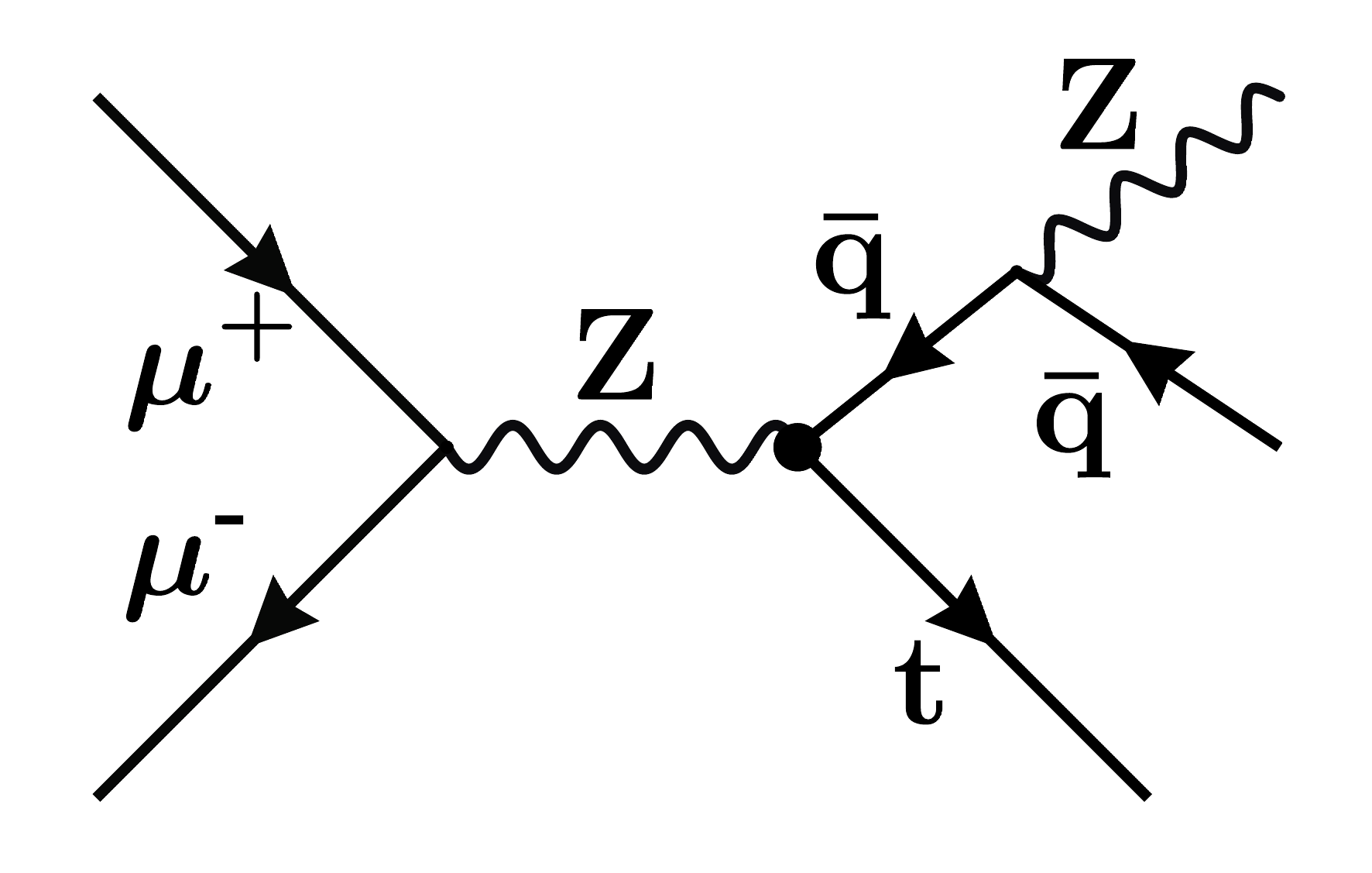}
    \raisebox{-0.5\height}{\makebox[\linewidth][c]{(i)}}
  \end{subfigure}
  \begin{subfigure}{0.22\textwidth}
    \centering

    \includegraphics[width=\linewidth]{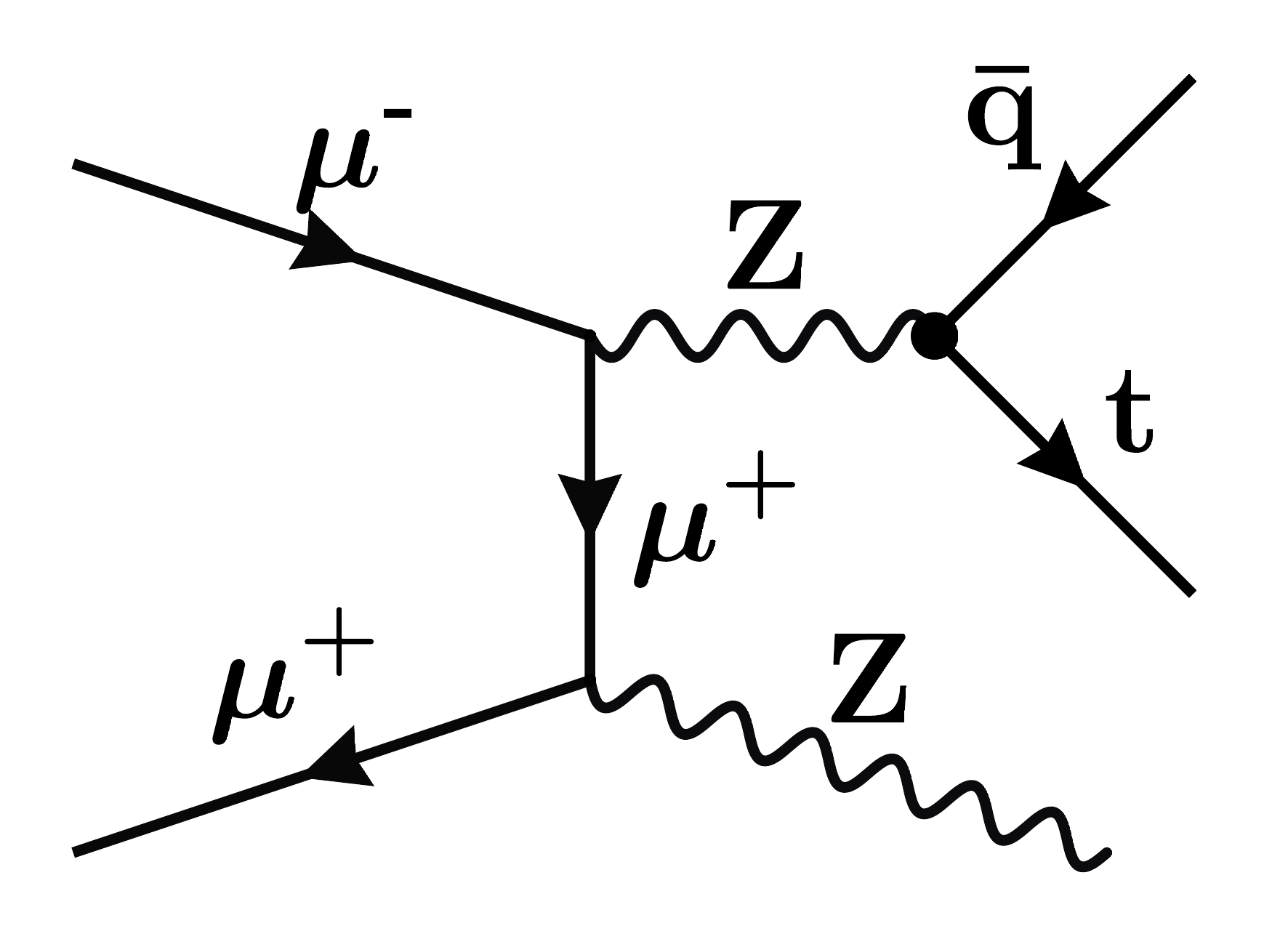}
    \raisebox{-0.5\height}{\makebox[\linewidth][c]{(j)}}
  \end{subfigure}
  \begin{subfigure}{0.22\textwidth}
    \centering
    \includegraphics[width=\linewidth]{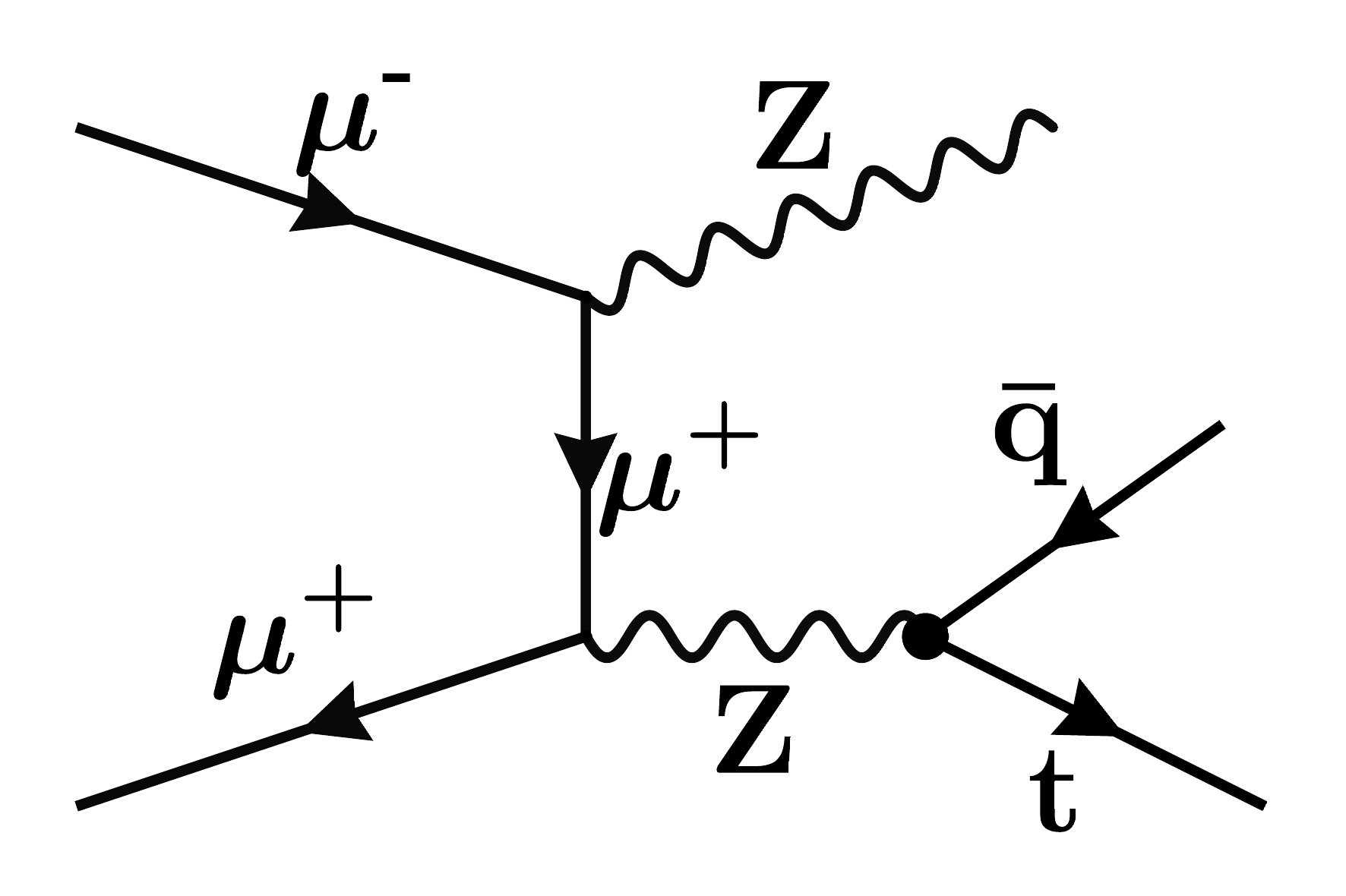}
    \raisebox{-0.5\height}{\makebox[\linewidth][c]{(k)}}
  \end{subfigure}

\caption{Feynman diagrams for $\mu^+\mu^- \to t\bar{q}Z$. Here, the black dot $\bullet$ denotes the $t\bar{q}Z$ vertex while $H$ is the SM-like Higgs boson.}
  \label{fig:feynman_diagrams}
\end{figure}

According to the given effective Lagrangian, the Feynman diagrams at tree level for top quark pair production followed by the FCNC contribution $\mu^+ \mu^- \to t\bar{q}Z$ at a muon collider are presented in Fig.~\ref{fig:feynman_diagrams}.

\subsection{Cross Sections}

For the simulations needed for the ensuing phenomenology analysis, we first use the \texttt{FeynRules} package~\cite{FeynRules_package} to generate the Universal FeynRules Output~(UFO) files~\cite{UFO_files}. The cross sections are then  obtained by using \texttt{MG5\_aMC@NLO v2.9.12}~\cite{MadGraph5_NLO}. The numerical values of the input parameters are taken as follows~\cite{PDG_2024}:
\begin{align}
m_t=172.57{\rm ~GeV}&, \ \ m_Z=91.18{\rm ~GeV}, \ \ m_H=125{\rm ~GeV}, \\ \nonumber
\alpha_{EM}^{-1}(m_Z)=127.9&, \ \ \alpha_{s}(m_Z)=0.1181, \ \ G_F=1.1664\times 10^{-5}\ {\rm GeV^{-2}}.
\end{align}

\begin{figure}[htbp]
  \centering
  \includegraphics[width=0.60\textwidth]{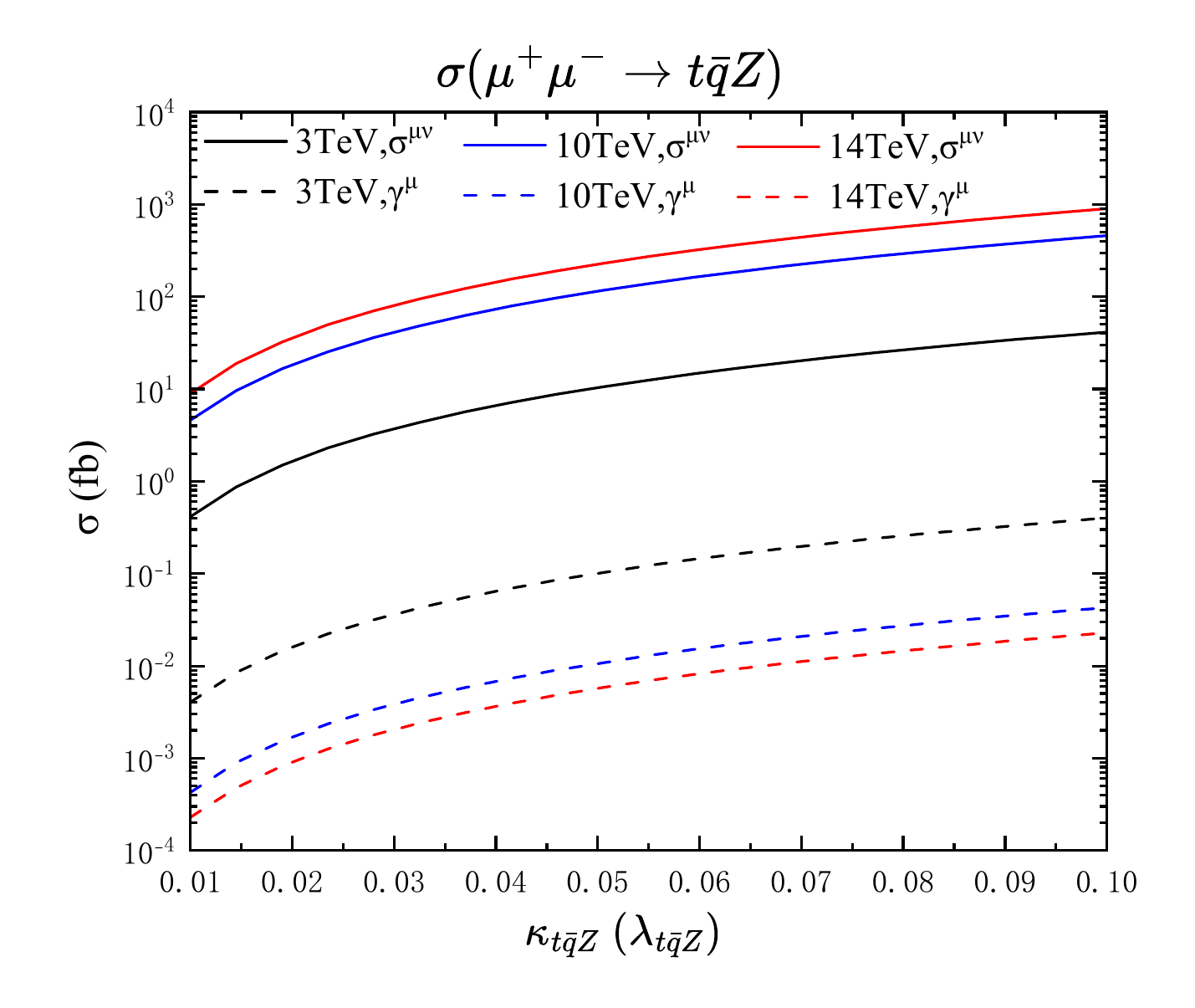}
  \caption{In $\mu^+\mu^-\to t{\bar q}Z$ process, cross section curves of the tensor term (solid lines) and vector term (dashed lines) for different collision energy and coupling parameters $\kappa_{t{\bar q}Z}$ and $\lambda_{t{\bar q}Z}$, considering $\kappa_{L}=\kappa_{R}$ and $\lambda_{L}=\lambda_{R}$.}
  \label{fig:ktqZ_vs_ltqZ}
\end{figure}

In Fig.~\ref{fig:ktqZ_vs_ltqZ}, we present the cross section of $\mu^+\mu^-\to t{\bar q}Z$ for two coupling parameters, $\kappa_{t{\bar q}Z}$ (associated with the tensor structure $\sigma^{\mu\nu}$) and $\lambda_{t{\bar q}Z}$ (associated with the vector structure $\gamma^\mu$), at 3, 10 and 14 TeV. Herein, the
tensor and vector contributions to the cross section are considered separately. When varying the value of $\kappa_{t{\bar q}Z}$, we fix $\lambda_{t{\bar q}Z}=0$. Similarly, $\kappa_{t{\bar q}Z}$ is set to zero when scanning over $\lambda_{t{\bar q}Z}$. As shown in the figure, for a fixed collision energy, the cross sections of both the tensor and vector components gradually increase with $\kappa_{t{\bar q}Z}$ and $\lambda_{t{\bar q}Z}$. With the rise of the collision energy, the cross section curve of the tensor term increases, while the vector term exhibits an opposite (i.e., decreasing) trend. Altogether, the cross section of the tensor term is considerably larger than that of the vector term.

The explanation is as follows. The Lagrangian of the tensor term contains the structure $\sigma^{\mu\nu}Z_{\mu\nu}$, where (as intimated) $Z_{\mu\nu}=\partial_\mu Z_\nu-\partial_\nu Z_\mu$ is the field stress tensor of the $Z$ boson. This structure introduces a factor proportional to the momentum of the $Z$ boson, implying that the scattering amplitude of the tensor term grows linearly with the collision energy~\cite{GIM_mechanism_1}. At 3 TeV, for example, the tensor cross section is about two orders of magnitude larger than the vector one, with the difference reaching 4 orders of magnitude at 10 TeV and 14 TeV. Given such a huge difference in their contributions, we mainly focus on the tensor term in the subsequent analysis.

\begin{figure}[htbp]
  \centering
  \includegraphics[width=0.60\textwidth]{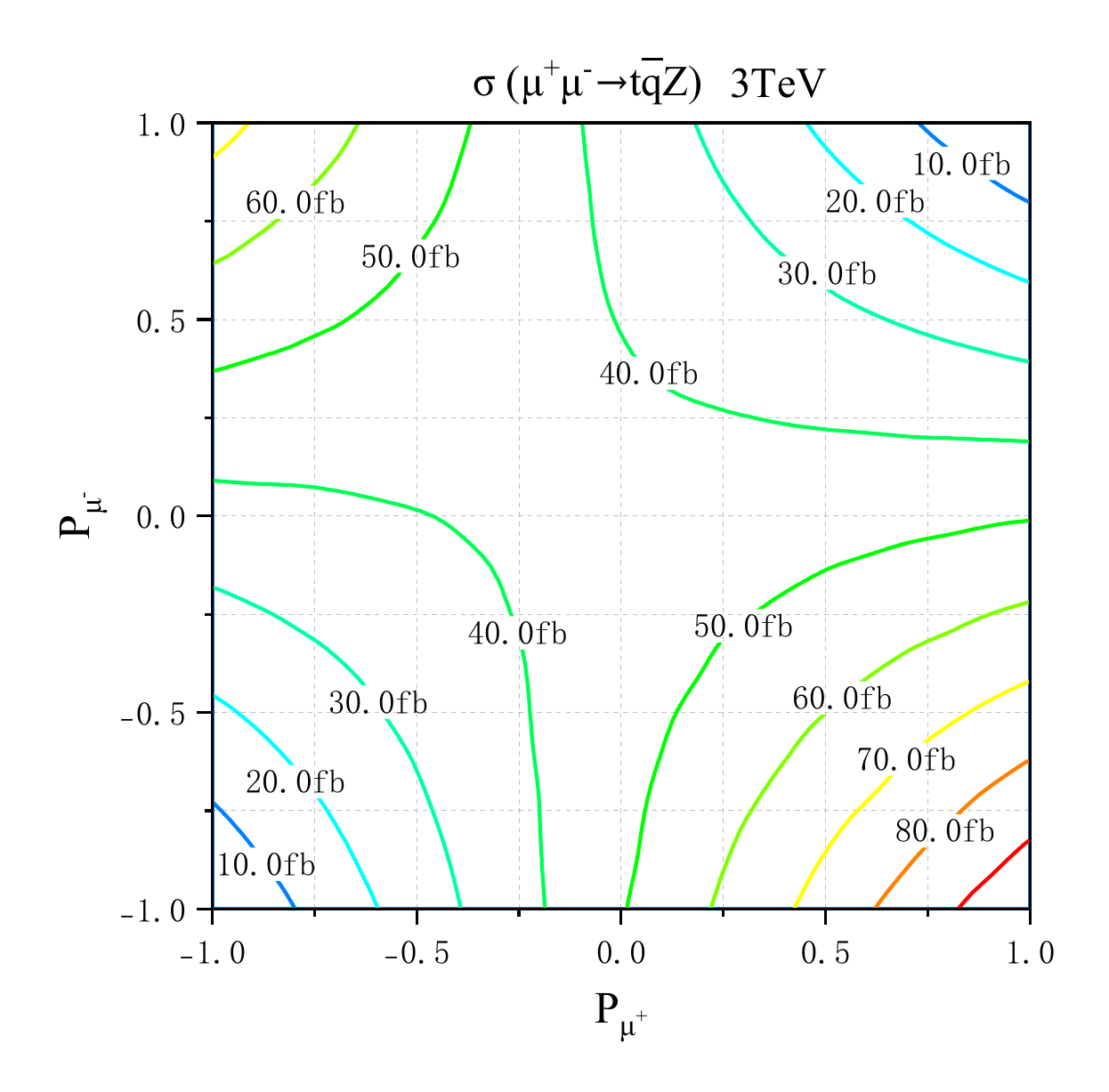}
  \caption{In $\mu^+\mu^-\to t{\bar q}Z$ process, contour plot of the cross section for different polarizations of $\mu^+$ and $\mu^-$, considering $\kappa_{t{\bar q}Z}=0.1$ and $\kappa_{L}=\kappa_{R}$.}
  \label{fig:ji_hua}
\end{figure}

Beam polarization is very helpful for improving the detection limit of $\text{BR}(t \to qZ)$~\cite{polar_2}. With double-beam polarization, the expected upper limit can be enhanced by about a factor of 2.5 compared with the unpolarized case~\cite{polar_1}. Considering the effect of polarization of the positive and negative muon beams on the signal cross section, we perform a scan over the polarization combinations of the muon beams in the polarization range $[-1,+1]$. The results are shown in Fig.~\ref{fig:ji_hua}. It can be observed that the cross section reaches its minimum value when $P_{\mu^+}=P_{\mu^-}=\pm 1$, while the cross section achieves its maximum value for $P_{\mu^+}=+1$ and $P_{\mu^-}=-1$. Taking the current technical capabilities into account, we adopt the polarization configuration $P_{\mu^+}=+0.8$ and $P_{\mu^-}=-0.8$~\cite{polar_3}.

\section{MC Simulation and Statistical Analysis}

According to the decay combinations of the $Z$ and $W$ bosons, the signal is classified into six decay scenarios, as shown in Tab.~\ref{tab:signal_cases}.

\begin{table}[!b]
\centering
\caption{Signal classification based on the decay modes of $W$ and $Z$ bosons.}
\label{tab:signal_cases}
\begin{tabular}{l l}
\hline
\hline
Case A & $\mu^+\mu^- \to t\bar{q}Z, \quad (t \to W^+b,\ W^+ \to \ell^+ \nu_\ell), (Z \to \ell^+ \ell^-)$ \\
Case B & $\mu^+\mu^- \to t\bar{q}Z, \quad (t \to W^+b,\ W^+ \to jj), (Z \to jj)$ \\
Case C & $\mu^+\mu^- \to t\bar{q}Z, \quad (t \to W^+b,\ W^+ \to \ell^+ \nu_\ell), (Z \to jj)$ \\
Case D & $\mu^+\mu^- \to t\bar{q}Z, \quad (t \to W^+b,\ W^+ \to jj), (Z \to \ell^+ \ell^-)$ \\
Case E & $\mu^+\mu^- \to t\bar{q}Z, \quad (t \to W^+b,\ W^+ \to \ell^+ \nu_\ell), (Z \to \nu_\ell \bar{\nu}_\ell)$ \\
Case F & $\mu^+\mu^- \to t\bar{q}Z, \quad (t \to W^+b,\ W^+ \to jj), (Z \to \nu_\ell \bar{\nu}_\ell)$ \\
\hline
\hline
\end{tabular}
\end{table} 
The signal mainly consists of a jet originating from a $u$ or $c$ quark, a top quark, decaying into a $W$ boson and a $b$-jet, as well as a $Z$ boson. Considering the  characteristics of the signals considered in such a table, we analyze the following SM backgrounds: $tWb$, $WW$, $WWZ$, $ZZ$, $ZH$, $WWH$ and $\nu_\ell\bar{\nu_\ell} H$. In addition, we include the $Z\to b\bar{b}$ decay channel within the $Z\to jj$ channel. 

Based on Ref.~\cite{top_quark_fcnc_at_muc}, we consider the following collider scenarios:
\begin{itemize}
    \item  $\sqrt{s} = 3$ TeV  with $L = 1\,\text{ab}^{-1}$,
    \item  $\sqrt{s} = 10$ TeV  with $L = 10\,\text{ab}^{-1}$,
    \item  $\sqrt{s} = 14$ TeV  with $L = 20\,\text{ab}^{-1}$.
\end{itemize}

We also use \texttt{MG5\_aMC@NLO v2.9.12} to generate MC signal and background events at parton level. Then, we pass the parton level events to \texttt{Pythia8}~\cite{pythia8} for showering and hadronization, followed by a fast detector simulation using \texttt{Delphes}~\cite{delphes}. In this analysis, we employ the \texttt{FASTJET} package~\cite{FastJet_user_manual} for jet reconstruction, utilizing the \texttt{Valencia Linear Collider (VLC)} algorithm~\cite{VLC_algorithm_1,VLC_algorithm_2} in exclusive mode with the optimized parameters $\beta=1$ and $\gamma=1$, specific  for the environment of high-energy lepton colliders. The number of jets $N$ is chosen according to the different decay modes, corresponding to the expected number of partons in the final state. Standard jets are reconstructed with a jet radius parameter $R=0.5$. For fat jet reconstruction, a larger jet radius parameter is adopted. In fact, consider that, at high energies, the $Z$ boson acquires a large longitudinal momentum, in turn leading to a significant Lorentz boost effect, hence, in this scenario, the angular separation between the two jets from the hadronic decay of the $Z$ boson in the laboratory frame is strongly compressed plus their angular distance $\Delta R \approx \frac{2m_Z}{p_T^Z}$~\cite{delta_R_1,delta_R_2,delta_R_3} decreases with the increasing transverse momentum of the $Z$ boson. Therefore, we set the fat jet radius $R=1.5$ at 3 TeV, $R=1.0$ at 10 TeV and $R=0.7$ at 14 TeV. Furthermore, to avoid excessively reducing the signal efficiency, a loose $b$-tagging criterion with a 90\% efficiency working point is applied (whereas no $c$-tagging is enforced). Finally, we use \texttt{EasyScanHEP}~\cite{easyscan} to interface these programs and perform scans of the parameter space while the final event statistics is analyzed using \texttt{MadAnalysis5}~\cite{MadAnalysis5_1,MadAnalysis5_2}. 

In order to simulate the acceptance and detection thresholds of the detector, we impose the following basic cuts for both signals and backgrounds:
\[
\begin{aligned}
\Delta R(x,y) &> 0.4, & x,y &= \ell~(\equiv e,\mu),\ j,\ b, \\
p_T^\ell &> 10\ \text{GeV}, & |\eta_\ell| &< 2.5, \\
p_T^b &> 20\ \text{GeV}, & |\eta_b| &< 5.0, \\
p_T^j &> 20\ \text{GeV}, & |\eta_j| &< 5.0,
\end{aligned}
\]
where $j$, $b$ and $p_T$ denote a light-flavor jet, a $b$-tagged one and a generic  transverse momentum (of any final state object). Furthermore, $\Delta R(x,y) = \sqrt{(\Delta\phi)^2 + (\Delta\eta)^2}$ is the spatial distance in the rapidity–azimuthal plane. Here, $\kappa_{L}=1, \ \kappa_{R}=0$ for LH interactions and $\kappa_{L}=0, \ \kappa_{R}=1$ for RH interactions.

\subsection{Selection Cuts for Case A \textnormal{\small{($W^+ \to \ell^+ \nu_\ell$ , $Z \to \ell^+ \ell^-$)}}}

\begin{figure}[htbp]
  \centering
  \includegraphics[width=0.48\textwidth]{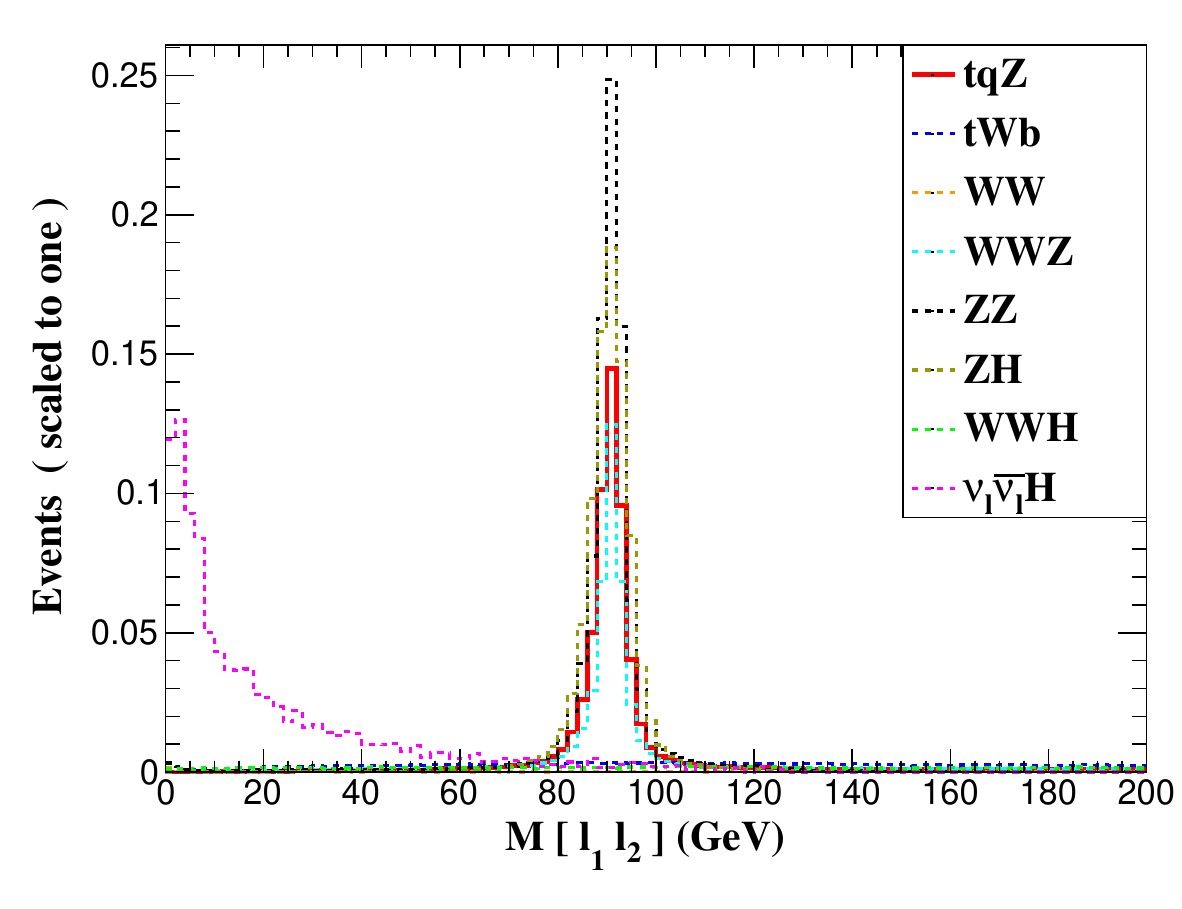}
  \hfill
  \includegraphics[width=0.48\textwidth]{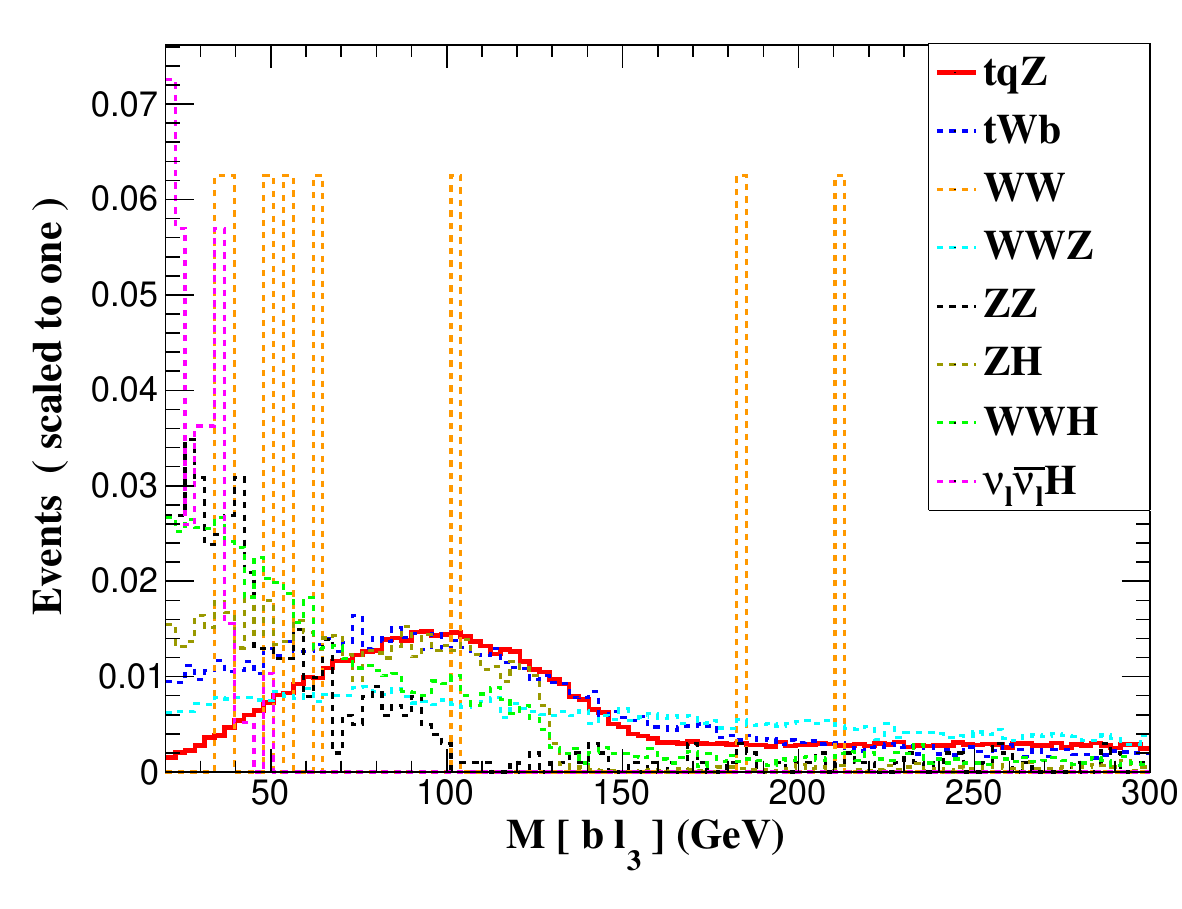}
  \caption{Normalized $M(\ell_1 \ \ell_2)$ and $M(b \ l_3)$ distributions for signal (solid red line) and backgrounds (dashed lines) with $\kappa_{tcZ}=0.5$  (LH) and $\sqrt{s}=3\,{\rm TeV}$ in Case A.}
\label{fig:caseA_yun}
\end{figure}

The production and decay processes of signal and background  for Case A are presented in Tab.~\ref{tab:caseA_processes}. Based on the event characteristics of the signal, we use the dilepton invariant mass $M(\ell_1 \ \ell_2)$ to reconstruct the $Z$ boson, which is employed as a kinematic selection variable. Since the remaining $WWZ$, $ZZ$ and $ZH$ backgrounds do not involve top quark decays, we partially reconstruct the invariant mass of the top quark $M(b \ \ell_3)$ to reject these backgrounds. The normalized distributions of the signal and backgrounds for the aforementioned kinematic variables at 3 TeV are displayed in Fig.~\ref{fig:caseA_yun} (the distribution behaviors at 10 and 14 TeV are analogous). Since the distributions of the LH and RH components for $tcZ$ and $tuZ$ are analogous, we only plot the LH component of $tcZ$ (red solid line) in Fig.~\ref{fig:caseA_yun}\footnote{Figs. 5--9 below are processed in the same way.}. According to the characteristics of the kinematic distributions, we summarize the cut selections at different collider energies in Tab.~\ref{tab:caseA_cuts}.

\begin{table}[htbp]
  \centering
  \caption{Processes for signal and backgrounds in Case A.}
  \label{tab:caseA_processes}
  \begin{ruledtabular}
  \begin{tabular}{ll}
    \hphantom{AAA}Case A & $W^+ \to \ell^+ \nu_\ell$,\quad $Z \to \ell^+ \ell^-$ \\
    \hline
    \hphantom{AAA}Signal & $\mu^+ \mu^- \to t \bar{q} Z$, $(t \to W^+ b,\ W^+ \to \ell^+ \nu_\ell),\ (Z \to \ell^+ \ell^-)$ \\
    \hphantom{AAA}Final states & $3 \ell + 1 j + 1 b + \slashed{E}$ \\
    \hline
    \hphantom{AAA}Backgrounds & $\mu^+ \mu^- \to \bar{t} W^+ b$, $(\bar{t} \to W^- \bar{b},\ W^- \to \ell^- \bar{\nu}_\ell),\ (W^+ \to \ell^+ \nu_\ell)$ \\
               & $\mu^+ \mu^- \to W^+ W^-$, $(W^+ \to \ell^+ \nu_\ell),\ (W^- \to \ell^- \bar{\nu}_\ell)$ \\
               & $\mu^+ \mu^- \to W^+ W^- Z$, $(W^+ \to \ell^+ \nu_\ell),\ (W^- \to \ell^- \bar{\nu}_\ell),\ (Z \to \ell^+ \ell^-)$ \\
               & $\mu^+ \mu^- \to Z Z$, $(Z \to j j),\ (Z \to \ell^+ \ell^-)$ \\
               & $\mu^+ \mu^- \to Z H$, $(Z \to \ell^+ \nu_\ell),\ (H \to b \bar{b})$ \\
               & $\mu^+ \mu^- \to W^+ W^- H$, $(W^+ \to \ell^+ \nu_\ell),\ (W^- \to \ell^- \bar{\nu}_\ell),\ (H \to b \bar{b})$ \\
               & $\mu^+ \mu^- \to \nu_\ell \bar{\nu}_\ell H$, $(H \to b \bar{b})$ \\
  \end{tabular}
  \end{ruledtabular}
\end{table}

\begin{table}[htbp]
  \centering
  \caption {Summary of cuts in Case A. }
  \vspace{-0.9em}
  \resizebox{1.0\textwidth}{!}{
  \begin{tabular}{l|ccc}
    \hline\hline
    \hphantom{A}Case A\hphantom{A} & $\sqrt{s} = 3$ TeV & $\sqrt{s} = 10$ TeV & $\sqrt{s} = 14$ TeV \\
    \hline
    \hphantom{A}Trigger & \hphantom{$N(b)=1, \ N(\ell)=3$} & $N(b)=1, \ N(\ell)=3$ & \hphantom{$N(b)=1, \ N(\ell)=3$} \\
    \hline
    \hphantom{A}Cut-1 & \hphantom{$80 \ \gev < M(\ell_1 \ \ell_2) < 100 \ \gev$} & $80 \ \gev < M(\ell_1 \ \ell_2) < 100 \ \gev$ & \hphantom{$80 \ \gev < M(\ell_1 \ \ell_2) < 100 \ \gev$} \\
    \hline
    \hphantom{A}Cut-2 & \hphantom{$50 \ \gev < M(b \ \ell_3) < 150 \ \gev$} & $50 \ \gev < M(b \ \ell_3) < 150 \ \gev$ & \hphantom{$50 \ \gev < M(b \ \ell_3) < 150 \ \gev$} \\
    \hline\hline
  \end{tabular}
  }
\label{tab:caseA_cuts}
\end{table}

We then recap the cut flow for the signals and backgrounds in Tab.~\ref{tab:caseA_all_cut_eff}. In what follows, we take the data at $\sqrt{s}=3\ \mathrm{TeV}$ as an example in all cases. As can be seen from the table, the majority of the $WW$ , $ZH$ and $\nu_{\ell}\bar{\nu_{\ell}}H$ backgrounds are removed by the trigger due to the lack of $b$-jets and a sufficient number of leptons. From the results of Cut-1, we can see that the $tWb$ background is effectively rejected by reconstructing the invariant mass of the $Z$ boson, from $8.79 \times 10^{-2} \ \text{fb}$ to $6.58 \times 10^{-3} \ \text{fb}$. For the same reason, the $WW$, $WWH$ and $\nu_{\ell}\bar{\nu_{\ell}}H$ backgrounds (thus, except for $ZZ$) are also significantly suppressed. The $ZZ$ background is however  rejected by Cut-2, with $1.02 \times 10^{-3} \ \text{fb}$ to $6.62 \times 10^{-4} \ \text{fb}$. In short, the proposed cut scheme effectively suppresses the backgrounds and enhances the signal significance.

\begin{table}[htbp]
  \centering
  \caption{Cut flow of the cross sections for the signals and backgrounds with $\kappa_{tcZ}=\kappa_{tuZ}=0.5$ in Case A.}
  \vspace{-0.5em}
  \footnotesize
  \setlength{\tabcolsep}{2.8pt}
  \resizebox{\textwidth}{!}{%
  \begin{tabular}{ l *{4}{c} c *{7}{c} }
    \hline\hline
    Process & \multicolumn{4}{c}{Signal (fb)} & & \multicolumn{7}{c}{Backgrounds (fb)} \\
    \cline{2-5} \cline{7-13}
            & $tcZ(LH)$ & $tcZ(RH)$ & $tuZ(LH)$ & $tuZ(RH)$ & & $tWb$ & $WW$ & $WWZ$ & $ZZ$ & $ZH$ & $WWH$ & $\nu_{\ell}\bar{\nu_{\ell}}H$ \\
    \hline
    \multicolumn{13}{c}{\boldmath$\sqrt{s}=3$ TeV} \\
    \hline
    Basic cuts & $1.95$ & $1.36$ & $1.95$ & $1.35$ & & $1.40$ & $19.89$ & $0.28$ & $0.44$ & $1.02 \times 10^{-3}$ & $3.46 \times 10^{-2}$ & $1.18 \times 10^{3}$ \\
    Trigger & $0.73$ & $0.47$ & $0.88$ & $0.54$ & & $8.79 \times 10^{-2}$ & $1.98 \times 10^{-3}$ & $2.80 \times 10^{-2}$ & $2.38 \times 10^{-3}$ & $5.00 \times 10^{-5}$ & $1.38 \times 10^{-3}$ & $1.65$ \\
    Cut-1 & $0.43$ & $0.25$ & $0.51$ & $0.29$ & & $6.58 \times 10^{-3}$ & $0$ & $6.15 \times 10^{-3}$ & $1.02 \times 10^{-3}$ & $2.03 \times 10^{-5}$ & $6.93 \times 10^{-6}$ & $0$ \\
    Cut-2 & $0.31$ & $0.16$ & $0.42$ & $0.23$ & & $4.90 \times 10^{-3}$ & $0$ & $1.03 \times 10^{-3}$ & $6.62 \times 10^{-4}$ & $1.72 \times 10^{-5}$ & $0$ & $0$ \\[5pt]
    Total Eff. & 16.00\% & 11.70\% & 21.50\% & 17.10\% & & 0.35\% & 0\% & 0.36\% & 0.15\% & 1.68\% & 0\% & 0\% \\[3pt]
    \hline
    \multicolumn{13}{c}{\boldmath$\sqrt{s}=10$ TeV} \\
    \hline
    Basic cuts & $0.15$ & $5.80 \times 10^{-2}$ & $0.14$ & $5.88 \times 10^{-2}$ & & $3.12 \times 10^{-2}$ & $1.83$ & $3.71 \times 10^{-3}$ & $3.46 \times 10^{-4}$ & $4.66 \times 10^{-8}$ & $1.08 \times 10^{-3}$ & $1.98 \times 10^{3}$ \\
    Trigger & $4.89 \times 10^{-2}$ & $1.85 \times 10^{-2}$ & $6.00 \times 10^{-2}$ & $2.26 \times 10^{-2}$ & & $1.80 \times 10^{-3}$ & $1.09 \times 10^{-4}$ & $3.56 \times 10^{-4}$ & $1.19 \times 10^{-6}$ & $2.33 \times 10^{-9}$ & $4.46 \times 10^{-5}$ & $1.59$ \\
    Cut-1 & $3.34 \times 10^{-2}$ & $1.27 \times 10^{-2}$ & $4.35 \times 10^{-2}$ & $1.63 \times 10^{-2}$ & & $5.30 \times 10^{-6}$ & $0$ & $6.45 \times 10^{-6}$ & $4.39 \times 10^{-7}$ & $6.71 \times 10^{-10}$ & $7.90 \times 10^{-8}$ & $0.20$ \\
    Cut-2 & $1.73 \times 10^{-2}$ & $5.28 \times 10^{-3}$ & $2.58 \times 10^{-2}$ & $8.73 \times 10^{-3}$ & & $1.76 \times 10^{-6}$ & $0$ & $1.12 \times 10^{-6}$ & $2.63 \times 10^{-7}$ & $6.71 \times 10^{-10}$ & $0$ & $0$ \\[5pt]
    Total Eff. & 11.90\% & 9.10\% & 17.80\% & 14.90\% & & 0.01\% & 0\% & 0.03\% & 0.08\% & 1.44\% & 0\% & 0\% \\[3pt]
    \hline
    \multicolumn{13}{c}{\boldmath$\sqrt{s}=14$ TeV} \\
    \hline
    Basic cuts & $4.66 \times 10^{-2}$ & $1.16 \times 10^{-2}$ & $4.63 \times 10^{-2}$ & $1.16 \times 10^{-2}$ & & $1.16 \times 10^{-2}$ & $0.94$ & $7.11 \times 10^{-4}$ & $2.92 \times 10^{-5}$ & $2.17 \times 10^{-9}$ & $4.03 \times 10^{-4}$ & $2.20 \times 10^{3}$ \\
    Trigger & $1.57 \times 10^{-2}$ & $3.55 \times 10^{-3}$ & $1.84 \times 10^{-2}$ & $3.94 \times 10^{-3}$ & & $5.91 \times 10^{-4}$ & $4.70 \times 10^{-5}$ & $5.86 \times 10^{-5}$ & $7.85 \times 10^{-8}$ & $1.16 \times 10^{-10}$ & $1.54 \times 10^{-5}$ & $1.52$ \\
    Cut-1 & $1.17 \times 10^{-2}$ & $2.60 \times 10^{-3}$ & $1.44 \times 10^{-2}$ & $3.02 \times 10^{-3}$ & & $1.05 \times 10^{-6}$ & $0$ & $6.21 \times 10^{-7}$ & $1.96 \times 10^{-8}$ & $3.60 \times 10^{-11}$ & $3.11 \times 10^{-8}$ & $0$ \\
    Cut-2 & $4.57 \times 10^{-3}$ & $6.14 \times 10^{-4}$ & $7.43 \times 10^{-3}$ & $1.21 \times 10^{-3}$ & & $1.05 \times 10^{-6}$ & $0$ & $6.21 \times 10^{-8}$ & $7.85 \times 10^{-9}$ & $3.60 \times 10^{-11}$ & $0$ & $0$ \\[5pt]
    Total Eff. & 9.79\% & 5.29\% & 16.00\% & 10.50\% & & 0.01\% & 0\% & 0.01\% & 0.03\% & 1.66\% & 0\% & 0\% \\
    \hline\hline
  \end{tabular}%
  }
  \label{tab:caseA_all_cut_eff}
\end{table}

\subsection{Selection Cuts for Case B \textnormal{\small{($W^+ \to jj$ , $Z \to jj$)}}}

\begin{figure}[htbp]
  \centering
  \includegraphics[width=0.49\textwidth]{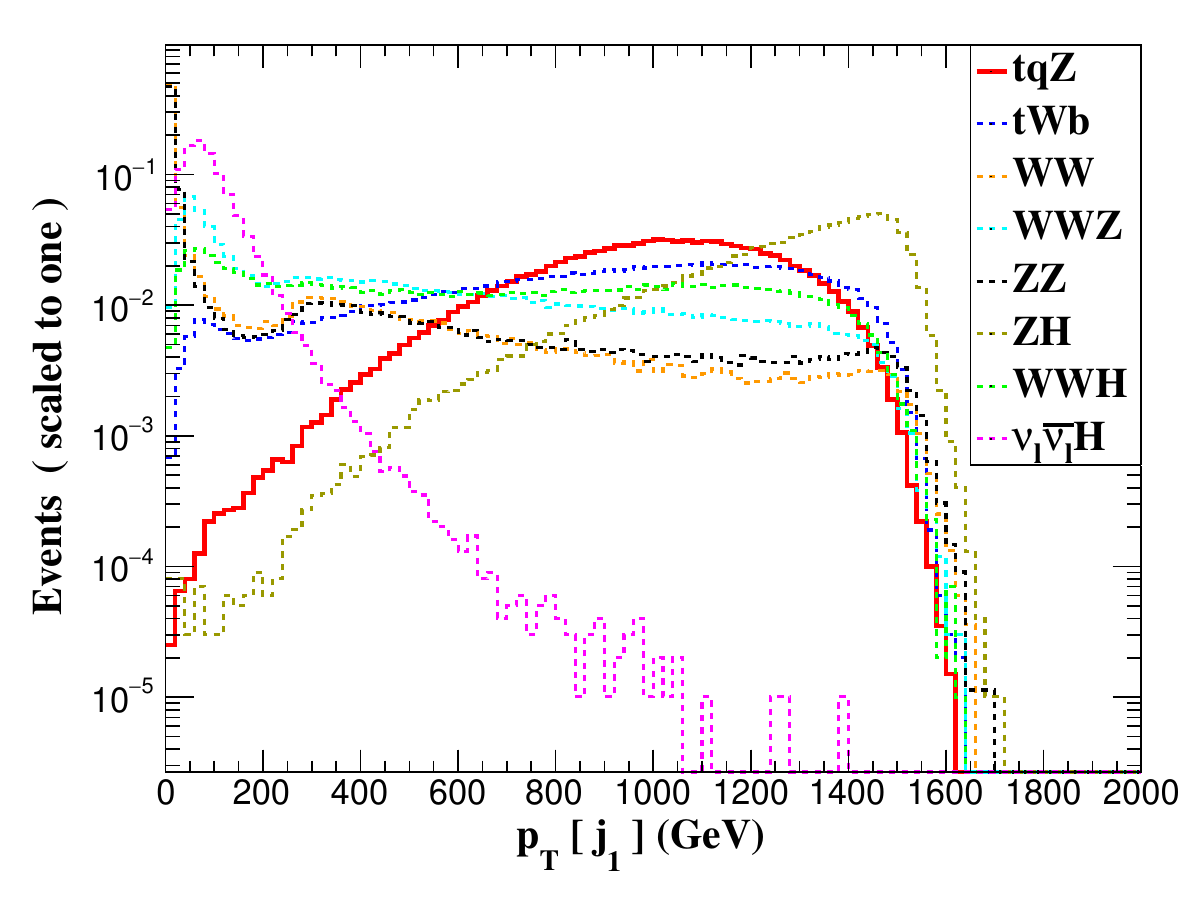}
  \hfill
  \includegraphics[width=0.49\textwidth]{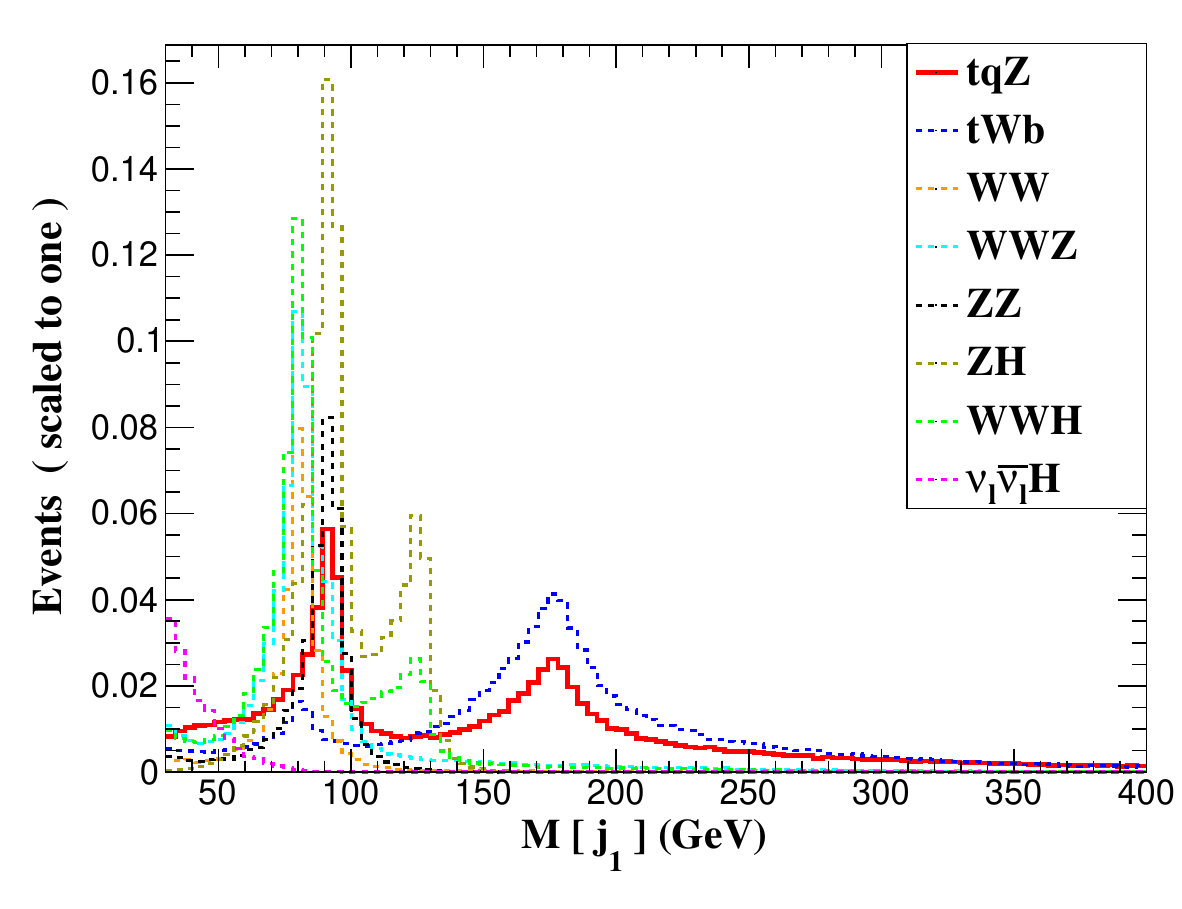}
  \vspace{1em}
  \includegraphics[width=0.49\textwidth]{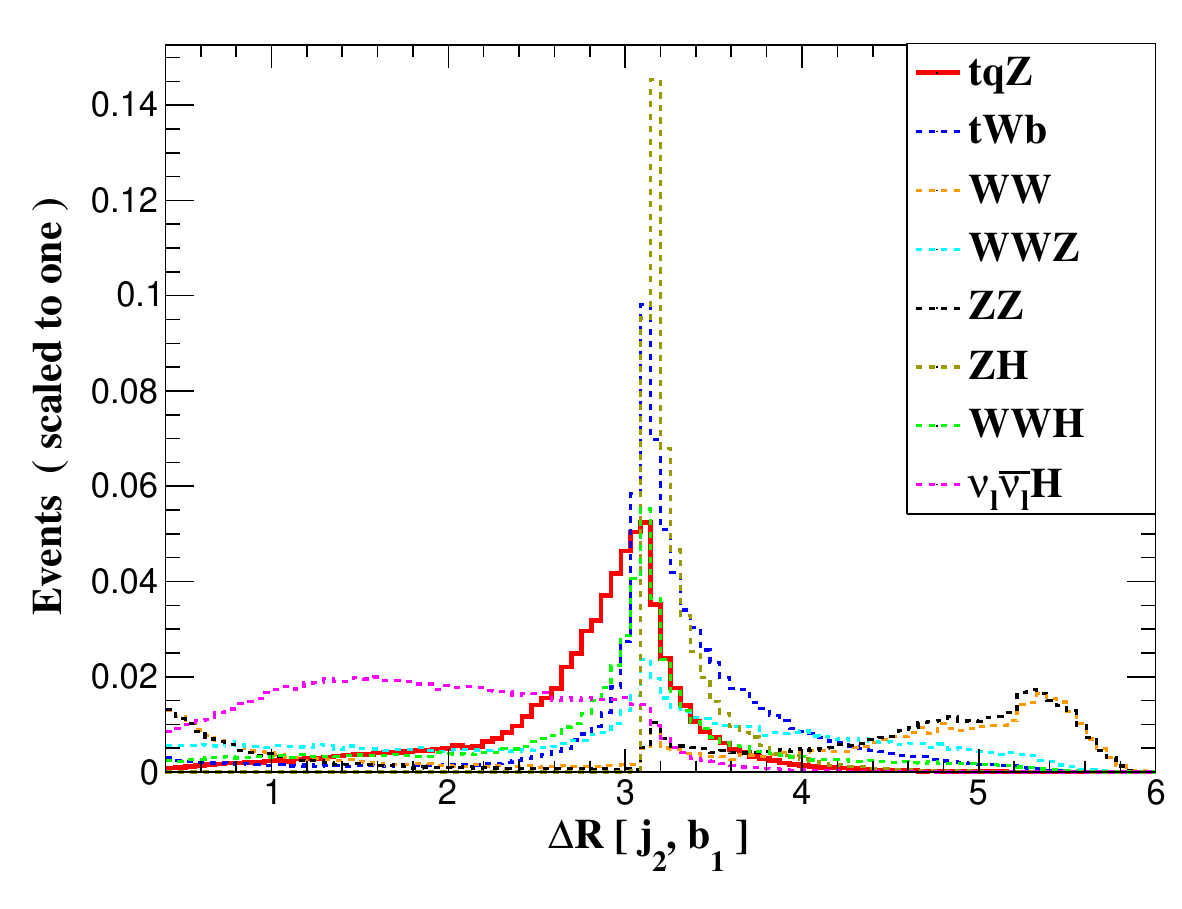}
  \caption{Normalized $p_T^{j_1}$, $M(j_1)$ and $\Delta R(j_2,b_1)$ distributions for signal (solid red line) and backgrounds (dashed lines) with $\kappa_{tcZ}=0.5$ (LH) and $\sqrt{s}=3\,{\rm TeV}$ in Case B.}
\label{fig:caseB_yun}
\end{figure}

The decay processes of signal and background events for Case B are presented in Tab.~\ref{tab:processes_caseB}. As shown in Tab.~\ref{tab:processes_caseB}, top quark decays do not appear in any backgrounds except the $tWb$ process. Therefore, we aim to reconstruct the signal of the top quark by employing the fat jet technique. First, we perform preliminary cuts on the transverse momentum of the leading fat jet. Next, we use $M(j_1)$ to reconstruct the invariant mass of the top quark, where $j_1$ denotes an un-tagged fat jet. To further suppress the $tWb$ background, we investigate $\Delta R(j_2, b_1)$ between the fat jet $j_2$ from the $Z$ boson and the b-jet from the top quark. In Fig.~\ref{fig:caseB_yun}, we plot the differential distributions for the signal and backgrounds at 3 TeV. Based on the characteristic features of these kinematic distributions, we list the selection criteria for different collision energies in Tab.~\ref{tab:caseB_cuts}.

We summarize the cut flow for the signals and backgrounds in Tab.~\ref{tab:caseB_all_cut_eff}. As discussed in our previous analysis, Cut-2 performs well in rejecting all backgrounds except the $tWb$ background. As an example, the $WW$ and $ZH$ cross sections drop from $11.31 \ \text{fb}$ and $8.57\times 10^{-3} \ \text{fb}$ to $2.85\times 10^{-2} \ \text{fb}$ and $4.77\times 10^{-5} \ \text{fb}$, respectively. From Fig.~\ref{fig:caseB_yun}, we can see that the $\Delta R(j_2, b_1)$ distributions for $tqZ$ (red solid) and $tWb$ (black dashed) are different, with the $tqZ$ shifting to lower values. We take $\Delta R(j_2, b_1) < 3$ as Cut-3 and find that this cut can retains roughly 60\% of the $tuZ$ and $tcZ$ signals, whereas only 30\% of the $tWb$ background.

\begin{table}[htbp]
  \centering
  \caption{Processes for signal and backgrounds in Case B.}
  \label{tab:processes_caseB}
  \begin{ruledtabular}
  \begin{tabular}{ll}
    \hphantom{AAA}Case B & $W^+ \to jj$,\quad $Z \to jj$, \\
    \hline
    \hphantom{AAA}Signal & $\mu^+ \mu^- \to t \bar{q} Z$, $(t \to W^+ b,\ W^+ \to jj),\ (Z \to jj)$ \\
    \hphantom{AAA}Final states & $5 j + 1 b$ \\
    \hline
    \hphantom{AAA}Backgrounds & $\mu^+ \mu^- \to \bar{t} W^+ b$, $(\bar{t} \to W^- \bar{b},\ W^- \to jj),\ (W^+ \to jj)$ \\
               & $\mu^+ \mu^- \to W^+ W^-$, $(W^+ \to jj),\ (W^- \to jj)$ \\
               & $\mu^+ \mu^- \to W^+ W^- Z$, $(W^+ \to jj),\ (W^- \to jj),\ (Z \to jj)$ \\
               & $\mu^+ \mu^- \to Z Z$, $(Z \to jj),\ (Z \to jj)$ \\
               & $\mu^+ \mu^- \to Z H$, $(Z \to jj),\ (H \to b \bar{b})$ \\
               & $\mu^+ \mu^- \to W^+ W^- H$, $(W^+ \to jj),\ (W^- \to jj),\ (H \to b \bar{b})$ \\
               & $\mu^+ \mu^- \to \nu_\ell \bar{\nu}_\ell H$, $(H \to b \bar{b})$ \\
  \end{tabular}
  \end{ruledtabular}
\end{table}

\begin{table}[htbp]
  \centering
  \caption {Summary of cuts in Case B.}
  \vspace{-0.9em}
  \resizebox{1.0\textwidth}{!}{
  \begin{tabular}{l|ccc}
    \hline\hline
    \hphantom{A}Case B\hphantom{A} & $\sqrt{s} = 3$ TeV & $\sqrt{s} = 10$ TeV & $\sqrt{s} = 14$ TeV \\
    \hline
    \hphantom{A}Trigger & \hphantom{$1 \leq N(b) \leq 2, \ N(\ell)=0$} & $1 \leq N(b) \leq 2, \ N(\ell)=0$ & \hphantom{$1 \leq N(b) \leq 2, \ N(\ell)=0$} \\
    \hline
    \hphantom{A}Cut-1 & $p_{T}^{j_1} > 600 \ \gev$ \hphantom{1111} \vline & $p_{T}^{j_1} > 2000 \ \gev$ \hphantom{1111} \vline & $p_{T}^{j_1} > 3000 \ \gev$ \\
    \hline
    \hphantom{A}Cut-2 & \hphantom{$140 \ \gev < M(j_1) < 200 \ \gev$} & $140 \ \gev < M(j_1) < 200 \ \gev$ & \hphantom{$140 \ \gev < M(j_1) < 200 \ \gev$} \\
    \hline
    \hphantom{A}Cut-3 & \hphantom{$\Delta R(j_2,b_1) < 3$} & $\Delta R(j_2,b_1) < 3$ & \hphantom{$\Delta R(j_2,b_1) < 3$} \\
    \hline\hline
  \end{tabular}
  }
\label{tab:caseB_cuts}
\end{table}

\begin{table}[htbp]
  \centering
  \caption{Cut flow of the cross sections for the signals and backgrounds with $\kappa_{tcZ}=\kappa_{tuZ}=0.5$ in Case B.}
  \vspace{-0.5em}
  \footnotesize
  \setlength{\tabcolsep}{2.8pt}
  \resizebox{\textwidth}{!}{%
  \begin{tabular}{ l *{4}{c} c *{7}{c} }
    \hline\hline
    Process & \multicolumn{4}{c}{Signal (fb)} & & \multicolumn{7}{c}{Backgrounds (fb)} \\
    \cline{2-5} \cline{7-13}
            & $tcZ(LH)$ & $tcZ(RH)$ & $tuZ(LH)$ & $tuZ(RH)$ & & $tWb$ & $WW$ & $WWZ$ & $ZZ$ & $ZH$ & $WWH$ & $\nu_{\ell}\bar{\nu_{\ell}}H$ \\
    \hline
    \multicolumn{13}{c}{\boldmath$\sqrt{s}=3$ TeV} \\
    \hline
    Basic cuts & $14.07$ & $12.06$ & $13.94$ & $12.02$ & & $5.04$ & $2.85 \times 10^{2}$ & $6.85$ & $13.52$ & $1.09 \times 10^{-2}$ & $0.25$ & $1.18 \times 10^{3}$ \\
    Trigger & $9.79$ & $8.11$ & $10.06$ & $8.21$ & & $2.62$ & $90.41$ & $2.95$ & $5.23$ & $8.69 \times 10^{-3}$ & $0.17$ & $8.41 \times 10^{2}$ \\
    Cut-1 & $9.25$ & $7.60$ & $9.54$ & $7.82$ & & $1.90$ & $11.31$ & $1.19$ & $1.04$ & $8.57 \times 10^{-3}$ & $0.10$ & $0.83$ \\
    Cut-2 & $2.67$ & $2.16$ & $2.46$ & $2.10$ & & $0.91$ & $2.85 \times 10^{-2}$ & $4.88 \times 10^{-2}$ & $0$ & $4.77 \times 10^{-5}$ & $3.63 \times 10^{-3}$ & $0$ \\
    Cut-3 & $1.64$ & $1.31$ & $1.60$ & $1.36$ & & $0.31$ & $0$ & $2.61 \times 10^{-2}$ & $0$ & $4.34 \times 10^{-6}$ & $1.93 \times 10^{-3}$ & $0$ \\[5pt]
    Total Eff. & 11.70\% & 10.90\% & 11.50\% & 11.30\% & & 6.21\% & 0\% & 0.38\% & 0\% & 0.04\% & 0.78\% & 0\% \\[3pt]
    \hline
    \multicolumn{13}{c}{\boldmath$\sqrt{s}=10$ TeV} \\
    \hline
    Basic cuts & $0.94$ & $0.81$ & $0.94$ & $0.80$ & & $0.24$ & $10.30$ & $0.79$ & $0.58$ & $7.23 \times 10^{-6}$ & $1.64 \times 10^{-2}$ & $1.98 \times 10^{3}$ \\
    Trigger & $0.68$ & $0.58$ & $0.72$ & $0.60$ & & $0.12$ & $1.37$ & $0.32$ & $9.38 \times 10^{-2}$ & $5.89 \times 10^{-6}$ & $1.04 \times 10^{-2}$ & $1.29 \times 10^{3}$ \\
    Cut-1 & $0.64$ & $0.55$ & $0.68$ & $0.57$ & & $8.12 \times 10^{-2}$ & $0.36$ & $0.11$ & $3.15 \times 10^{-2}$ & $5.79 \times 10^{-6}$ & $6.16 \times 10^{-3}$ & $0.20$ \\
    Cut-2 & $0.10$ & $9.11 \times 10^{-2}$ & $0.11$ & $9.49 \times 10^{-2}$ & & $1.89 \times 10^{-2}$ & $1.03 \times 10^{-2}$ & $4.48 \times 10^{-3}$ & $1.28 \times 10^{-3}$ & $1.60 \times 10^{-7}$ & $1.85 \times 10^{-4}$ & $0$ \\
    Cut-3 & $6.59 \times 10^{-2}$ & $5.80 \times 10^{-2}$ & $7.63 \times 10^{-2}$ & $6.48 \times 10^{-2}$ & & $5.30 \times 10^{-3}$ & $7.19 \times 10^{-3}$ & $2.28 \times 10^{-3}$ & $6.38 \times 10^{-4}$ & $2.82 \times 10^{-8}$ & $7.88 \times 10^{-5}$ & $0$ \\[5pt]
    Total Eff. & 7.02\% & 7.18\% & 8.11\% & 8.06\% & & 2.23\% & 0.07\% & 0.29\% & 0.11\% & 0.39\% & 0.48\% & 0\% \\[3pt]
    \hline
    \multicolumn{13}{c}{\boldmath$\sqrt{s}=14$ TeV} \\
    \hline
    Basic cuts & $0.44$ & $0.37$ & $0.44$ & $0.37$ & & $8.10 \times 10^{-2}$ & $3.93$ & $0.31$ & $0.20$ & $9.53 \times 10^{-7}$ & $5.32 \times 10^{-3}$ & $2.20 \times 10^{3}$ \\
    Trigger & $0.31$ & $0.27$ & $0.33$ & $0.28$ & & $3.91 \times 10^{-2}$ & $0.42$ & $0.12$ & $2.60 \times 10^{-2}$ & $7.74 \times 10^{-7}$ & $3.35 \times 10^{-3}$ & $1.38 \times 10^{3}$ \\
    Cut-1 & $0.29$ & $0.25$ & $0.31$ & $0.26$ & & $2.43 \times 10^{-2}$ & $0.11$ & $3.75 \times 10^{-2}$ & $9.81 \times 10^{-3}$ & $7.57 \times 10^{-7}$ & $2.03 \times 10^{-3}$ & $0$ \\
    Cut-2 & $4.73 \times 10^{-2}$ & $3.90 \times 10^{-2}$ & $4.58 \times 10^{-2}$ & $4.05 \times 10^{-2}$ & & $5.49 \times 10^{-3}$ & $1.25 \times 10^{-2}$ & $3.25 \times 10^{-3}$ & $1.06 \times 10^{-3}$ & $4.37 \times 10^{-8}$ & $1.06 \times 10^{-4}$ & $0$ \\
    Cut-3 & $3.14 \times 10^{-2}$ & $2.58 \times 10^{-2}$ & $3.22 \times 10^{-2}$ & $2.98 \times 10^{-2}$ & & $1.55 \times 10^{-3}$ & $7.84 \times 10^{-3}$ & $2.10 \times 10^{-3}$ & $6.60 \times 10^{-4}$ & $9.44 \times 10^{-9}$ & $5.53 \times 10^{-5}$ & $0$ \\[5pt]
    Total Eff. & 7.22\% & 6.89\% & 7.39\% & 7.96\% & & 1.91\% & 0.20\% & 0.68\% & 0.33\% & 0.99\% & 1.04\% & 0\% \\
    \hline\hline
  \end{tabular}%
  }
  \label{tab:caseB_all_cut_eff}
\end{table}

\subsection{Selection Cuts for Case C \textnormal{\small{($W^+ \to \ell^+ \nu_\ell$ , $Z \to jj$)}}}

\begin{figure}[htbp]
  \centering
  \includegraphics[width=0.48\textwidth]{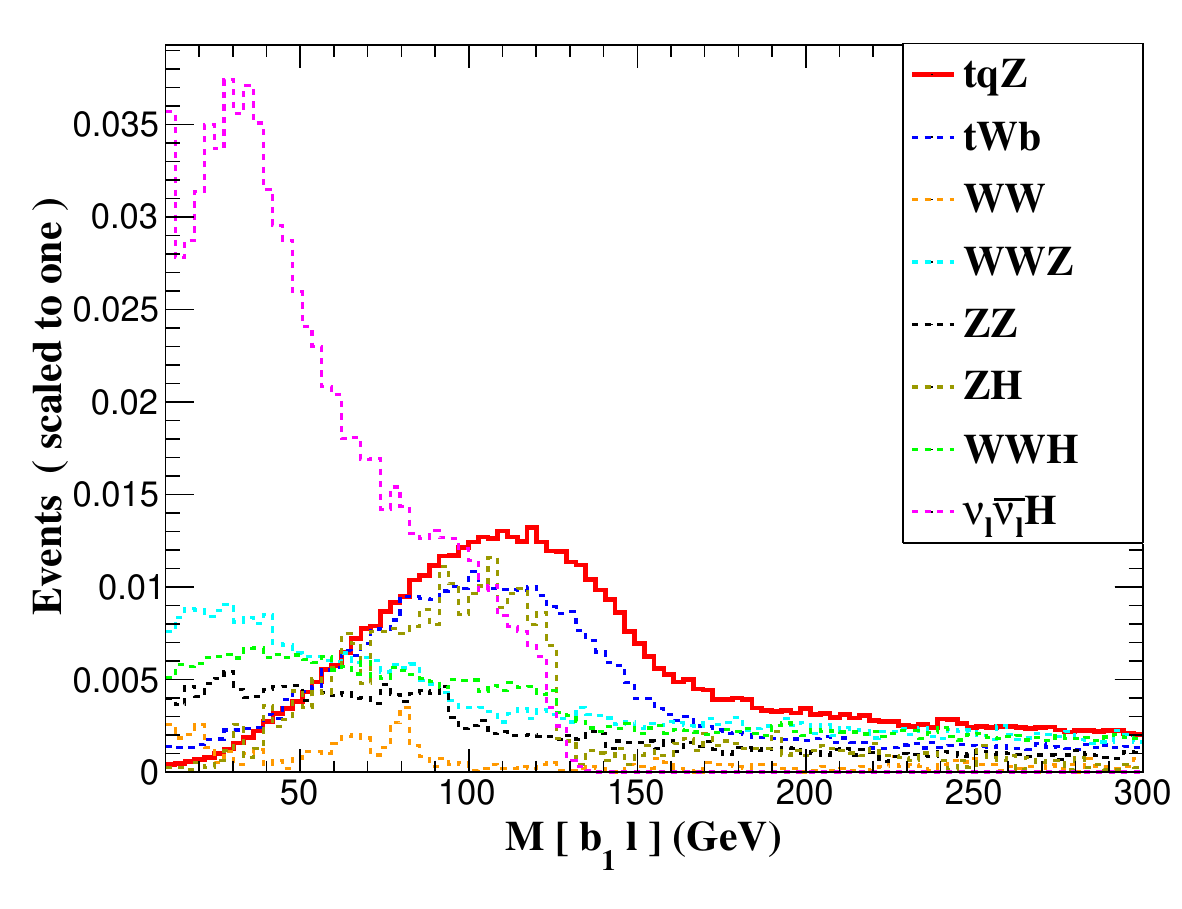}
  \hfill
  \includegraphics[width=0.48\textwidth]{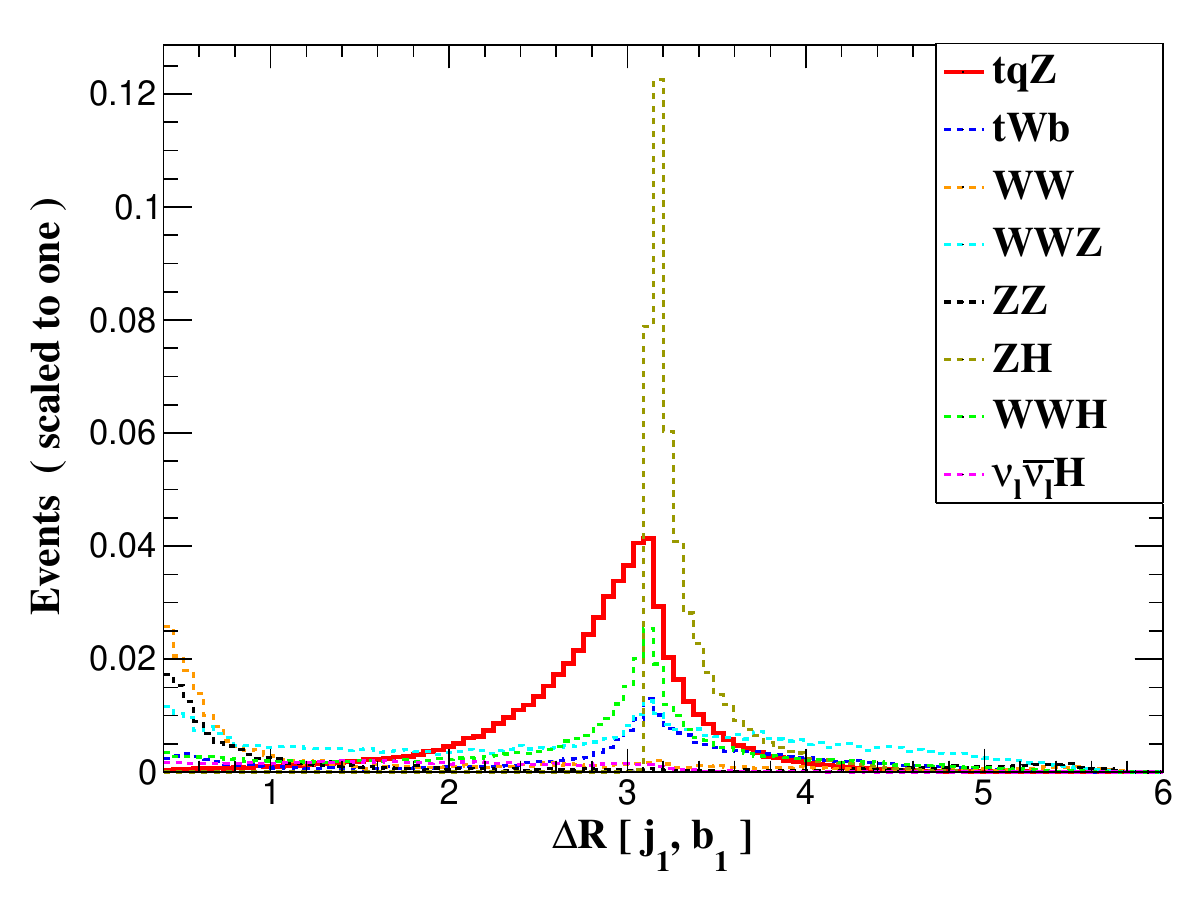}
  \caption{Normalized $M(b_1 \ \ell)$ and $\Delta R(j_1,b_1)$ distributions for signal (solid red line) and backgrounds (dashed lines) with $\kappa_{tcZ}=0.5$ \ (LH) and $\sqrt{s}=3\,{\rm TeV}$ in Case C.}
\label{fig:caseC_yun}
\end{figure}

The decay processes of the signal and backgrounds for Case C are presented in Tab.~\ref{tab:processes_caseC}. As shown in Tab.~\ref{tab:processes_caseC}, none of the backgrounds involve top quark decays except for the $tWb$. Therefore, we partially reconstruct the invariant mass of the top quark $M(b_1 \ \ell)$, considering the missing transverse energy ($\slashed{E}_T$) caused by the neutrino, to reject all backgrounds other than $tWb$. In addition, the jet formed by the light quark $q$ in the signal events exhibits a significant spatial correlation with the b-jet from the top quark decay, and the rapidity-azimuthal angle distance $\Delta R(j_1,b_1)$ between the two jets is concentrated in a narrow range. Thus, $\Delta R(j_1,b_1)$ is adopted to further reject the $tWb$. The normalized distributions of these kinematic variables for the signal and backgrounds at 3 TeV are illustrated in Fig.~\ref{fig:caseC_yun} Based on the above analysis, we summarize the selection cuts for different collision energies in Tab.~\ref{tab:caseC_cuts}.

We summarize the cut flow for the signals and backgrounds in Tab. \ref{tab:caseC_all_cut_eff}. Due to a mismatch in the number of b-jets and leptons, $WW$, $ZZ$, $ZH$ and $\nu_{\ell}\bar{\nu_{\ell}}H$ backgrounds are effectively suppressed by the trigger. We can see that their surviving cross sections are less than 10\% of the original values. Since the presence of missing transverse energy ($\slashed{E}_T$) shifts the peak position of the invariant mass $M(b_1 \ \ell)$ away from $175\ \text{GeV}$, we define Cut-1 as $50\ \text{GeV}<M(b_1 \ \ell)<150\ \text{GeV}$. We use Cut-1 to substantially suppress the $WW$, $WWZ$, $WWH$, and $\nu_\ell\bar{\nu}_\ell H$ backgrounds, whose cross sections decrease from $4.26\ \text{fb}$, $0.98\ \text{fb}$, $5.36\times10^{-2}\ \text{fb}$ and $1.43\times10^{2}\ \text{fb}$ to $3.84\times10^{-2}\ \text{fb}$, $0.12\ \text{fb}$, $6.05\times10^{-3}\ \text{fb}$ and $55.09\ \text{fb}$, respectively. Based on the $\Delta R(j_1,b_1)$ shown in Fig.~\ref{fig:caseC_yun}, we impose Cut-2 as $2 < \Delta R(j_1,b_1) < 4$ to retain a large fraction of signal events. After Cut-2, approximately 80\% of the $tuZ$ and $tcZ$ signal survive, while only around 45\% for the $tWb$ background.

\begin{table}[htbp]
  \centering
  \caption{Processes for signal and backgrounds in Case C.}
  \label{tab:processes_caseC}
  \begin{ruledtabular}
  \begin{tabular}{ll}
    \hphantom{AAA}Case C & $W^+ \to \ell^+ \nu_\ell$,\quad $Z \to jj$ \\
    \hline
    \hphantom{AAA}Signal & $\mu^+ \mu^- \to t \bar{q} Z$, $(t \to W^+ b,\ W^+ \to \ell^+ \nu_\ell),\ (Z \to jj)$ \\
    \hphantom{AAA}Final states & $1 \ell + 3 j + 1 b + \slashed{E}$ \\
    \hline
    \hphantom{AAA}Backgrounds & $\mu^+ \mu^- \to \bar{t} W^+ b$, $(\bar{t} \to W^- \bar{b},\ W^- \to \ell^- \bar{\nu}_\ell),\ (W^+ \to jj)$ \\
               & $\mu^+ \mu^- \to W^+ W^-$, $(W^+ \to \ell^+ \nu_\ell),\ (W^- \to jj)$ \\
               & $\mu^+ \mu^- \to W^+ W^- Z$, $(W^+ \to \ell^+ \nu_\ell),\ (W^- \to jj),\ (Z \to jj)$ \\
               & $\mu^+ \mu^- \to Z Z$, $(Z \to jj),\ (Z \to \ell^+ \ell^-)$ \\
               & $\mu^+ \mu^- \to Z H$, $(Z \to jj),\ (H \to b \bar{b})$ \\
               & $\mu^+ \mu^- \to W^+ W^- H$, $(W^+ \to \ell^+ \nu_\ell),\ (W^- \to jj),\ (H \to b \bar{b})$ \\
               & $\mu^+ \mu^- \to \nu_\ell \bar{\nu}_\ell H$, $(H \to b \bar{b})$ \\
  \end{tabular}
  \end{ruledtabular}
\end{table}

\begin{table}[htbp]
  \centering
  \caption {Summary of cuts in Case C. }
  \vspace{-0.9em} 
  \resizebox{1.0\textwidth}{!}{
  \begin{tabular}{l|ccc}
    \hline\hline
    \hphantom{A}Case C\hphantom{A} & $\sqrt{s} = 3$ TeV & $\sqrt{s} = 10$ TeV & $\sqrt{s} = 14$ TeV \\
    \hline
    \hphantom{A}Trigger & \hphantom{$1 \leq N(b) \leq 2, \ N(\ell)=1$} & $1 \leq N(b) \leq 2, \ N(\ell)=1$ & \hphantom{$1 \leq N(b) \leq 2, \ N(\ell)=1$} \\
    \hline
    \hphantom{A}Cut-1 & \hphantom{$50\gev < M(b_1 \ \ell) < 150\gev$} & $50 \ \gev < M(b_1 \ \ell) < 150 \ \gev$ & \hphantom{$50\gev < M(b_1 \ \ell) < 150\gev$} \\
    \hline
    \hphantom{A}Cut-2 & \hphantom{$2 < \Delta R(j_1,b_1) < 4$} & $2 < \Delta R(j_1,b_1) < 4$ & \hphantom{$2 < \Delta R(j_1,b_1) < 4$} \\
    \hline\hline
  \end{tabular}
  }
\label{tab:caseC_cuts}
\end{table}

\begin{table}[htbp]
  \centering
  \caption{Cut flow of the cross sections for the signals and backgrounds with $\kappa_{tcZ}=\kappa_{tuZ}=0.5$ in Case C.}
  \vspace{-0.5em}
  \footnotesize
  \setlength{\tabcolsep}{2.8pt}
  \resizebox{\textwidth}{!}{%
  \begin{tabular}{ l *{4}{c} c *{7}{c} }
    \hline\hline
    Process & \multicolumn{4}{c}{Signal (fb)} & & \multicolumn{7}{c}{Backgrounds (fb)} \\
    \cline{2-5} \cline{7-13}
            & $tcZ(LH)$ & $tcZ(RH)$ & $tuZ(LH)$ & $tuZ(RH)$ & & $tWb$ & $WW$ & $WWZ$ & $ZZ$ & $ZH$ & $WWH$ & $\nu_{\ell}\bar{\nu_{\ell}}H$ \\
    \hline
    \multicolumn{13}{c}{\boldmath$\sqrt{s}=3$ TeV} \\
    \hline
    Basic cuts & $20.94$ & $14.67$ & $20.99$ & $14.72$ & & $2.62$ & $48.00$ & $3.68$ & $0.44$ & $1.09 \times 10^{-2}$ & $0.15$ & $1.18 \times 10^{3}$ \\
    Trigger & $11.28$ & $7.62$ & $12.00$ & $7.81$ & & $0.68$ & $4.26$ & $0.98$ & $4.28 \times 10^{-2}$ & $9.45 \times 10^{-4}$ & $5.36 \times 10^{-2}$ & $1.43 \times 10^{2}$ \\
    Cut-1 & $6.16$ & $3.62$ & $8.00$ & $4.76$ & & $0.22$ & $3.84 \times 10^{-2}$ & $0.12$ & $1.02 \times 10^{-2}$ & $4.60 \times 10^{-4}$ & $6.05 \times 10^{-3}$ & $55.09$ \\
    Cut-2 & $4.74$ & $3.07$ & $6.51$ & $4.10$ & & $0.11$ & $0$ & $3.89 \times 10^{-2}$ & $3.08 \times 10^{-4}$ & $2.94 \times 10^{-4}$ & $2.87 \times 10^{-3}$ & $0.14$ \\[5pt]
    Total Eff. & 22.60\% & 20.90\% & 31.00\% & 27.80\% & & 4.06\% & 0\% & 1.06\% & 0.07\% & 2.71\% & 1.90\% & 0.01\% \\[3pt]
    \hline
    \multicolumn{13}{c}{\boldmath$\sqrt{s}=10$ TeV} \\
    \hline
    Basic cuts & $3.29$ & $1.61$ & $3.29$ & $1.62$ & & $9.70 \times 10^{-2}$ & $0.22$ & $0.16$ & $3.46 \times 10^{-4}$ & $7.23 \times 10^{-6}$ & $3.72 \times 10^{-3}$ & $1.98 \times 10^{3}$ \\
    Trigger & $1.62$ & $0.76$ & $1.86$ & $0.86$ & & $2.32 \times 10^{-2}$ & $1.71 \times 10^{-2}$ & $4.09 \times 10^{-2}$ & $1.20 \times 10^{-5}$ & $7.07 \times 10^{-7}$ & $1.01 \times 10^{-3}$ & $3.51 \times 10^{2}$ \\
    Cut-1 & $0.83$ & $0.27$ & $1.24$ & $0.47$ & & $1.19 \times 10^{-2}$ & $1.51 \times 10^{-4}$ & $4.46 \times 10^{-3}$ & $9.79 \times 10^{-7}$ & $9.90 \times 10^{-8}$ & $1.88 \times 10^{-4}$ & $1.30 \times 10^{2}$ \\
    Cut-2 & $0.55$ & $0.17$ & $0.95$ & $0.34$ & & $2.85 \times 10^{-3}$ & $1.89 \times 10^{-5}$ & $1.17 \times 10^{-3}$ & $0$ & $5.75 \times 10^{-8}$ & $8.63 \times 10^{-5}$ & $6.48 \times 10^{-2}$ \\[5pt]
    Total Eff. & 16.80\% & 10.70\% & 29.00\% & 21.10\% & & 2.94\% & 0.01\% & 0.74\% & 0\% & 0.80\% & 2.32\% & $3.27 \times 10^{-3}$\% \\[3pt]
    \hline
    \multicolumn{13}{c}{\boldmath$\sqrt{s}=14$ TeV} \\
    \hline
    Basic cuts & $1.46$ & $0.62$ & $1.46$ & $0.61$ & & $3.11 \times 10^{-2}$ & $3.20 \times 10^{-2}$ & $4.96 \times 10^{-2}$ & $2.92 \times 10^{-5}$ & $9.53 \times 10^{-7}$ & $8.74 \times 10^{-4}$ & $2.20 \times 10^{3}$ \\
    Trigger & $0.72$ & $0.29$ & $0.81$ & $0.32$ & & $7.43 \times 10^{-3}$ & $1.76 \times 10^{-3}$ & $1.20 \times 10^{-2}$ & $9.50 \times 10^{-7}$ & $8.38 \times 10^{-8}$ & $2.12 \times 10^{-4}$ & $3.49 \times 10^{2}$ \\
    Cut-1 & $0.36$ & $0.12$ & $0.52$ & $0.18$ & & $4.21 \times 10^{-3}$ & $2.17 \times 10^{-5}$ & $1.31 \times 10^{-3}$ & $7.98 \times 10^{-8}$ & $1.47 \times 10^{-8}$ & $4.09 \times 10^{-5}$ & $1.34 \times 10^{2}$ \\
    Cut-2 & $0.23$ & $8.10 \times 10^{-2}$ & $0.38$ & $0.13$ & & $8.93 \times 10^{-4}$ & $0$ & $3.09 \times 10^{-4}$ & $3.80 \times 10^{-9}$ & $8.23 \times 10^{-9}$ & $1.44 \times 10^{-5}$ & $0$ \\[5pt]
    Total Eff. & 15.90\% & 13.20\% & 26.20\% & 21.10\% & & 2.88\% & 0\% & 0.62\% & 0.01\% & 0.86\% & 1.65\% & 0\% \\
    \hline\hline
  \end{tabular}%
  }
  \label{tab:caseC_all_cut_eff}
\end{table}

\subsection{Selection Cuts for Case D \textnormal{\small{($W^+ \to jj$ , $Z \to \ell^+ \ell^-$)}}}

\begin{figure}[htbp]
  \centering
  \includegraphics[width=0.48\textwidth]{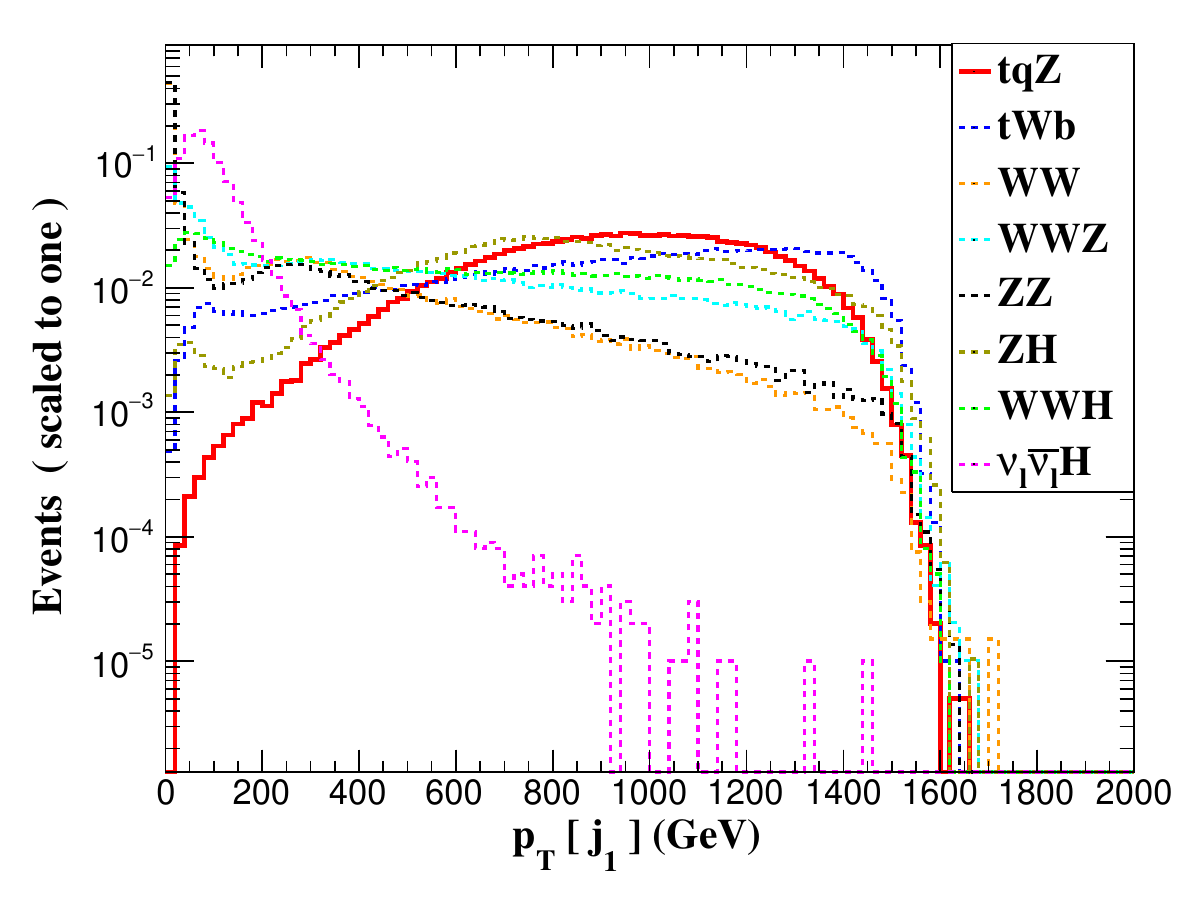}
  \hfill
  \includegraphics[width=0.48\textwidth]{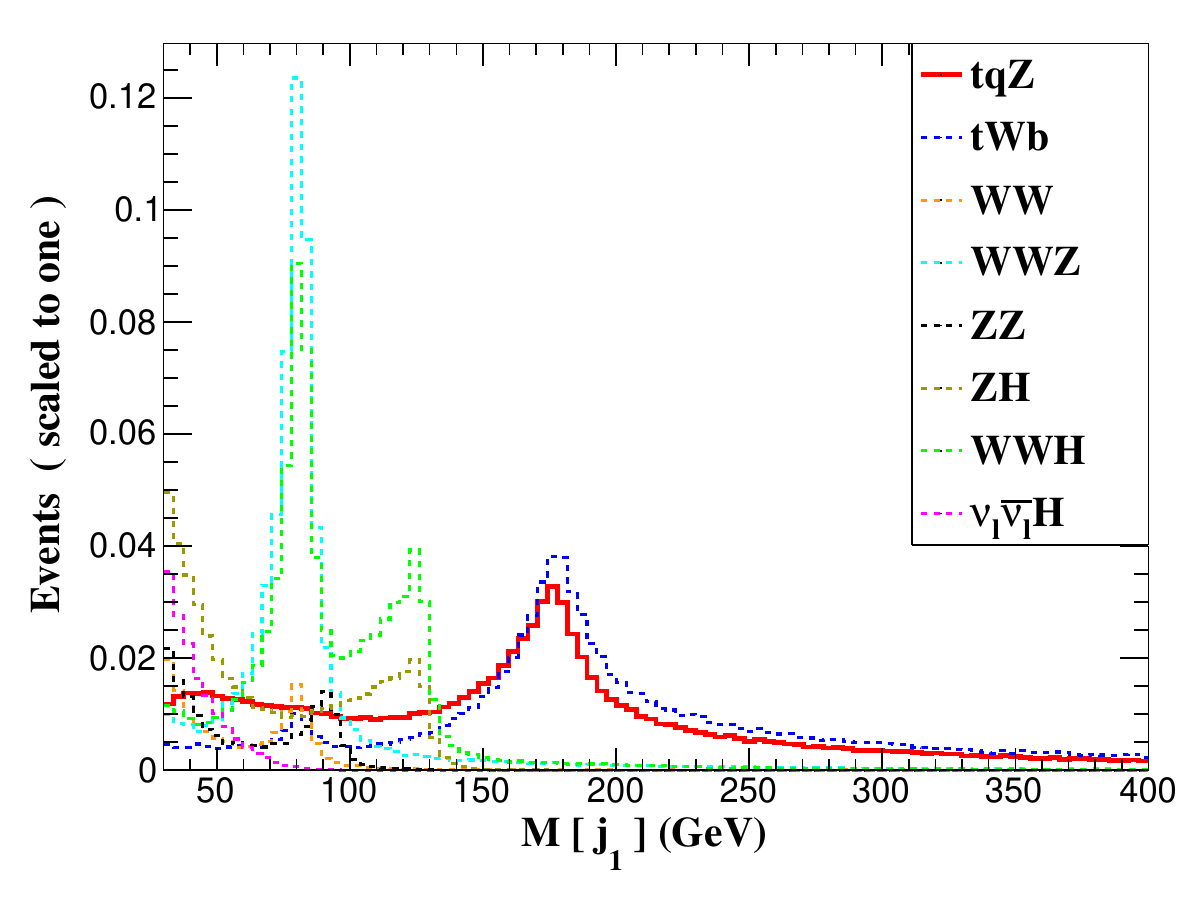}
  \vspace{1em}
  \includegraphics[width=0.49\textwidth]{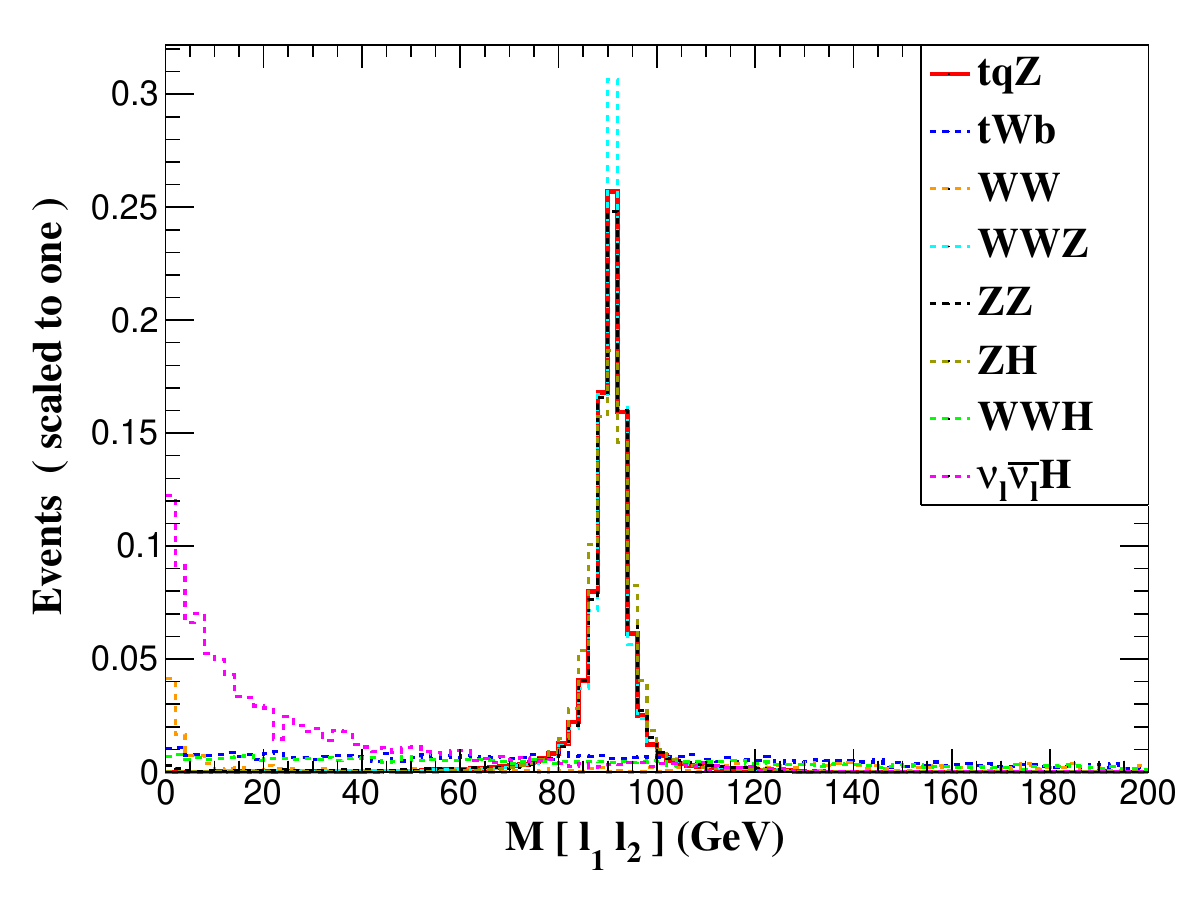}
  \caption{Normalized $p_T^{j_1}$, $M(j_1)$ and $M(\ell_1 \ \ell_2)$ distributions for signal (solid red line) and backgrounds (dashed lines) with $\kappa_{tcZ}=0.5$  (LH) and $\sqrt{s}=3\,{\rm TeV}$ in Case D.}
\label{fig:caseD_yun}
\end{figure}

The decay processes of the signal and background for Case D are presented in Tab.~\ref{tab:processes_caseD}. Since all backgrounds except for the $tWb$ do not contain top quark decays, we adopt the fat jet method to reconstruct the top quark invariant mass to reject these backgrounds. First, we perform a preliminary selection on the transverse momentum of the leading fat jet. Subsequently, we use $M(j_1)$ to reconstruct the invariant mass of the top quark. Since the $tWb$ background does not involve $Z$ boson decays, we reconstruct the invariant mass of the $Z$ boson via $M(\ell_1 \ \ell_2)$ to specifically suppress the $tWb$ background. The normalized distributions of these kinematic variables for the signal and backgrounds at 3 TeV are presented in Fig.\ref{fig:caseD_yun}. Based on these kinematic features, we list the selection schemes at different collision energies in Tab.\ref{tab:caseD_cuts} to improve the signal significance.

\begin{table}[htbp]
  \centering
  \caption{Processes for signal and backgrounds in Case D.}
  \label{tab:processes_caseD}
  \begin{ruledtabular}
  \begin{tabular}{ll}
    \hphantom{AAA}Case D & $W^+ \to jj$,\quad $Z \to \ell^+ \ell^-$ \\
    \hline
    \hphantom{AAA}Signal & $\mu^+ \mu^- \to t \bar{q} Z$, $(t \to W^+ b,\ W^+ \to jj),\ (Z \to \ell^+ \ell^-)$ \\
    \hphantom{AAA}Final states & $2 \ell + 3 j + 1 b $ \\
    \hline
    \hphantom{AAA}Backgrounds & $\mu^+ \mu^- \to \bar{t} W^+ b$, $(\bar{t} \to W^- \bar{b},\ W^- \to jj),\ (W^+ \to \ell^+ \nu_\ell)$ \\
               & $\mu^+ \mu^- \to W^+ W^-$, $(W^+ \to jj),\ (W^- \to \ell^- \bar{\nu}_\ell)$ \\
               & $\mu^+ \mu^- \to W^+ W^- Z$, $(W^+ \to jj),\ (W^- \to jj),\ (Z \to \ell^+ \ell^-)$ \\
               & $\mu^+ \mu^- \to Z Z$, $(Z \to jj),\ (Z \to \ell^+ \ell^-)$ \\
               & $\mu^+ \mu^- \to Z H$, $(Z \to \ell^+ \ell^-),\ (H \to b \bar{b})$ \\
               & $\mu^+ \mu^- \to W^+ W^- H$, $(W^+ \to \ell^+ \nu_\ell),\ (W^- \to jj),\ (H \to b \bar{b})$ \\
               & $\mu^+ \mu^- \to \nu_\ell \bar{\nu}_\ell H$, $(H \to b \bar{b})$ \\
  \end{tabular}
  \end{ruledtabular}
\end{table}

We summarize the cut flow for the signals and backgrounds in Tab.\ref{tab:caseD_all_cut_eff}. As expected from the kinematic analysis, Cut-2 exhibits significant suppression performance for all backgrounds except $tWb$. The $WW$ and $ZZ$ backgrounds are completely suppressed to zero, while the $WWZ$, $ZH$, and $WWH$ cross sections are reduced from $4.03\times 10^{-2}\ \text{fb}$, $1.64\times 10^{-4}\ \text{fb}$ and $3.26\times 10^{-3}\ \text{fb}$ to $1.26\times 10^{-3}\ \text{fb}$, $8.17\times 10^{-7}\ \text{fb}$, and $1.21\times 10^{-4}\ \text{fb}$, respectively. Meanwhile, we find that Cut-3 provides strong discrimination against the $tWb$ background. After Cut-3, around 90\% of the $tuZ$ and $tcZ$ signal cross sections remain, while only 15\% of the $tWb$ survives.

\begin{table}[htbp]
  \centering
  \caption {Summary of cuts in Case D.}
  \vspace{-0.9em}
  \resizebox{1.0\textwidth}{!}{
  \begin{tabular}{l|ccc}
    \hline\hline
    \hphantom{A}Case D\hphantom{A} & $\sqrt{s} = 3$ TeV & $\sqrt{s} = 10$ TeV & $\sqrt{s} = 14$ TeV \\
    \hline
    \hphantom{A}Trigger & \hphantom{$N(b)=1, \ N(\ell)=2$} & $N(b)=1, \ N(\ell)=2$ & \hphantom{$N(b)=1, \ N(\ell)=2$} \\
    \hline
    \hphantom{A}Cut-1 & $p_{T}^{j_1} > 600 \ \gev$ \hphantom{1111} \vline & $p_{T}^{j_1} > 2000 \ \gev$ \hphantom{1111} \vline & $p_{T}^{j_1} > 3000 \ \gev$ \\
    \hline
    \hphantom{A}Cut-2 & \hphantom{$140 \ \gev < M(j_1) < 200 \ \gev$} & $140 \ \gev < M(j_1) < 200 \ \gev$ & \hphantom{$140 \ \gev < M(j_1) < 200 \ \gev$} \\
    \hline
    \hphantom{A}Cut-3 & \hphantom{$80 \ \gev < M(\ell_1 \ \ell_2) < 100 \ \gev$} & $80 \ \gev < M(\ell_1 \ \ell_2) < 100 \ \gev$ & \hphantom{$80 \ \gev < M(\ell_1 \ \ell_2) < 100 \ \gev$} \\
    \hline\hline
  \end{tabular}
  }
\label{tab:caseD_cuts}
\end{table}

\begin{table}[htbp]
  \centering
  \caption{Cut flow of the cross sections for the signals and backgrounds with $\kappa_{tcZ}=\kappa_{tuZ}=0.5$ in Case D.}
  \vspace{-0.5em}
  \footnotesize
  \setlength{\tabcolsep}{2.8pt}
  \resizebox{\textwidth}{!}{%
  \begin{tabular}{ l *{4}{c} c *{7}{c} }
    \hline\hline
    Process & \multicolumn{4}{c}{Signal (fb)} & & \multicolumn{7}{c}{Backgrounds (fb)} \\
    \cline{2-5} \cline{7-13}
            & $tcZ(LH)$ & $tcZ(RH)$ & $tuZ(LH)$ & $tuZ(RH)$ & & $tWb$ & $WW$ & $WWZ$ & $ZZ$ & $ZH$ & $WWH$ & $\nu_{\ell}\bar{\nu_{\ell}}H$ \\
    \hline
    \multicolumn{13}{c}{\boldmath$\sqrt{s}=3$ TeV} \\
    \hline
    Basic cuts & $1.15$ & $0.98$ & $1.16$ & $0.97$ & & $7.53$ & $48.00$ & $0.52$ & $0.44$ & $1.02 \times 10^{-3}$ & $0.15$ & $1.18 \times 10^{3}$ \\
    Trigger & $0.48$ & $0.40$ & $0.60$ & $0.48$ & & $0.28$ & $0.15$ & $9.13 \times 10^{-2}$ & $2.29 \times 10^{-2}$ & $1.95 \times 10^{-4}$ & $7.85 \times 10^{-3}$ & $19.59$ \\
    Cut-1 & $0.44$ & $0.36$ & $0.56$ & $0.45$ & & $0.19$ & $1.44 \times 10^{-2}$ & $4.03 \times 10^{-2}$ & $7.64 \times 10^{-3}$ & $1.64 \times 10^{-4}$ & $3.26 \times 10^{-3}$ & $0$ \\
    Cut-2 & $0.16$ & $0.13$ & $0.19$ & $0.16$ & & $7.23 \times 10^{-2}$ & $0$ & $1.26 \times 10^{-3}$ & $0$ & $8.17 \times 10^{-7}$ & $1.21 \times 10^{-4}$ & $0$ \\
    Cut-3 & $0.15$ & $0.12$ & $0.18$ & $0.14$ & & $1.13 \times 10^{-2}$ & $0$ & $1.15 \times 10^{-3}$ & $0$ & $6.13 \times 10^{-7}$ & $0$ & $0$ \\[5pt]
    Total Eff. & 12.80\% & 12.50\% & 15.20\% & 14.60\% & & 0.15\% & 0\% & 0.22\% & 0\% & 0.06\% & 0\% & 0\% \\[3pt]
    \hline
    \multicolumn{13}{c}{\boldmath$\sqrt{s}=10$ TeV} \\
    \hline
    Basic cuts & $9.40 \times 10^{-3}$ & $8.00 \times 10^{-3}$ & $9.39 \times 10^{-3}$ & $7.97 \times 10^{-3}$ & & $0.52$ & $0.22$ & $3.81 \times 10^{-2}$ & $3.46 \times 10^{-4}$ & $4.66 \times 10^{-8}$ & $3.72 \times 10^{-3}$ & $1.98 \times 10^{3}$ \\
    Trigger & $3.66 \times 10^{-3}$ & $2.98 \times 10^{-3}$ & $4.63 \times 10^{-3}$ & $3.80 \times 10^{-3}$ & & $4.03 \times 10^{-3}$ & $8.70 \times 10^{-4}$ & $6.40 \times 10^{-3}$ & $1.36 \times 10^{-5}$ & $8.25 \times 10^{-9}$ & $2.53 \times 10^{-4}$ & $26.26$ \\
    Cut-1 & $3.31 \times 10^{-3}$ & $2.71 \times 10^{-3}$ & $4.23 \times 10^{-3}$ & $3.48 \times 10^{-3}$ & & $2.36 \times 10^{-3}$ & $1.34 \times 10^{-4}$ & $2.05 \times 10^{-3}$ & $1.67 \times 10^{-6}$ & $3.24 \times 10^{-9}$ & $1.08 \times 10^{-4}$ & $0$ \\
    Cut-2 & $7.21 \times 10^{-4}$ & $5.41 \times 10^{-4}$ & $8.38 \times 10^{-4}$ & $6.78 \times 10^{-4}$ & & $5.18 \times 10^{-4}$ & $0$ & $8.69 \times 10^{-5}$ & $0$ & $2.31 \times 10^{-10}$ & $1.61 \times 10^{-6}$ & $0$ \\
    Cut-3 & $6.56 \times 10^{-4}$ & $4.87 \times 10^{-4}$ & $7.87 \times 10^{-4}$ & $5.83 \times 10^{-4}$ & & $1.79 \times 10^{-5}$ & $0$ & $8.07 \times 10^{-5}$ & $0$ & $7.70 \times 10^{-11}$ & $0$ & $0$ \\[5pt]
    Total Eff. & 6.98\% & 6.10\% & 8.38\% & 7.32\% & & $3.47 \times 10^{-3}$\% & 0\% & 0.21\% & 0\% & 0.17\% & 0\% & 0\% \\[3pt]
    \hline
    \multicolumn{13}{c}{\boldmath$\sqrt{s}=14$ TeV} \\
    \hline
    Basic cuts & $2.12 \times 10^{-3}$ & $1.82 \times 10^{-3}$ & $2.14 \times 10^{-3}$ & $1.83 \times 10^{-3}$ & & $0.26$ & $3.20 \times 10^{-2}$ & $1.31 \times 10^{-2}$ & $2.92 \times 10^{-5}$ & $2.17 \times 10^{-9}$ & $8.74 \times 10^{-4}$ & $2.20 \times 10^{3}$ \\
    Trigger & $7.14 \times 10^{-4}$ & $5.38 \times 10^{-4}$ & $9.57 \times 10^{-4}$ & $8.31 \times 10^{-4}$ & & $1.59 \times 10^{-3}$ & $7.89 \times 10^{-5}$ & $1.90 \times 10^{-3}$ & $9.85 \times 10^{-7}$ & $3.57 \times 10^{-10}$ & $5.59 \times 10^{-5}$ & $27.93$ \\
    Cut-1 & $6.25 \times 10^{-4}$ & $4.80 \times 10^{-4}$ & $8.71 \times 10^{-4}$ & $7.49 \times 10^{-4}$ & & $7.74 \times 10^{-4}$ & $1.22 \times 10^{-5}$ & $5.40 \times 10^{-4}$ & $5.88 \times 10^{-8}$ & $1.28 \times 10^{-10}$ & $2.50 \times 10^{-5}$ & $0$ \\
    Cut-2 & $1.09 \times 10^{-4}$ & $9.83 \times 10^{-5}$ & $1.63 \times 10^{-4}$ & $1.54 \times 10^{-4}$ & & $2.00 \times 10^{-4}$ & $0$ & $3.61 \times 10^{-5}$ & $7.35 \times 10^{-9}$ & $9.15 \times 10^{-12}$ & $1.51 \times 10^{-6}$ & $0$ \\
    Cut-3 & $9.19 \times 10^{-5}$ & $8.85 \times 10^{-5}$ & $1.48 \times 10^{-4}$ & $1.41 \times 10^{-4}$ & & $9.12 \times 10^{-6}$ & $0$ & $3.38 \times 10^{-5}$ & $7.35 \times 10^{-9}$ & $4.58 \times 10^{-12}$ & $0$ & $0$ \\[5pt]
    Total Eff. & 4.33\% & 4.86\% & 6.92\% & 7.70\% & & $3.49 \times 10^{-3}$\% & 0\% & 0.26\% & 0.03\% & 0.21\% & 0\% & 0\% \\
    \hline\hline
  \end{tabular}%
  }
  \label{tab:caseD_all_cut_eff}
\end{table}

\subsection{Selection Cuts for Case E \textnormal{\small{($W^+ \to \ell^+ \nu_\ell$ , $Z \to \nu_\ell \bar{\nu}_\ell$)}}}

\begin{figure}[htbp]
\vspace{4mm}
  \centering
  \includegraphics[width=0.48\textwidth]{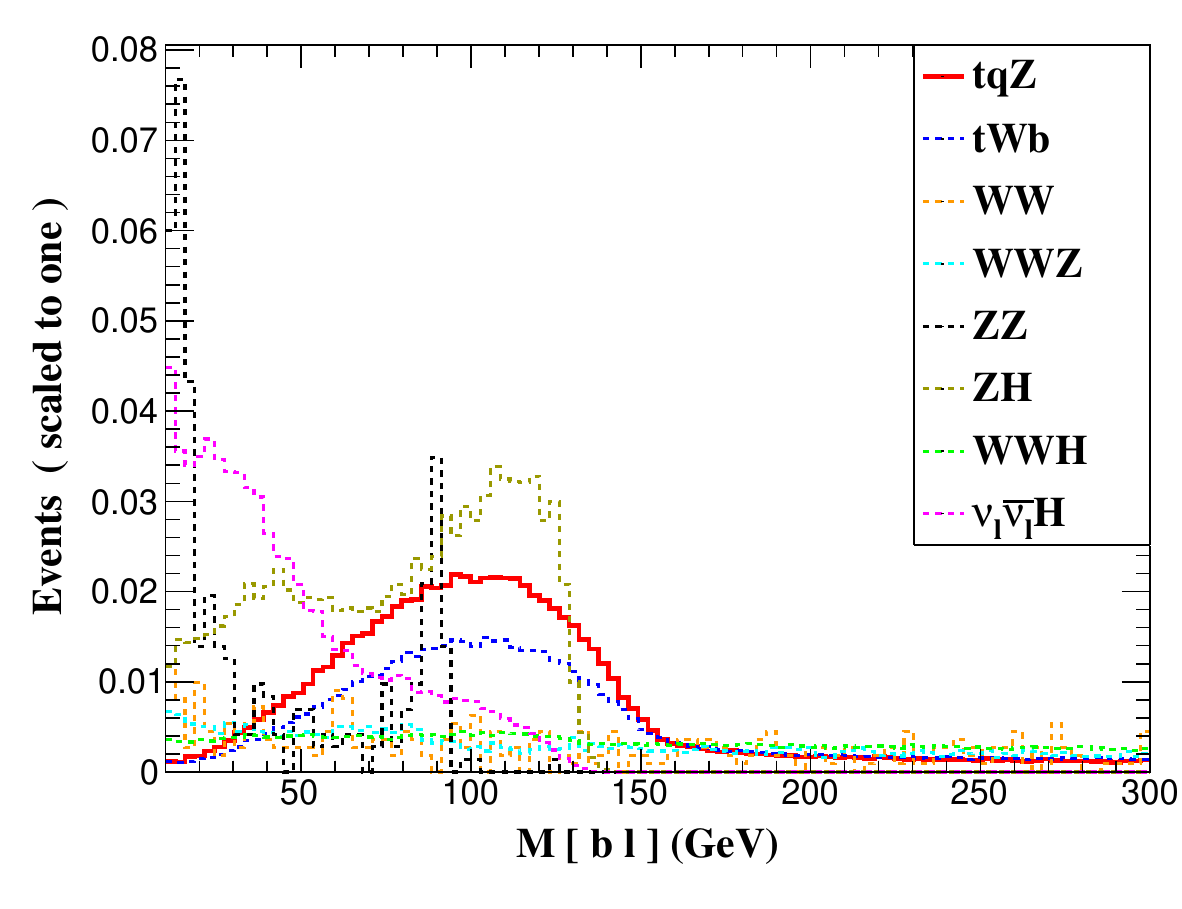}
  \hfill
  \includegraphics[width=0.48\textwidth]{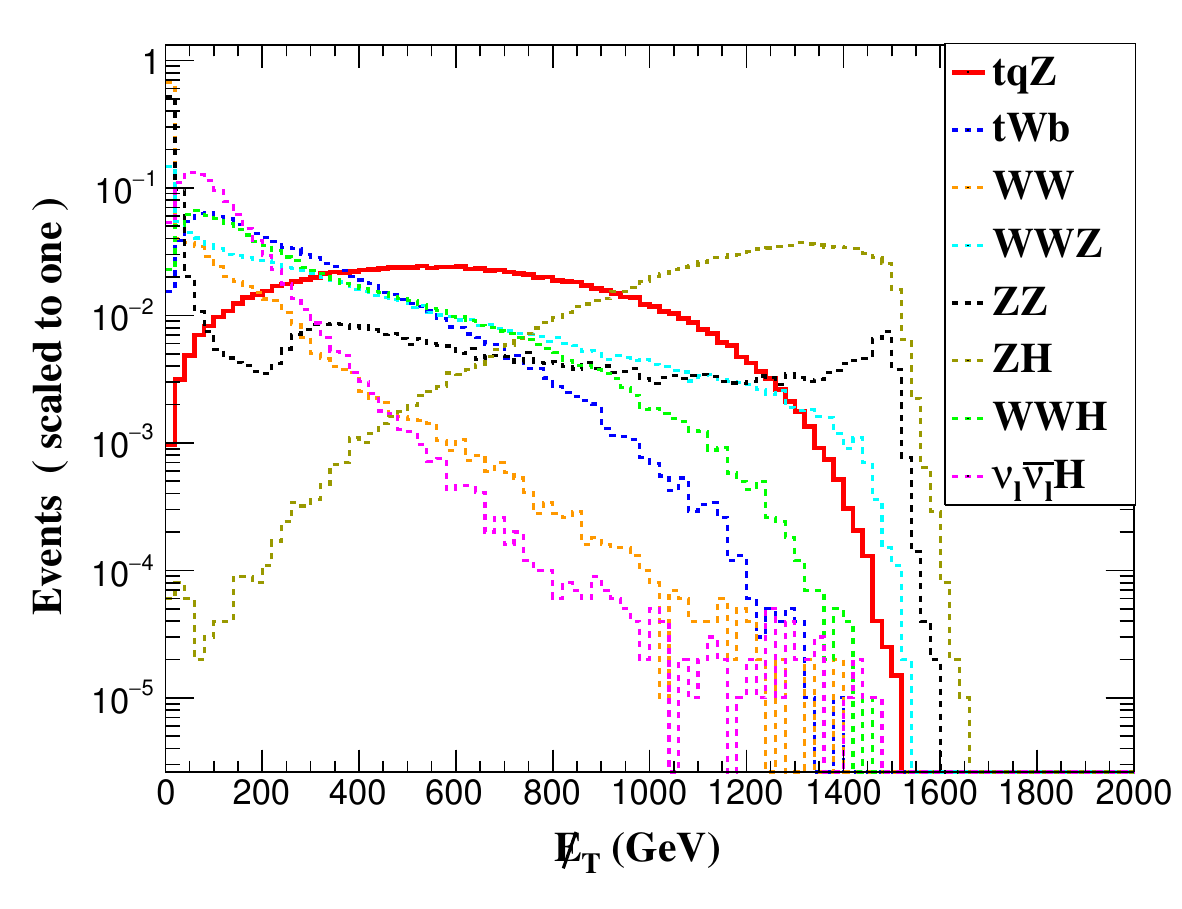}
  \caption{Normalized $M(b \ \ell)$ and $\slashed{E}_T$ distributions for signal (solid red line) and backgrounds (dashed lines) with $\kappa_{tcZ}=0.5$ (LH) and $\sqrt{s}=3\,{\rm TeV}$ in Case E.}
\label{fig:caseE_yun}
\vspace{4mm}
\end{figure}

The decay processes of the signal and background for Case E are presented in Tab.~\ref{tab:processes_caseE}. Based on the characteristics of the signal and various backgrounds, we choose $M(b \ \ell)$ and the missing transverse energy $\slashed{E}_T$ as the selection criteria. Among these variables, $M(b \ \ell)$ is adopted to partially reconstruct the invariant mass of the top quark to suppress all backgrounds except for $tWb$. The $\slashed{E}_T$ one is selected because neutrinos arising from the decays of the $Z$ and $W$ bosons can produce a rather large missing transverse momentum/energy. The normalized distributions of the above kinematic variables for the signal and all backgrounds at 3 TeV are displayed in Fig.~\ref{fig:caseE_yun}. Based on the features of each kinematic distribution, we provide the event selection schemes at different collider energies in Tab.~\ref{tab:caseE_cuts}, for the purpose of enhancing the signal significance.

\begin{table}[htbp]
\vspace{4mm}
  \centering
  \caption{Processes for signal and backgrounds in Case E.}
  \label{tab:processes_caseE}
  \begin{ruledtabular}
  \begin{tabular}{ll}
    \hphantom{AAA}Case E & $W^+ \to \ell^+ \nu_\ell$,\quad $Z \to \nu_\ell \bar{\nu}_\ell$ \\
    \hline
    \hphantom{AAA}Signal & $\mu^+ \mu^- \to t \bar{q} Z$, $(t \to W^+ b,\ W^+ \to \ell^+ \nu_\ell),\ (Z \to \nu_\ell \bar{\nu}_\ell)$ \\
    \hphantom{AAA}Final states & $1 \ell + 1 j + 1 b + \slashed{E}$ \\
    \hline
    \hphantom{AAA}Backgrounds & $\mu^+ \mu^- \to \bar{t} W^+ b$, $(\bar{t} \to W^- \bar{b},\ W^- \to \ell^- \bar{\nu}_\ell),\ (W^+ \to \ell^+ \nu_\ell)$ \\
               & $\mu^+ \mu^- \to W^+ W^-$, $(W^+ \to \ell^+ \nu_\ell),\ (W^- \to \ell^- \bar{\nu}_\ell)$ \\
               & $\mu^+ \mu^- \to W^+ W^- Z$, $(W^+ \to \ell^+ \nu_\ell),\ (W^- \to jj),\ (Z \to \nu_\ell \bar{\nu}_\ell)$ \\
               & $\mu^+ \mu^- \to Z Z$, $(Z \to \ell^+ \ell^-),\ (Z \to \nu_\ell \bar{\nu}_\ell)$ \\
               & $\mu^+ \mu^- \to Z H$, $(Z \to \nu_\ell \bar{\nu}_\ell),\ (H \to b \bar{b})$ \\
               & $\mu^+ \mu^- \to W^+ W^- H$, $(W^+ \to \ell^+ \nu_\ell),\ (W^- \to \ell^- \bar{\nu}_\ell),\ (H \to b \bar{b})$ \\
               & $\mu^+ \mu^- \to \nu_\ell \bar{\nu}_\ell H$, $(H \to b \bar{b})$ \\
  \end{tabular}
  \end{ruledtabular}
\vspace{4mm}
\end{table}

We summarize the cut flow for the signals and backgrounds in Tab.\ref{tab:caseE_all_cut_eff}. Owing to a mismatch in the number of b-jets and leptons, we effectively suppress the backgrounds of $tWb$, $WW$, $ZZ$, $ZH$ and $\nu_{\ell}\bar{\nu_{\ell}}H$ using the trigger. Their surviving cross sections are less than 10\% of the original values. Since missing transverse energy ($\slashed{E}_T$) arises from top quark decays and shifts the peak of $M(b_1 \ \ell)$ away from $175\ \text{GeV}$, Cut-1 is chosen as $50\ \text{GeV}<M(b_1 \ \ell)<150\ \text{GeV}$. Cut-1 can effectively reject the $WW$, $WWZ$, $WWH$ and $\nu_{\ell}\bar{\nu_{\ell}}H$ backgrounds, whose cross sections drop from $7.36 \times 10^{-2} \ \text{fb}$, $0.28 \ \text{fb}$, $3.66 \times 10^{-3} \ \text{fb}$ and $1.16 \times 10^{2} \ \text{fb}$ to $5.97 \times 10^{-3} \ \text{fb}$, $3.38 \times 10^{-2} \ \text{fb}$, $5.49 \times 10^{-4} \ \text{fb}$ and $45.46 \ \text{fb}$, respectively. Cut-2 further suppresses the $tWb$ and $\nu_{\ell}\bar{\nu_{\ell}}H$ backgrounds, whose cross sections decrease from $3.99\times 10^{-2}\ \text{fb}$ to $3.51\times 10^{-3}\ \text{fb}$ and from $45.46\ \text{fb}$ to $0.17\ \text{fb}$, respectively.

\begin{table}[htbp]
  \centering
  \caption {Summary of cuts in Case E.}
  \vspace{-0.9em}
  \resizebox{1.0\textwidth}{!}{
  \begin{tabular}{l|ccc}
    \hline\hline
    \hphantom{A}Case E\hphantom{A} & $\sqrt{s} = 3$ TeV & $\sqrt{s} = 10$ TeV & $\sqrt{s} = 14$ TeV \\
    \hline
    \hphantom{A}Trigger & \hphantom{$N(b)=1, \ N(\ell)=1$} & $N(b)=1, \ N(\ell)=1$ & \hphantom{$N(b)=1, \ N(\ell)=1$} \\
    \hline
    \hphantom{A}Cut-1 & \hphantom{$50 \ \gev < M(b \ \ell) < 150 \ \gev$} & $50 \ \gev < M(b \ \ell) < 150 \ \gev$ & \hphantom{$50 \ \gev < M(b \ \ell) < 150 \ \gev$} \\
    \hline
    \hphantom{A}Cut-2 & $\slashed{E}_T > 600 \ \gev$ \hphantom{1111} \vline & $\slashed{E}_T > 1500 \ \gev$ \hphantom{1111} \vline & $\slashed{E}_T > 2500 \ \gev$ \\
    \hline\hline
  \end{tabular}
  }
\label{tab:caseE_cuts}
\end{table}

\begin{table}[htbp]
  \centering
  \caption{Cut flow of the cross sections for the signals and backgrounds with $\kappa_{tcZ}=\kappa_{tuZ}=0.5$ in Case E.}
  \vspace{-0.5em}
  \footnotesize
  \setlength{\tabcolsep}{2.8pt}
  \resizebox{\textwidth}{!}{%
  \begin{tabular}{ l *{4}{c} c *{7}{c} }
    \hline\hline
    Process & \multicolumn{4}{c}{Signal (fb)} & & \multicolumn{7}{c}{Backgrounds (fb)} \\
    \cline{2-5} \cline{7-13}
            & $tcZ(LH)$ & $tcZ(RH)$ & $tuZ(LH)$ & $tuZ(RH)$ & & $tWb$ & $WW$ & $WWZ$ & $ZZ$ & $ZH$ & $WWH$ & $\nu_{\ell}\bar{\nu_{\ell}}H$ \\
    \hline
    \multicolumn{13}{c}{\boldmath$\sqrt{s}=3$ TeV} \\
    \hline
    Basic cuts & $20.09$ & $14.92$ & $19.95$ & $15.02$ & & $1.40$ & $19.89$ & $1.96$ & $0.21$ & $4.86 \times 10^{-2}$ & $3.46 \times 10^{-2}$ & $1.18 \times 10^{3}$ \\
    Trigger & $9.22$ & $6.49$ & $11.56$ & $7.47$ & & $0.12$ & $7.36 \times 10^{-2}$ & $0.28$ & $4.52 \times 10^{-4}$ & $2.96 \times 10^{-3}$ & $3.66 \times 10^{-3}$ & $1.16 \times 10^{2}$ \\
    Cut-1 & $6.82$ & $4.52$ & $9.59$ & $6.09$ & & $3.99 \times 10^{-2}$ & $5.97 \times 10^{-3}$ & $3.38 \times 10^{-2}$ & $1.02 \times 10^{-4}$ & $2.80 \times 10^{-3}$ & $5.49 \times 10^{-4}$ & $45.46$ \\
    Cut-2 & $3.00$ & $2.04$ & $4.57$ & $2.96$ & & $3.51 \times 10^{-3}$ & $0$ & $8.45 \times 10^{-3}$ & $4.10 \times 10^{-5}$ & $2.71 \times 10^{-3}$ & $1.04 \times 10^{-5}$ & $0.17$ \\[5pt]
    Total Eff. & 14.90\% & 13.70\% & 22.90\% & 19.70\% & & 0.25\% & 0\% & 0.43\% & 0.02\% & 5.57\% & 0.03\% & 0.01\% \\[3pt]
    \hline
    \multicolumn{13}{c}{\boldmath$\sqrt{s}=10$ TeV} \\
    \hline
    Basic cuts & $39.45$ & $23.01$ & $39.42$ & $23.06$ & & $3.12 \times 10^{-2}$ & $1.83$ & $0.22$ & $1.57 \times 10^{-3}$ & $5.71 \times 10^{-5}$ & $1.08 \times 10^{-3}$ & $1.98 \times 10^{3}$ \\
    Trigger & $17.07$ & $9.72$ & $21.53$ & $11.85$ & & $2.80 \times 10^{-3}$ & $7.41 \times 10^{-3}$ & $3.20 \times 10^{-2}$ & $3.50 \times 10^{-6}$ & $3.89 \times 10^{-6}$ & $9.77 \times 10^{-5}$ & $1.71 \times 10^{2}$ \\
    Cut-1 & $8.55$ & $4.14$ & $12.49$ & $6.22$ & & $6.19 \times 10^{-4}$ & $6.18 \times 10^{-4}$ & $1.72 \times 10^{-3}$ & $1.00 \times 10^{-6}$ & $2.84 \times 10^{-6}$ & $6.48 \times 10^{-6}$ & $83.29$ \\
    Cut-2 & $6.16$ & $3.05$ & $9.74$ & $4.85$ & & $4.17 \times 10^{-5}$ & $0$ & $4.86 \times 10^{-4}$ & $2.50 \times 10^{-7}$ & $1.67 \times 10^{-6}$ & $8.75 \times 10^{-8}$ & $0$ \\[5pt]
    Total Eff. & 15.60\% & 13.30\% & 24.70\% & 21.00\% & & 0.13\% & 0\% & 0.22\% & 0.02\% & 2.93\% & 0.01\% & 0\% \\[3pt]
    \hline
    \multicolumn{13}{c}{\boldmath$\sqrt{s}=14$ TeV} \\
    \hline
    Basic cuts & $41.14$ & $22.28$ & $41.43$ & $22.41$ & & $1.16 \times 10^{-2}$ & $0.94$ & $0.11$ & $3.45 \times 10^{-4}$ & $7.24 \times 10^{-6}$ & $4.03 \times 10^{-4}$ & $2.20 \times 10^{3}$ \\
    Trigger & $17.64$ & $9.10$ & $22.60$ & $11.92$ & & $1.12 \times 10^{-3}$ & $3.42 \times 10^{-3}$ & $1.47 \times 10^{-2}$ & $5.96 \times 10^{-7}$ & $4.70 \times 10^{-7}$ & $3.57 \times 10^{-5}$ & $1.73 \times 10^{2}$ \\
    Cut-1 & $9.21$ & $4.10$ & $14.69$ & $6.47$ & & $1.53 \times 10^{-4}$ & $5.02 \times 10^{-5}$ & $7.21 \times 10^{-4}$ & $1.73 \times 10^{-7}$ & $3.10 \times 10^{-7}$ & $1.39 \times 10^{-6}$ & $91.58$ \\
    Cut-2 & $5.82$ & $2.71$ & $10.06$ & $4.40$ & & $3.12 \times 10^{-6}$ & $0$ & $1.67 \times 10^{-4}$ & $1.73 \times 10^{-7}$ & $3.01 \times 10^{-7}$ & $2.73 \times 10^{-8}$ & $0$ \\[5pt]
    Total Eff. & 14.10\% & 12.10\% & 24.30\% & 19.60\% & & 0.03\% & 0\% & 0.15\% & 0.05\% & 4.16\% & 0.01\% & 0\% \\
    \hline\hline
  \end{tabular}%
  }
  \label{tab:caseE_all_cut_eff}
\end{table}

\subsection{Selection Cuts for Case F \textnormal{\small{($W^+ \to jj$ , $Z \to \nu_\ell \bar{\nu}_\ell$)}}}

\begin{figure}[htbp]
  \centering
  \includegraphics[width=0.48\textwidth]{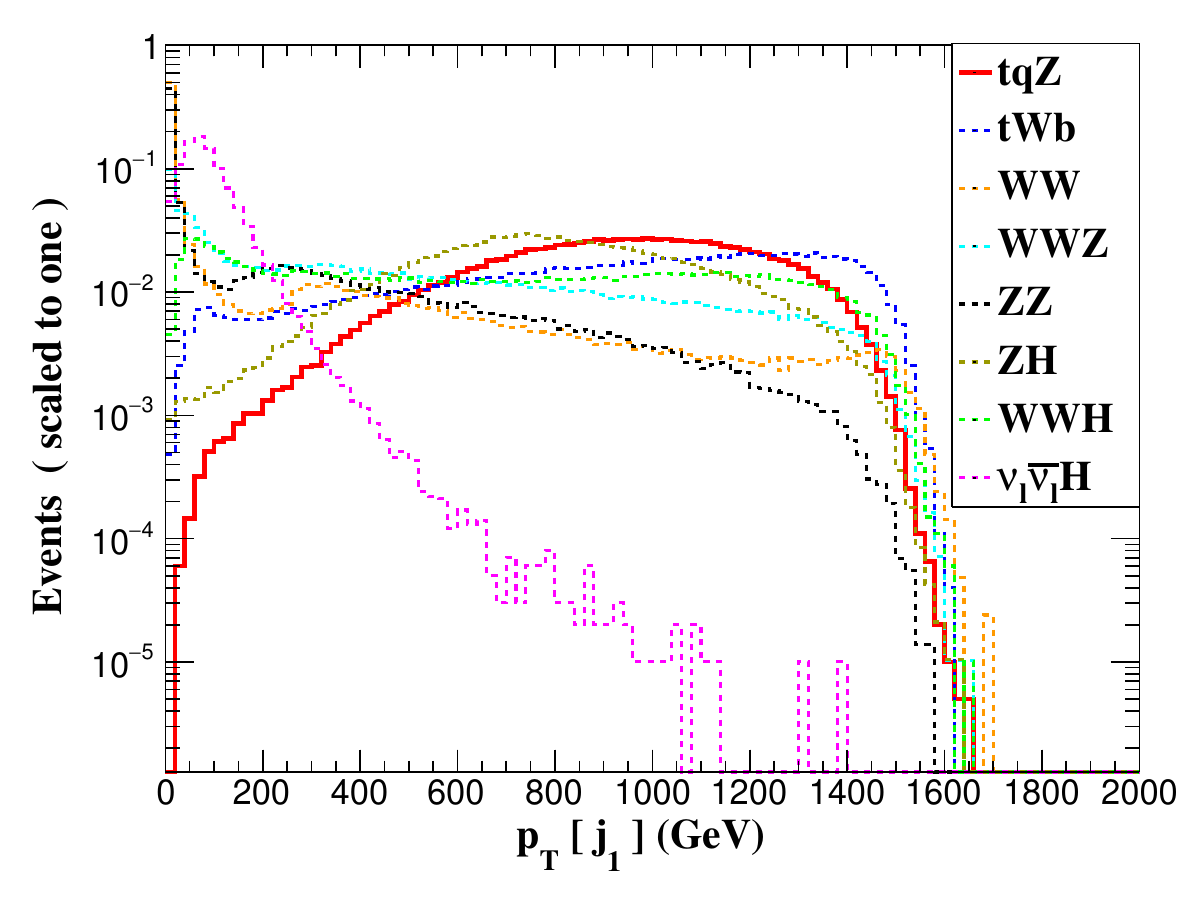}
  \hfill
  \includegraphics[width=0.48\textwidth]{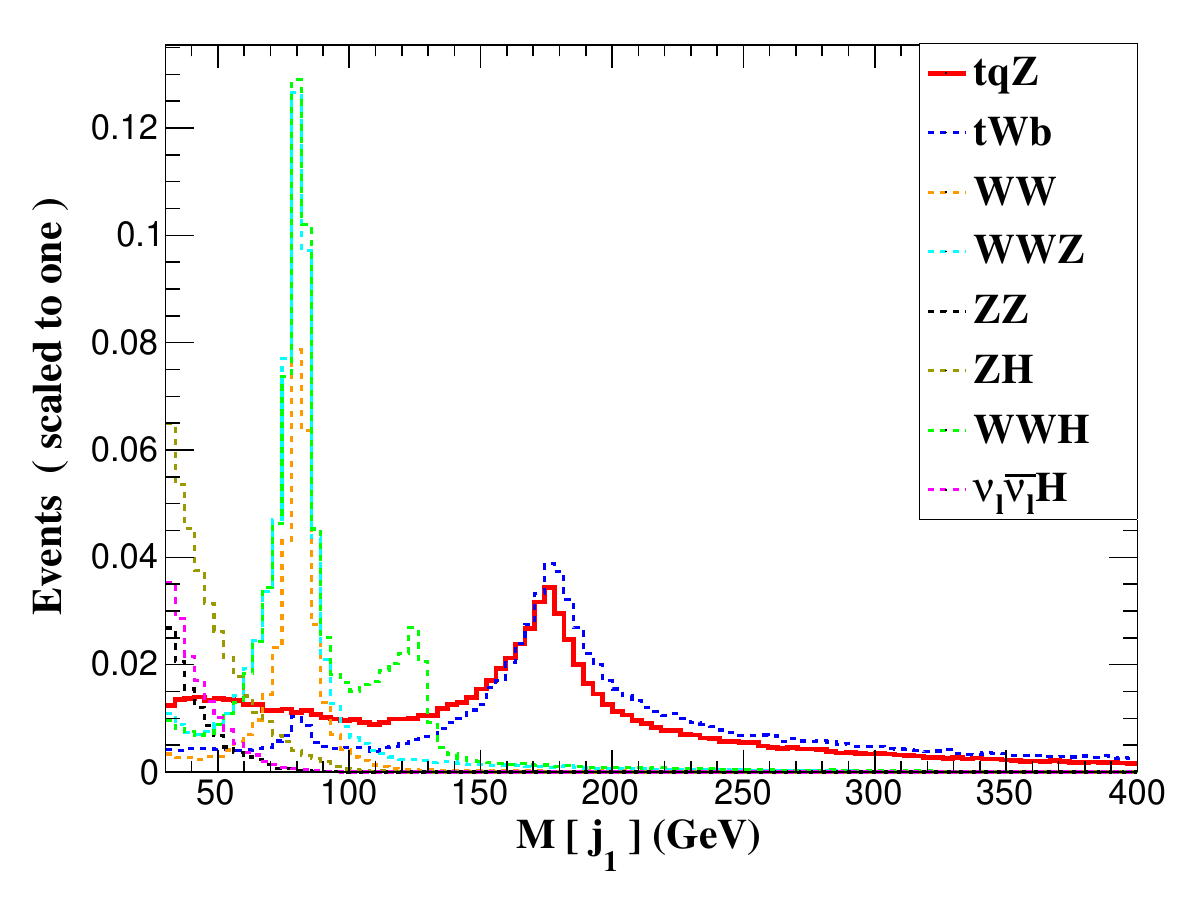}
  \vspace{1em}
  \includegraphics[width=0.49\textwidth]{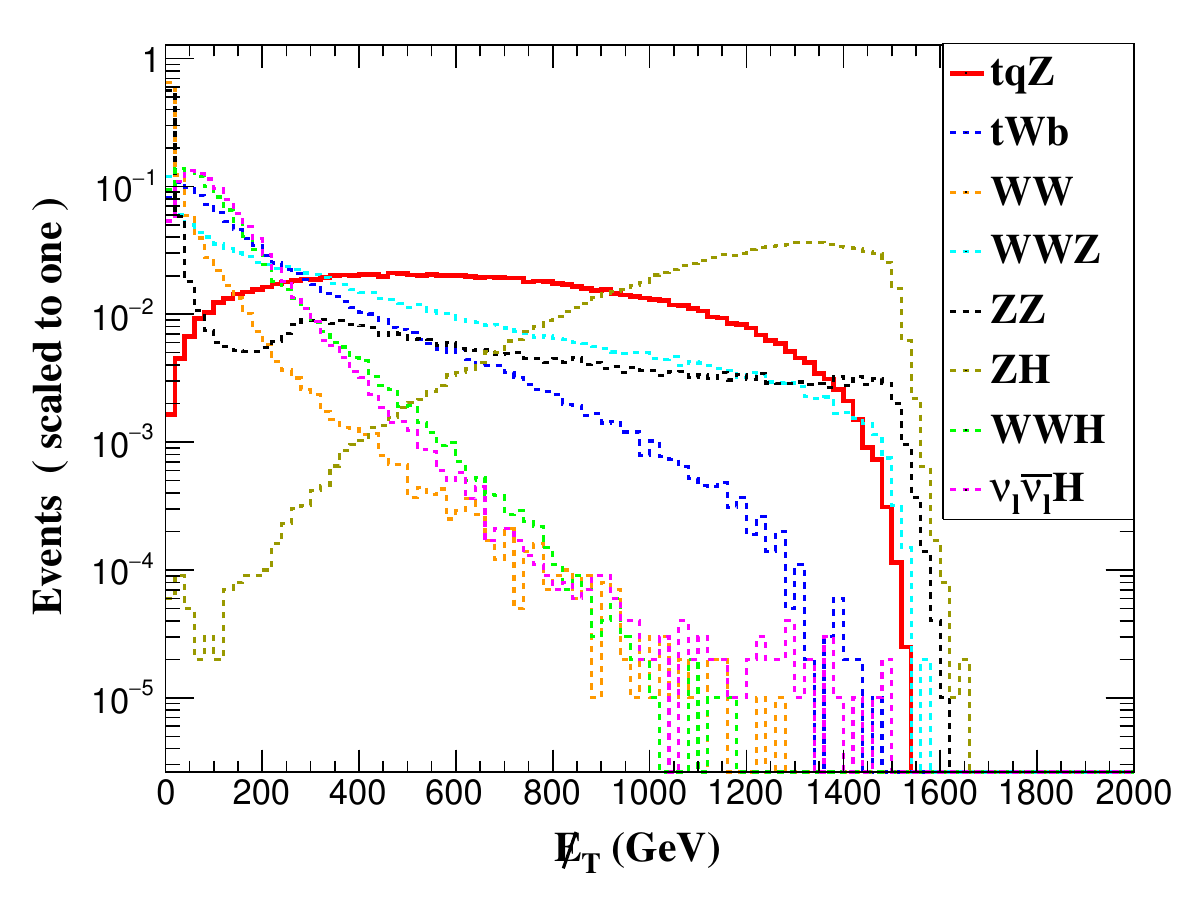}
  \caption{Normalized $p_T^{j_1}$, $M(j_1)$ and $\slashed{E}_T$ distributions for signal (solid red line) and backgrounds (dashed lines) with $\kappa_{tcZ}=0.5$  (LH) and $\sqrt{s}=3\,{\rm TeV}$ in Case F.}
\label{fig:caseF_yun}
\end{figure}

The decay processes of signal and background for Case F are presented in Tab.~\ref{tab:processes_caseF}. Because, none of the backgrounds contain top quark decay except the $tWb$ background in Case F, we employ the fat jet method to reconstruct the invariant mass of the top quark to suppress $WW$, $WWZ$, $ZZ$, $ZH$, $WWH$ and $\nu_{\ell}\bar{\nu_{\ell}}H$. Here, we use $p_T^{j_1}$ and $M(j_1)$ to perform selections for the top quark invariant mass. The $tWb$ background produces only one neutrino from the $W$ boson decay whereas the signal process yields two neutrinos from the decay of the $Z$ boson, thus enabling their separation through a missing transverse momentum/energy cut. In addition, the constraint on the lepton multiplicity $N(\ell)$ is also helpful for rejecting the $tWb$ background. The normalized distributions of these kinematic variables for the signal and background at 3 TeV are shown in Fig.~\ref{fig:caseF_yun}. Based on the characteristic features of the analyzed kinematic distributions, we list the selection schemes at different collider energies in Tab.~\ref{tab:caseF_cuts}.

We summarize the cut flow for the signals and backgrounds in Tab.~\ref{tab:caseF_all_cut_eff}. As can be seen from it, Cut-2 performs well in suppressing all backgrounds except the $tWb$ one. For instance, the $ZZ$, $ZH$, and $\nu_\ell\bar{\nu}_\ell H$ backgrounds are fully eliminated and their cross sections vanish to zero. The cross sections of the $WW$, $WWZ$, and $WWH$ backgrounds are substantially reduced from $11.65\ \text{fb}$, $0.28\ \text{fb}$ and $6.22\times10^{-2}\ \text{fb}$ to $2.84\times10^{-2}\ \text{fb}$, $1.25\times10^{-2}\ \text{fb}$ and $2.01\times10^{-3}\ \text{fb}$, respectively. We use Cut-3 to further suppress the $tWb$ background, leaving less than 5\% of it while retaining a large fraction of $tuZ$ and $tcZ$ events.

\begin{table}[htbp]
  \centering
  \caption{Processes for signal and backgrounds in Case F.}
  \label{tab:processes_caseF}
  \begin{ruledtabular}
  \begin{tabular}{ll}
    \hphantom{AAA}Case F & $W^+ \to jj$,\quad $Z \to \nu_\ell \bar{\nu}_\ell$ \\
    \hline
    \hphantom{AAA}Signal & $\mu^+ \mu^- \to t \bar{q} Z$, $(t \to W^+ b,\ W^+ \to jj),\ (Z \to \nu_\ell \bar{\nu}_\ell)$ \\
    \hphantom{AAA}Final states & $ 3 j + 1 b + \slashed{E}$ \\
    \hline
    \hphantom{AAA}Backgrounds & $\mu^+ \mu^- \to \bar{t} W^+ b$, $(\bar{t} \to W^- \bar{b},\ W^- \to jj),\ (W^+ \to \ell^+ \nu_\ell)$ \\
               & $\mu^+ \mu^- \to W^+ W^-$, $(W^+ \to jj),\ (W^- \to jj)$ \\
               & $\mu^+ \mu^- \to W^+ W^- Z$, $(W^+ \to jj),\ (W^- \to jj),\ (Z \to \nu_\ell \bar{\nu}_\ell)$ \\
               & $\mu^+ \mu^- \to Z Z$, $(Z \to jj),\ (Z \to \nu_\ell \bar{\nu}_\ell)$ \\
               & $\mu^+ \mu^- \to Z H$, $(Z \to \nu_\ell \bar{\nu}_\ell),\ (H \to b \bar{b})$ \\
               & $\mu^+ \mu^- \to W^+ W^- H$, $(W^+ \to jj),\ (W^- \to jj),\ (H \to b \bar{b})$ \\
               & $\mu^+ \mu^- \to \nu_\ell \bar{\nu}_\ell H$, $(H \to b \bar{b})$ \\
  \end{tabular}
  \end{ruledtabular}
\end{table}

\begin{table}[htbp]
  \centering
  \caption {Summary of cuts in Case F.}
  \vspace{-0.9em}
  \resizebox{1.0\textwidth}{!}{
  \begin{tabular}{l|ccc}
    \hline\hline
    \hphantom{A}Case F\hphantom{A} & $\sqrt{s} = 3$ TeV & $\sqrt{s} = 10$ TeV & $\sqrt{s} = 14$ TeV \\
    \hline
    \hphantom{A}Trigger & \hphantom{$N(b)=1, \ N(\ell)=0$} & $N(b)=1, \ N(\ell)=0$ & \hphantom{$N(b)=1, \ N(\ell)=0$} \\
    \hline
    \hphantom{A}Cut-1 & $p_{T}^{j_1} > 600 \ \gev$ \hphantom{1111} \vline & $p_{T}^{j_1} > 2000 \ \gev$ \hphantom{1111} \vline & $p_{T}^{j_1} > 3000 \ \gev$ \\
    \hline
    \hphantom{A}Cut-2 & \hphantom{$140 \ \gev < M(j_1) < 200 \ \gev$} & $140 \ \gev < M(j_1) < 200 \ \gev$ & \hphantom{$140 \ \gev < M(j) < 200 \ \gev$} \\
    \hline
    \hphantom{A}Cut-3 & $\slashed{E}_T > 600 \ \gev$ \hphantom{1111} \vline & $\slashed{E}_T > 1500 \ \gev$ \hphantom{1111} \vline & $\slashed{E}_T > 2500 \ \gev$ \\
    \hline\hline
  \end{tabular}
  }
\label{tab:caseF_cuts}
\end{table}

\begin{table}[htbp]
  \centering
  \caption{Cut flow of the cross sections for the signals and backgrounds with $\kappa_{tcZ}=\kappa_{tuZ}=0.5$ in Case F.}
  \vspace{-0.5em}
  \footnotesize
  \setlength{\tabcolsep}{2.8pt}
  \resizebox{\textwidth}{!}{%
  \begin{tabular}{ l *{4}{c} c *{7}{c} }
    \hline\hline
    Process & \multicolumn{4}{c}{Signal (fb)} & & \multicolumn{7}{c}{Backgrounds (fb)} \\
    \cline{2-5} \cline{7-13}
            & $tcZ(LH)$ & $tcZ(RH)$ & $tuZ(LH)$ & $tuZ(RH)$ & & $tWb$ & $WW$ & $WWZ$ & $ZZ$ & $ZH$ & $WWH$ & $\nu_{\ell}\bar{\nu_{\ell}}H$ \\
    \hline
    \multicolumn{13}{c}{\boldmath$\sqrt{s}=3$ TeV} \\
    \hline
    Basic cuts & $18.58$ & $16.37$ & $18.62$ & $16.36$ & & $7.53$ & $2.85 \times 10^{2}$ & $2.83$ & $10.54$ & $4.86 \times 10^{-2}$ & $0.25$ & $1.18 \times 10^{3}$ \\
    Trigger & $9.87$ & $8.25$ & $12.05$ & $9.87$ & & $1.50$ & $77.33$ & $0.78$ & $1.93$ & $7.71 \times 10^{-3}$ & $9.86 \times 10^{-2}$ & $2.37 \times 10^{2}$ \\
    Cut-1 & $8.84$ & $7.46$ & $11.07$ & $9.15$ & & $1.02$ & $11.65$ & $0.28$ & $0.20$ & $6.05 \times 10^{-3}$ & $6.22 \times 10^{-2}$ & $0.12$ \\
    Cut-2 & $3.27$ & $2.74$ & $3.86$ & $3.31$ & & $0.35$ & $2.84 \times 10^{-2}$ & $1.25 \times 10^{-2}$ & $0$ & $0$ & $2.01 \times 10^{-3}$ & $0$ \\
    Cut-3 & $1.79$ & $1.47$ & $2.06$ & $1.76$ & & $9.04 \times 10^{-3}$ & $0$ & $9.07 \times 10^{-3}$ & $0$ & $0$ & $0$ & $0$ \\[5pt]
    Total Eff. & 9.65\% & 8.99\% & 11.00\% & 10.70\% & & 0.12\% & 0\% & 0.32\% & 0\% & 0\% & 0\% & 0\% \\[3pt]
    \hline
    \multicolumn{13}{c}{\boldmath$\sqrt{s}=10$ TeV} \\
    \hline
    Basic cuts & $19.71$ & $17.11$ & $19.71$ & $16.97$ & & $0.52$ & $10.30$ & $0.36$ & $0.66$ & $5.71 \times 10^{-5}$ & $1.64 \times 10^{-2}$ & $1.98 \times 10^{3}$ \\
    Trigger & $10.46$ & $8.90$ & $13.58$ & $11.47$ & & $0.11$ & $0.97$ & $8.59 \times 10^{-2}$ & $4.03 \times 10^{-2}$ & $9.10 \times 10^{-6}$ & $5.51 \times 10^{-3}$ & $3.95 \times 10^{2}$ \\
    Cut-1 & $9.37$ & $7.94$ & $12.38$ & $10.53$ & & $6.31 \times 10^{-2}$ & $0.26$ & $2.66 \times 10^{-2}$ & $8.67 \times 10^{-3}$ & $6.85 \times 10^{-6}$ & $3.50 \times 10^{-3}$ & $0.20$ \\
    Cut-2 & $2.05$ & $1.88$ & $2.52$ & $2.22$ & & $1.02 \times 10^{-2}$ & $9.64 \times 10^{-3}$ & $8.58 \times 10^{-4}$ & $0$ & $0$ & $1.20 \times 10^{-4}$ & $0$ \\
    Cut-3 & $1.38$ & $1.29$ & $1.74$ & $1.51$ & & $8.73 \times 10^{-4}$ & $8.77 \times 10^{-4}$ & $4.29 \times 10^{-4}$ & $0$ & $0$ & $4.92 \times 10^{-6}$ & $0$ \\[5pt]
    Total Eff. & 6.97\% & 7.56\% & 8.83\% & 8.91\% & & 0.17\% & 0.01\% & 0.12\% & 0\% & 0\% & 0.03\% & 0\% \\[3pt]
    \hline
    \multicolumn{13}{c}{\boldmath$\sqrt{s}=14$ TeV} \\
    \hline
    Basic cuts & $19.82$ & $17.14$ & $19.91$ & $17.18$ & & $0.26$ & $3.93$ & $0.15$ & $0.19$ & $7.24 \times 10^{-6}$ & $5.32 \times 10^{-3}$ & $2.20 \times 10^{3}$ \\
    Trigger & $10.52$ & $9.04$ & $13.45$ & $11.55$ & & $5.74 \times 10^{-2}$ & $0.34$ & $3.50 \times 10^{-2}$ & $1.02 \times 10^{-2}$ & $1.11 \times 10^{-6}$ & $1.76 \times 10^{-3}$ & $4.02 \times 10^{2}$ \\
    Cut-1 & $9.02$ & $7.72$ & $11.73$ & $10.10$ & & $3.09 \times 10^{-2}$ & $9.49 \times 10^{-2}$ & $1.02 \times 10^{-2}$ & $2.21 \times 10^{-3}$ & $7.41 \times 10^{-7}$ & $1.13 \times 10^{-3}$ & $0$ \\
    Cut-2 & $1.91$ & $1.60$ & $2.24$ & $2.07$ & & $4.82 \times 10^{-3}$ & $8.67 \times 10^{-3}$ & $8.67 \times 10^{-4}$ & $0$ & $0$ & $5.89 \times 10^{-5}$ & $0$ \\
    Cut-3 & $1.12$ & $0.97$ & $1.28$ & $1.22$ & & $3.39 \times 10^{-4}$ & $0$ & $4.33 \times 10^{-4}$ & $0$ & $0$ & $2.65 \times 10^{-6}$ & $0$ \\[5pt]
    Total Eff. & 5.65\% & 5.64\% & 6.41\% & 7.11\% & & 0.13\% & 0\% & 0.28\% & 0\% & 0\% & 0.05\% & 0\% \\
    \hline\hline
  \end{tabular}%
  }
  \label{tab:caseF_all_cut_eff}
\end{table}

\subsection{Exclusion Limits}

To estimate the exclusion significance, we use the following formula from Ref.~\cite{z_excl_1,z_excl_2}:
\begin{align}
  Z_\text{excl} =\sqrt{2\left[S-B\ln\left(\frac{B+S+x}{2B}\right)
  - \frac{1}{\delta^2 }\ln\left(\frac{B-S+x}{2B}\right)\right] -
  \left(B+S-x\right)\left(1+\frac{1}{\delta^2 B}\right)},
\end{align}
with
\begin{align}
 x=\sqrt{(S+B)^2- 4 \delta^2 S B^2/(1+\delta^2 B)}.
\end{align}
Here, $S$ and $B$ represent the total signal and  background events, respectively. $\delta$ is the percentage systematic error on the background estimate. Following Refs.~\cite{z_excl<1.645,z_excl_1}, we define the regions with $Z_\text{excl} \leq 1.645$ as those that can be excluded at 95\% CL. In the case of $\delta \to 0$,  the above expressions  are simplified as
\begin{align}
 Z_\text{excl} = \sqrt{2[S-B\ln(1+S/B)]}.
\end{align}

\begin{figure}[htbp]
 \centering
 \hspace{-10mm}
   \begin{subfigure}{0.36\textwidth}
   \centering
   \includegraphics[width=\linewidth]{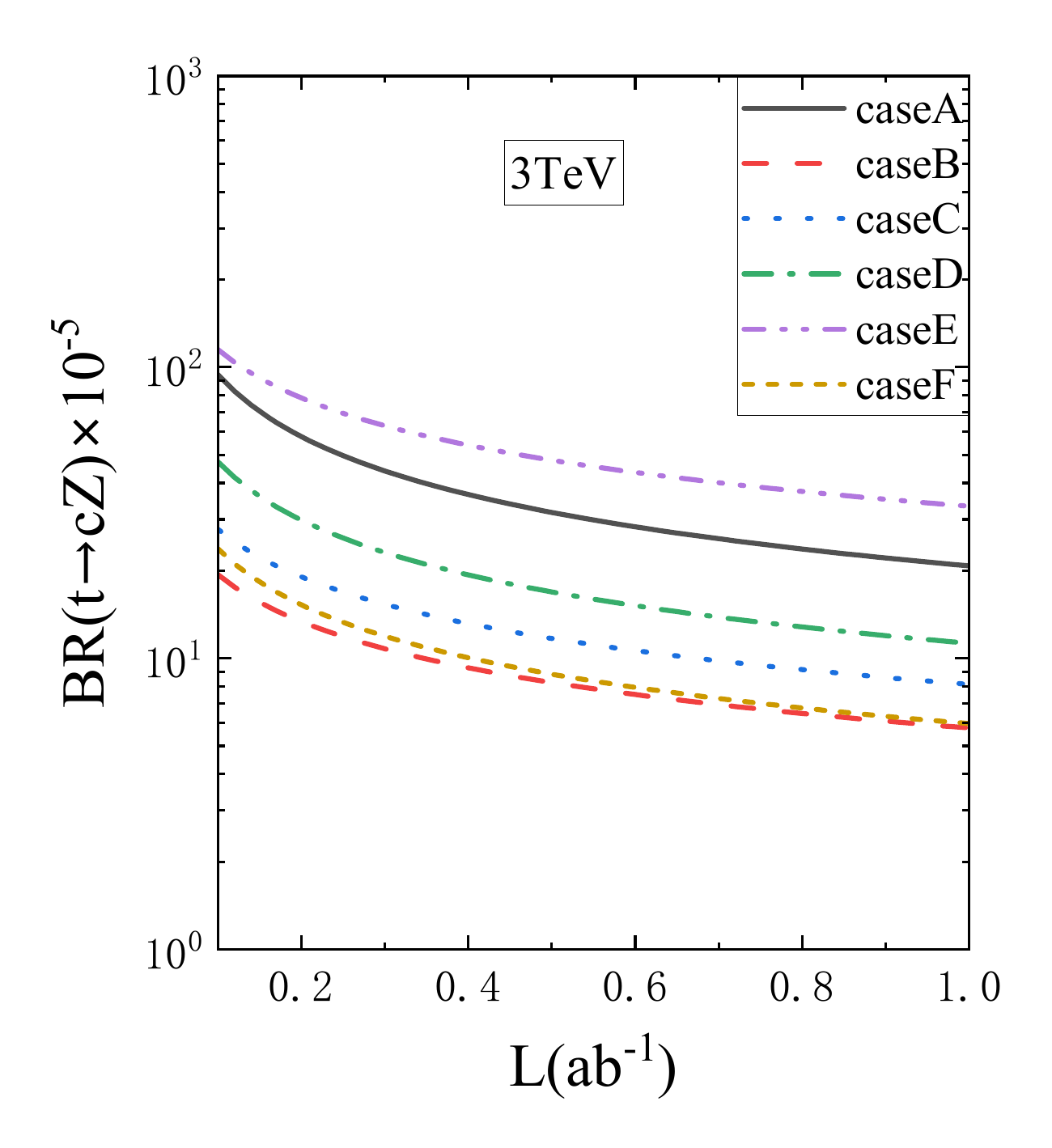}
   \end{subfigure}
  \hspace{-6mm}
   \begin{subfigure}{0.36\textwidth}
   \centering
   \includegraphics[width=\linewidth]{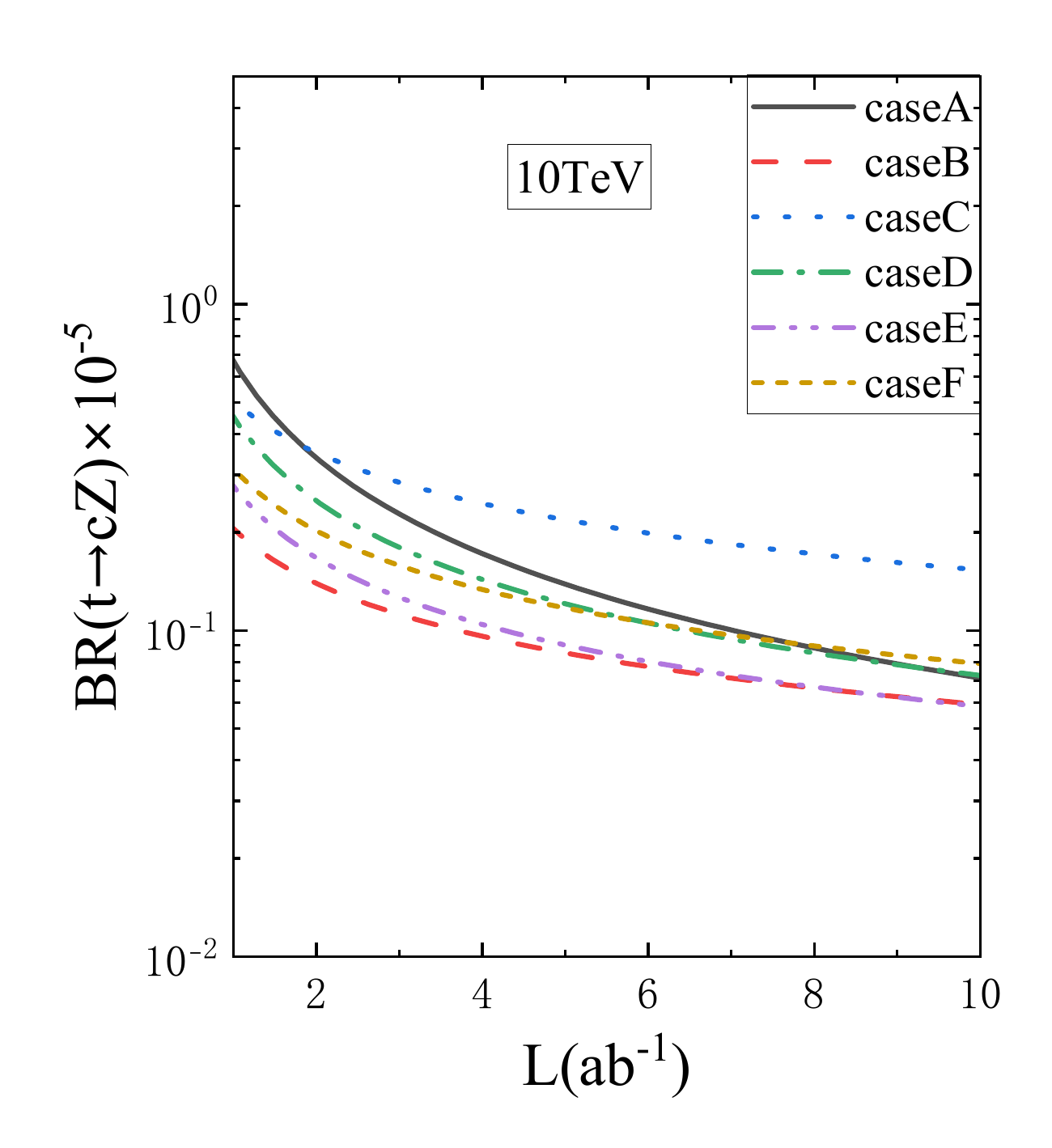}
   \end{subfigure}
  \hspace{-6mm}
   \begin{subfigure}{0.36\textwidth}
   \centering
   \includegraphics[width=\linewidth]{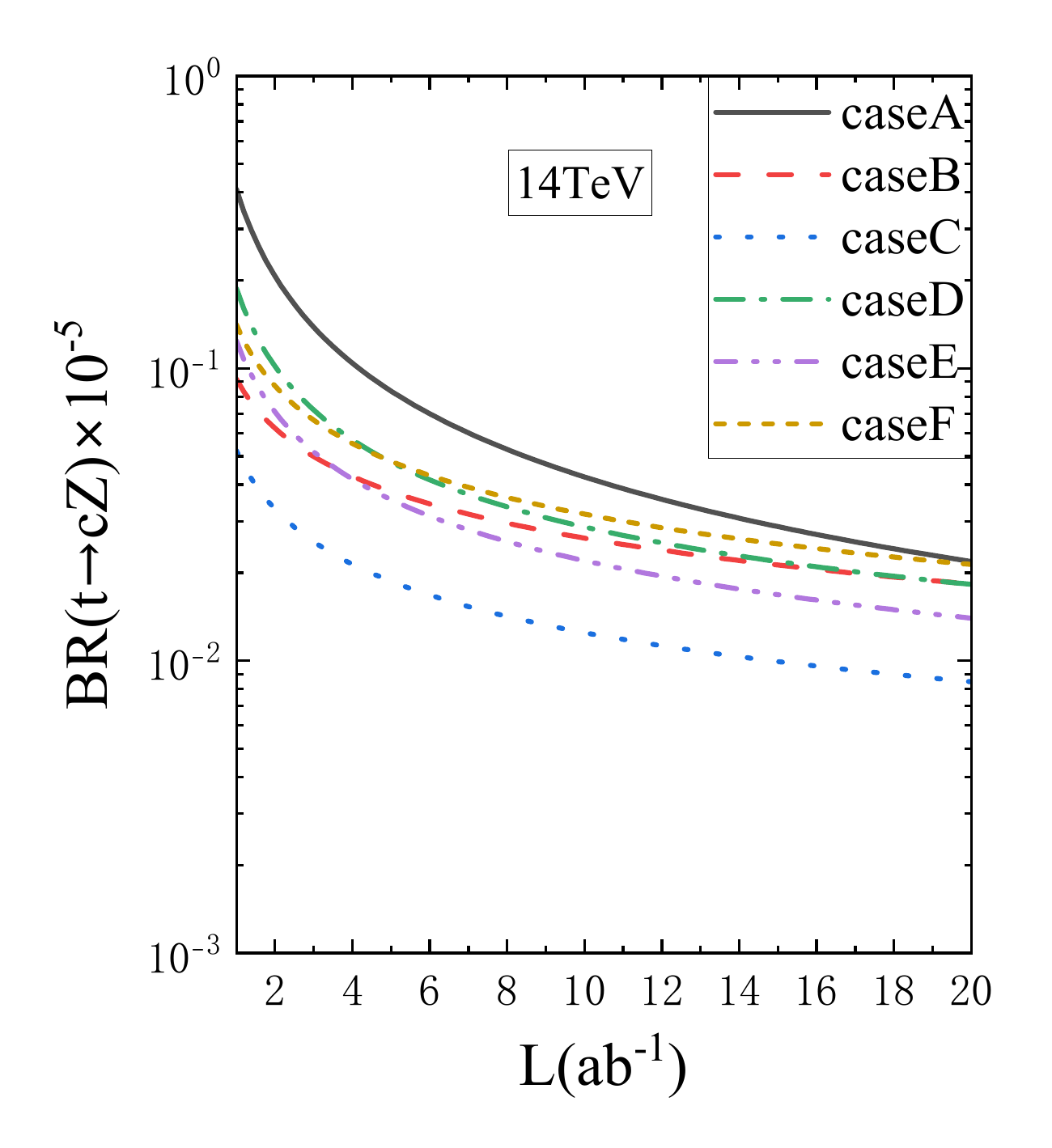}
   \end{subfigure}
  \vspace{-1mm}
  \hspace{-10mm}
   \begin{subfigure}{0.36\textwidth}
   \centering
   \includegraphics[width=\linewidth]{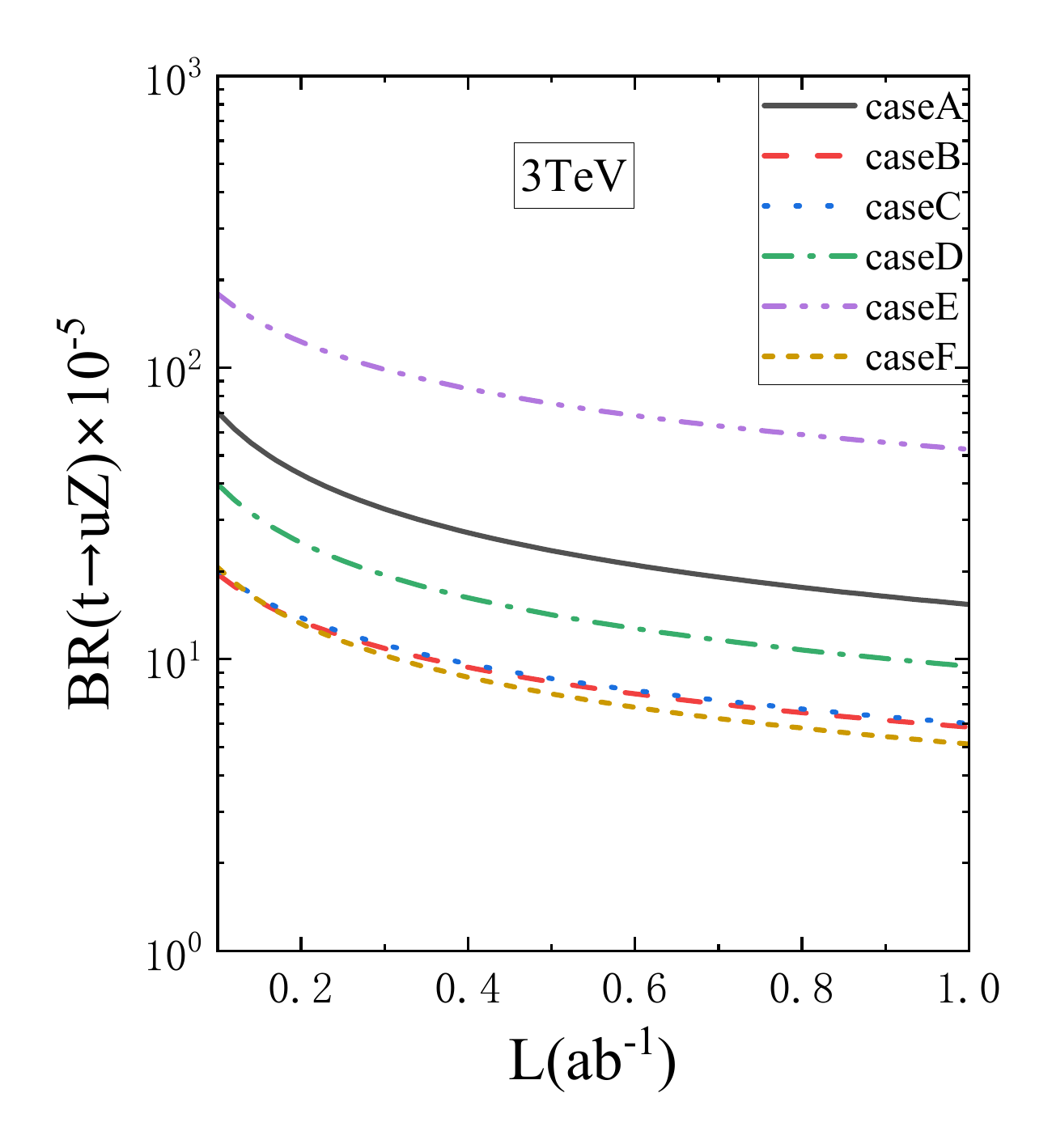}
   \end{subfigure}
  \hspace{-6mm}
   \begin{subfigure}{0.36\textwidth}
   \centering
   \includegraphics[width=\linewidth]{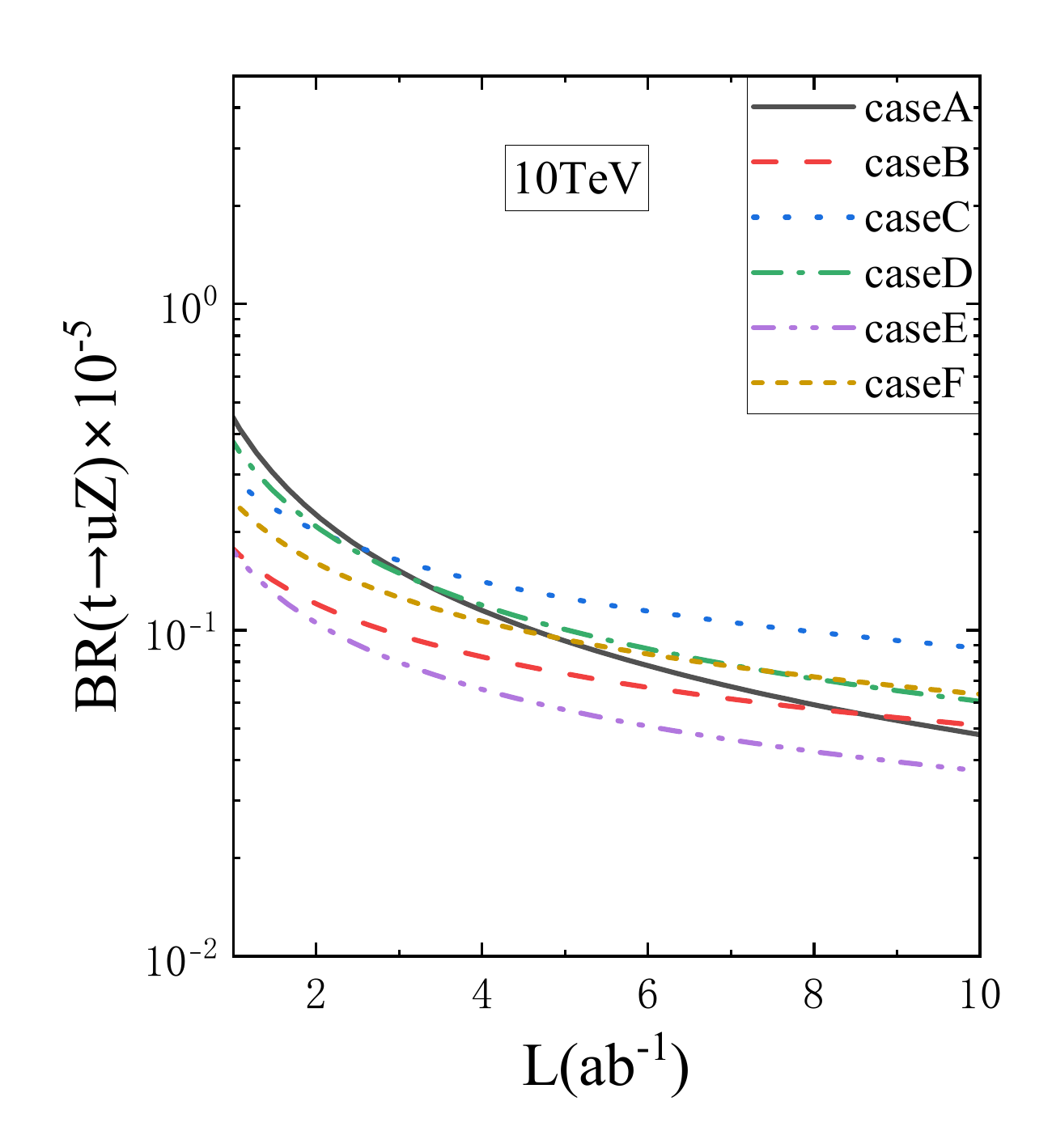}
   \end{subfigure}
  \hspace{-6mm}
   \begin{subfigure}{0.36\textwidth}
   \centering
   \includegraphics[width=\linewidth]{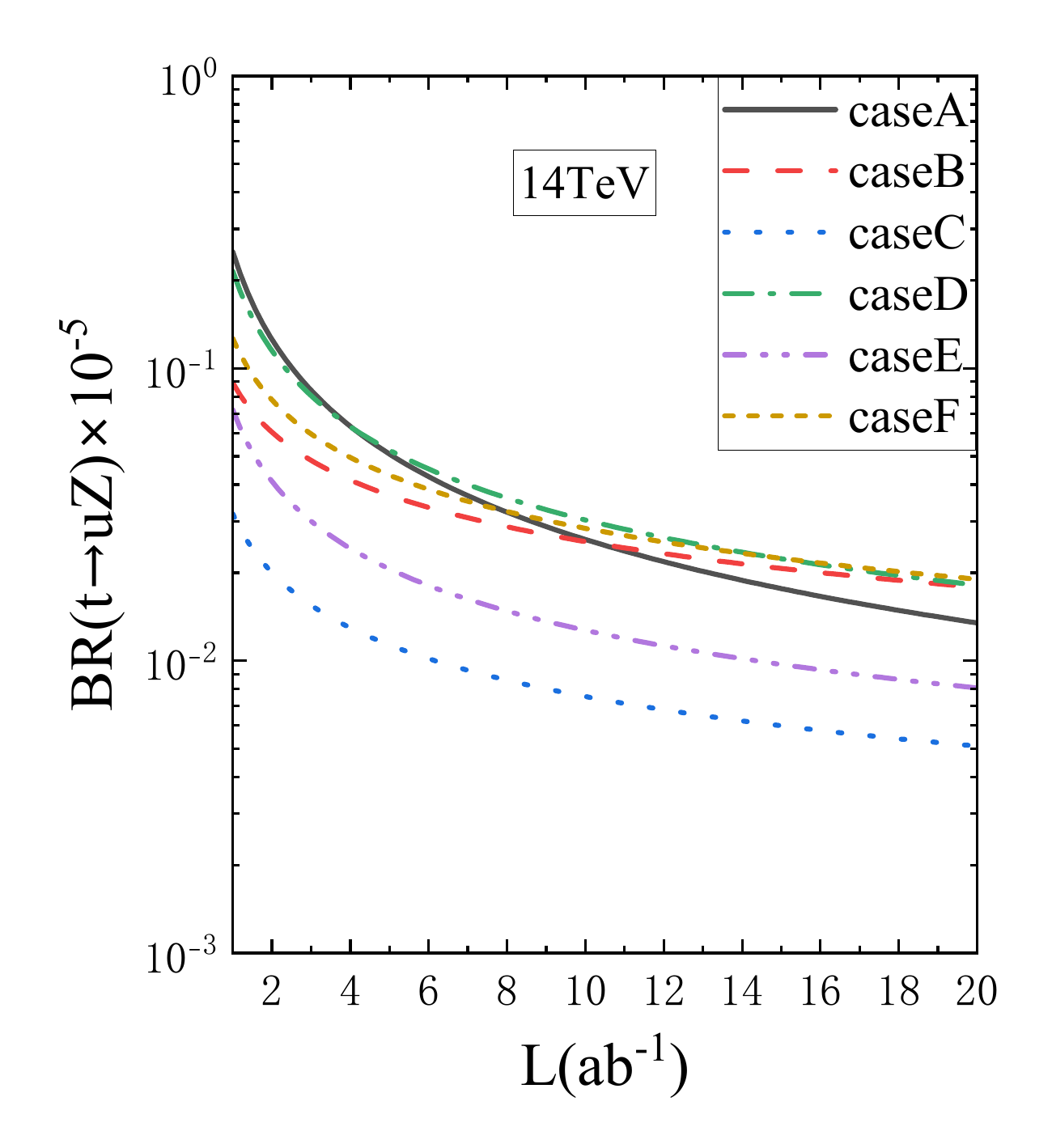}
   \end{subfigure}
\caption{95\% CL contours in $L$--$\mathrm{BR}(t\to cZ)$ planes (top) and $L$--$\mathrm{BR}(t\to uZ)$ planes (bottom) for the six decay cases.}
\label{fig:pai_tcz_tuz}
\end{figure}

In Fig.~\ref{fig:pai_tcz_tuz}, we plot the 95\% CL exclusion limits as functions of the integrated luminosity, $L$, on $\text{BR}(t\to cZ)$ and $\text{BR}(t\to uZ)$, involving the $tcZ$ and $tuZ$ couplings, respectively. According to the simulation results of the six decay cases mentioned above, the LH and RH chiral components exhibit negligible differences in the kinematic distributions, cross sections and cut efficiencies. Therefore, only the curves of the LH components for $tcZ$ and $tuZ$ are presented in Fig.~\ref{fig:pai_tcz_tuz}. At $3\ \text{TeV}$, Case B and Case F yield the largest exclusion regions, with BR$(t \to qZ)$ down to the $10^{-5}$. At $10\ \text{TeV}$, Case B and Case E provide the optimal exclusion performance, with BR$(t \to qZ)$ reaching the $10^{-7}$ level. At $14\ \text{TeV}$, Case C exhibits the strongest exclusion capability and BR$(t \to qZ)$ can probe down to $10^{-8}$. The most stringent experimental exclusion limits currently given by the CMS and ATLAS collaborations are at the level of $10^{-4}$ and $10^{-5}$. Hence, even under conservative assumptions, the muon collider at 10 TeV with $10\ \text{ab}^{-1}$ still provides exclusion capabilities superior to those of the LHC by two to three orders of magnitude.

In Tab.~\ref{tab:br_tcz_tuz}, we present the 95\% CL upper exclusion limits at a  muon collider for three usual machine setups: 3 TeV with $1\ \text{ab}^{-1}$, 10 TeV with $10\ \text{ab}^{-1}$ and 14 TeV with $20\ \text{ab}^{-1}$. To realistically evaluate the detection sensitivity of the collider, we explicitly consider the impact of systematic uncertainties. The main sources of systematic errors include: background normalization, integrated luminosity, lepton identification efficiency, misidentification from $b$-tagging and jet reconstruction uncertainty. To quantify their effects, two benchmark scenarios are adopted in this work, corresponding to the relative systematic uncertainties of  $\delta=0\%$ and $\delta=10\%$. From Tab.~\ref{tab:br_tcz_tuz}, we can draw the following conclusions: the detection sensitivity for the $tuZ$ coupling is better than that for the $tcZ$ coupling. When considering the 10\% systematic uncertainty, the 95\% CL upper limit for the $tuZ$ coupling can still reach the order of $10^{-8}$.

\begin{table}[htbp]
\centering
\setlength{\tabcolsep}{8pt}
\caption{BR$(t\to cZ)$ and BR$(t\to uZ)$ at 95\% CL with systematic errors $\delta=0\%$ and $10\%$ on the SM background events.}
\begin{tabular}{l|l|cccccc}
\hline
\multicolumn{2}{c|}{BR$(t\to cZ)$} & Case A & Case B & Case C & Case D & Case E & Case F \\
\hline
\multirow{2}{*}{\makecell[c]{$3\ \text{TeV},\ 1\ \text{ab}^{-1}$ \\ (in $\times10^{-5}$)}} 
& $\delta=0\%$  & 20.75 & 5.76  & 8.13  & 11.25 & 33.27 & 5.96 \\
& $\delta=10\%$ & 21.13 & 11.37 & 15.12 & 11.69 & 52.87 & 6.31 \\
\hline
\multirow{2}{*}{\makecell[c]{$10\ \text{TeV},\ 10\ \text{ab}^{-1}$ \\ (in $\times10^{-7}$)}}
& $\delta=0\%$  & 7.14 & 5.93 & 15.30 & 7.28 & 5.85 & 7.91 \\
& $\delta=10\%$ & 7.14 & 8.93 & 40.30 & 7.29 & 5.93 & 8.49 \\
\hline
\multirow{2}{*}{\makecell[c]{$14\ \text{TeV},\ 20\ \text{ab}^{-1}$ \\ (in $\times10^{-8}$)}}
& $\delta=0\%$  & 21.83 & 18.28 & 8.47 & 18.22 & 13.97 & 21.32 \\
& $\delta=10\%$ & 21.83 & 31.98 & 9.16 & 18.24 & 14.08 & 22.37 \\
\hline\hline
\multicolumn{2}{c|}{BR$(t\to uZ)$} & Case A & Case B & Case C & Case D & Case E & Case F \\
\hline
\multirow{2}{*}{\makecell[c]{$3\ \text{TeV},\ 1\ \text{ab}^{-1}$ \\ (in $\times10^{-5}$)}} 
& $\delta=0\%$  & 15.41 & 5.85 & 6.02 & 9.45 & 52.44 & 5.11 \\
& $\delta=10\%$ & 15.69 & 11.50& 11.06 & 9.82 & 83.01 & 5.42 \\
\hline
\multirow{2}{*}{\makecell[c]{$10\ \text{TeV},\ 10\ \text{ab}^{-1}$ \\ (in $\times10^{-7}$)}}
& $\delta=0\%$  & 4.78 & 5.11 & 8.82 & 6.05 & 3.70 & 6.37 \\
& $\delta=10\%$ & 4.78 & 7.71 & 23.22 & 6.06 & 3.76 & 6.82 \\
\hline
\multirow{2}{*}{\makecell[c]{$14\ \text{TeV},\ 20\ \text{ab}^{-1}$ \\ (in $\times10^{-8}$)}}
& $\delta=0\%$  & 13.46 & 17.83 & 5.10 & 18.12 & 8.06 & 18.99 \\
& $\delta=10\%$ & 13.46 & 31.08 & 5.52 & 18.14 & 8.12 & 19.94 \\
\hline
\end{tabular}
\label{tab:br_tcz_tuz}
\end{table}

\section{Conclusions}

In this work, we have performed a full MC simulation of the production process $\mu^+ \mu^- \to t{\bar q}Z$ in  six decay channels (see Tab.~\ref{tab:signal_cases}) and investigated the effects of the intervening $t\bar{q}Z$ anomalous couplings, triggering FCNCs,  at a future muon collider. Based on  independent selection criteria for each of the six signatures, we have carried  out a detector level analysis for the signals and relevant SM backgrounds. As the main goal of our studies, we have established the 95\% CL  upper limit on the $\text{BR}(t\to qZ)$. Among the six decay scenarios, Case C yields the most stringent exclusion limits. The corresponding lowest BRs allowed obtained in this work are: BR$(t \to cZ)=8.47 \times 10^{-8}\ (\delta=0\%)$, BR$(t \to cZ)=9.16 \times 10^{-8}\ (\delta=10\%)$, BR$(t \to uZ)=5.10 \times 10^{-8}\ (\delta=0\%)$ and BR$(t \to uZ)=5.52 \times 10^{-8}\ (\delta=10\%)$. Crucially, these are significantly better than the limits from CMS, i.e., BR$(t \to cZ) < 4.5 \times 10^{-4}$ and BR$(t \to uZ) < 2.4 \times 10^{-4}$, as well as ATLAS, i.e., BR$(t \to cZ) < 1.2 \times 10^{-4}$ and  BR$(t \to uZ) < 6.2 \times 10^{-5}$. Overall, these limits are nearly 2 to 3 orders of magnitude more sensitive than the current experimental results from the LHC at 13 TeV. Therefore, we expect that the signal channels studied in this work will provide competitive and complementary information for probing the FCNCs induced by $t\bar{q}Z$ anomalous couplings at future muon colliders, owing to their high-energy reach, significant luminosity, clean environment as well as, last but not least, ability to use polarized beams.

Numerous phenomenological studies in the literature have extensively investigated the anomalous couplings of top quark (inducing FCNCs) at various future high-energy colliders, including an HE-LHC, FCC-hh and $\mu^+ \mu^-$ collider. Recent comprehensive reviews on this topic can be found in Refs.~\cite{dif_chan_1,dif_chan_2,dif_chan_3,dif_chan_4,dif_chan_5,dif_chan_6,dif_chan_7,dif_chan_9,dif_chan_10,dif_chan_11,dif_chan_12}. It is therefore necessary to compare the upper limits on BR$(t\to qZ)$ obtained in this work with the results from such studies: the relevant comparisons are summarized in Tab.~\ref{tab:limits_tqZ_future}. It can be seen that the expected limits from such alternative future machines on the two studied BRs are of the order $O(10^{-4}\text{--}10^{-7})$.

\begin{table}[htbp]
  \centering
  \caption{Projected 95\% CL limits on $\text{BR}(t \to qZ)~(q = u,c)$ from different channels at various future colliders.}
  \label{tab:limits_tqZ_future}
  \footnotesize
  \renewcommand{\arraystretch}{1.4}
  \setlength{\tabcolsep}{3pt}

  \begin{tabular}{c|c|c}
    \hline
    Channels & Data Set & Limits \\
    \hline

      \multirow{3}{*}{\makecell{$pp \to t(\to bW^+ \to b\ell^+\nu)Z(\to \ell^+\ell^-),$\\ $pp \to t(\to bW^+ \to b\ell^+\nu)\bar{t}(\to \bar{q}Z(\to \ell^+\ell^-))$\\~\cite{dif_chan_0}}}
    & \makecell{HL-LHC, $3\ \text{ab}^{-1}$ \\ @ 14 TeV}
    & $\begin{aligned}
        \text{BR}(t \to uZ) &< 7.3  \times 10^{-6}\ (\sigma^{\mu\nu}) \\[-7pt]
        \text{BR}(t \to cZ) &< 2.3  \times 10^{-5}\ (\sigma^{\mu\nu}) \\[-7pt]
        \text{BR}(t \to uZ) &< 2.34 \times 10^{-5}\ (\gamma^{\mu}) \\[-7pt]
        \text{BR}(t \to cZ) &< 3.13 \times 10^{-5}\ (\gamma^{\mu})
      \end{aligned}$ \\
    \cline{2-3}
    & \makecell{HE-LHC, $15\ \text{ab}^{-1}$ \\ @ 27 TeV}
    & $\begin{aligned}
        \text{BR}(t \to uZ) &< 1.83 \times 10^{-6}\ (\sigma^{\mu\nu}) \\[-7pt]
        \text{BR}(t \to cZ) &< 3.64 \times 10^{-6}\ (\sigma^{\mu\nu}) \\[-7pt]
        \text{BR}(t \to uZ) &< 4.28 \times 10^{-6}\ (\gamma^{\mu}) \\[-7pt]
        \text{BR}(t \to cZ) &< 5.22 \times 10^{-6}\ (\gamma^{\mu})
      \end{aligned}$ \\
    \cline{2-3}
    & \makecell{FCC-hh, $30\ \text{ab}^{-1}$ \\ @ 100 TeV}
    & $\begin{aligned}
        \text{BR}(t \to uZ) &< 4.35 \times 10^{-7}\ (\sigma^{\mu\nu}) \\[-7pt]
        \text{BR}(t \to cZ) &< 6.54 \times 10^{-7}\ (\sigma^{\mu\nu}) \\[-7pt]
        \text{BR}(t \to uZ) &< 6.86 \times 10^{-7}\ (\gamma^{\mu}) \\[-7pt]
        \text{BR}(t \to cZ) &< 8.87 \times 10^{-7}\ (\gamma^{\mu})
      \end{aligned}$ \\
    \hline

    $p p \to t q$~\cite{BDT_SAJA}
    & \makecell{LHC, $138\ \ \text{fb}^{-1}$ \\ @ 13 TeV}
    & $\begin{aligned}
        \text{BR}(t \to ug) &< 6.73 \times 10^{-6}\ (\text{BDT}) \\[-7pt]
        \text{BR}(t \to ug) &< 5.61 \times 10^{-6}\ (\text{SAJA}) \\[-7pt]
        \text{BR}(t \to cg) &< 5.89 \times 10^{-6}\ (\text{BDT}) \\[-7pt]
        \text{BR}(t \to cg) &< 4.40 \times 10^{-6}\ (\text{SAJA})
      \end{aligned}$ \\
    \hline

    $q g \to \ell \nu b$~\cite{BDT_2} %BDT_2
    & \makecell{FCC-hh, $10\ \text{ab}^{-1}$ \\ @ 100 TeV}
    & $\begin{aligned}
        \text{BR}(t \to ug) &< 5.18 \times 10^{-7} \hphantom{(\sigma^{\mu\nu})} \\[-7pt]
        \text{BR}(t \to cg) &< 4.45 \times 10^{-7} \hphantom{(\sigma^{\mu\nu})}
      \end{aligned}$ \\
    \hline

    $e^+ e^- \to tq$~\cite{dif_chan_6}
    & \makecell{FCC-ee, $300\ \ \text{fb}^{-1}$ \\ @ 350 GeV}
    & $\begin{aligned}
        \text{BR}(t \to qZ) &< 3.12 \times 10^{-5}\ (\sigma^{\mu\nu}) \\[-7pt]
        \text{BR}(t \to qZ) &< 1.22 \times 10^{-4}\ (\gamma^{\mu})
      \end{aligned}$ \\
    \hline

      $\mu^+ \mu^- \to \nu_\mu \mu^+ b j$~\cite{dif_chan_11}
    & \makecell{MuC, $10\ \text{ab}^{-1}$ \\ @ 10 TeV}
    & $\begin{aligned}
        \text{BR}(t \to qZ) &< 1.03 \times 10^{-6}\ (\sigma^{\mu\nu})
      \end{aligned}$ \\
    \hline

      \multirow{3}{*}{\makecell{$\mu^+ \mu^- \to t \bar{q} Z$\\
      (This work)\\}}
    & \makecell{MuC, $1\ \text{ab}^{-1}$ \\ @ 3 TeV}
    & $\begin{aligned}
        \text{BR}(t \to uZ) &< 5.11  \times 10^{-5}\ (\sigma^{\mu\nu}) \\[-7pt]
        \text{BR}(t \to cZ) &< 5.76  \times 10^{-5}\ (\sigma^{\mu\nu})
      \end{aligned}$ \\
    \cline{2-3}
    & \makecell{MuC, $10\ \text{ab}^{-1}$ \\ @ 10 TeV}
    & $\begin{aligned}
        \text{BR}(t \to uZ) &< 3.70 \times 10^{-7}\ (\sigma^{\mu\nu}) \\[-7pt]
        \text{BR}(t \to cZ) &< 5.85 \times 10^{-7}\ (\sigma^{\mu\nu}) \\
      \end{aligned}$ \\
    \cline{2-3}
    & \makecell{MuC, $20\ \text{ab}^{-1}$ \\ @ 14 TeV}
    & $\begin{aligned}
        \text{BR}(t \to uZ) &< 5.10 \times 10^{-8}\ (\sigma^{\mu\nu}) \\[-7pt]
        \text{BR}(t \to cZ) &< 8.47 \times 10^{-8}\ (\sigma^{\mu\nu})
      \end{aligned}$ \\
    \hline

  \end{tabular}
\end{table}

\section*{Acknowledgments}
This work was supported by the High Performance Computing Platform of Henan Normal University (HNU). Henan Provincial International Science and Technology Cooperation Incubation Project 262102520027. SM is supported in part through the NExT Institute and STFC Consolidated Grant ST/X000583/1.

\clearpage
\bibliographystyle{apsrev}
\bibliography{references.bib}

@article{probe_for_BSM_physics,
   tttitle={Single top quark production as a window to physics beyond the standard model},
   author={Tait, Tim M. P. and Yuan, C.-P.},
   journal={\href{http://dx.doi.org/10.1103/PhysRevD.63.014018}{Phys. Rev. D \textbf{63}, 014018}},
   year={2000}}

@misc{GIM_mechanism_1,
      tttitle={Top flavour-changing neutral interactions: theoretical expectations and experimental detection}, 
      author={J. A. Aguilar-Saavedra},
      eprint={\href{https://arxiv.org/abs/hep-ph/0409342}{arXiv:hep-ph/0409342}},
      }

@article{GIM_mechanism_2,
   tttitle={A minimal set of top-Higgs anomalous couplings},
   journal={\href{http://dx.doi.org/10.1016/j.nuclphysb.2009.06.022}{Nucl. Phys. B \textbf{821}, 215-227}},
   author={J.A. Aguilar-Saavedra},
   year={2009},
   }

@article{the_quark_singlet_model,
  tttitle={Effects of mixing with quark singlets},
  author={Juan Antonio Aguilar-Saavedra},
  journal={\href{https://api.semanticscholar.org/CorpusID:119454607}{Phys. Rev. D \textbf{69}, 099901}},
  year={2002},
  }

@article{the_warped_extra_dimension_model,
   tttitle={Collider signals of top quark flavor violation from a warped extra dimension},
   author={Agashe, Kaustubh and Perez, Gilad and Soni, Amarjit},
   journal={\href{http://dx.doi.org/10.1103/PhysRevD.75.015002}{Phys. Rev. D \textbf{75}, 015002}},
   year={2007},
   }

@article{2HDM_without_flavor_changing_neutral_currents,
   tttitle={Phenomenology of two Higgs doublet models with flavor-changing neutral currents},
   author={Atwood, David and Reina, Laura and Soni, Amarjit},
   journal={\href{http://dx.doi.org/10.1103/PhysRevD.55.3156}{Phys. Rev. D \textbf{55}, 3156}},
   year={1997}
   }

@article{MSSM,
   tttitle={Supersymmetry-induced flavor-changing neutral-current top-quark processes at the CERN Large Hadron Collider},
   author={Cao, J. J. and Eilam, G. and Frank, M. and Hikasa, K. and Liu, G. L. and Turan, I. and Yang, J. M.},
   journal={\href{http://dx.doi.org/10.1103/PhysRevD.75.075021}{Phys. Rev. D \textbf{75}, 075021}},
   year={2007},
   }

@article{R-parity_violating_MSSM,
   tttitle={Flavor-changing top quark decays in<i>R</i>-parity-violating supersymmetric models},
   author={Yang, Jin Min and Young, Bing-Lin and Zhang, X.},
   journal={\href{http://dx.doi.org/10.1103/PhysRevD.58.055001}{Phys. Rev. D \textbf{58}, 055001}},
   year={1998},
   }

@article{mirror_fermion_model,
   tttitle={Top quark rare decays via loop-induced FCNC interactions in extended mirror fermion model},
   author={Hung, P.Q. and Lin, Yu-Xiang and Nugroho, Chrisna Setyo and Yuan, Tzu-Chiang},
   journal={\href{http://dx.doi.org/10.1016/j.nuclphysb.2017.12.014}{Nucl. Phys. B \textbf{927}, 166-183}},
   year={2018},
   }

@article{new_physics_beyond_the_Standard_Model,
   tttitle={Probing top flavour-changing neutral scalar couplings at the CERN LHC},
   author={Aguilar-Saavedra, J.A. and Branco, G.C.},
   journal={\href{http://dx.doi.org/10.1016/S0370-2693(00)01259-4}{Phys. Lett. B \textbf{495}, 347-356}},
   year={2000},
   }

@article{ATLAS_CMS_conduct_related_studies_1,
   title={Search for flavour-changing neutral current top-quark decays to $$\varvec{qZ}$$ q Z in $$\varvec{pp}$$ p p collision data collected with the ATLAS detector at $$\varvec{\sqrt{s}=8}$$ s = 8  TeV},
   author={ Aad, G. and Abbott, B. and Abdallah, J. and others} ,
   collaboration = {ATLAS Collaboration},
   journal={\href{http://dx.doi.org/10.1140/epjc/s10052-015-3851-5}{Eur. Phys. J. C \textbf{76}, 12}},
   year={2016},
   }

@article{ATLAS_CMS_conduct_related_studies_2,
   tttitle={Search for flavour-changing neutral current top-quark decays t → qZ in proton-proton collisions at $$ \sqrt{s}=13 $$ TeV with the ATLAS detector},
   author={ M. Aaboud and G. Aad and B. Abbott and others },
   collaboration = {ATLAS Collaboration},
   journal={\href{http://dx.doi.org/10.1007/JHEP07(2018)176}{JHEP \textbf{1807}, 176}},
   year = {2018},
   }

@article{ATLAS_CMS_conduct_related_studies_3,
   title = {Search for flavour-changing neutral-current couplings between the top quark and the photon with the ATLAS detector at s=13 TeV},
   author = {G. Aad and B. Abbott and D.C. Abbott and others},
   collaboration = {ATLAS Collaboration},
   journal={\href{https://doi.org/10.1016/j.physletb.2022.137379}{Phys. Lett. B \textbf{842}, 137379}},
   year = {2023},
   }

@article{ATLAS_CMS_conduct_related_studies_4,
   tttitle={Search for Flavor-Changing Neutral Currents in Top-Quark Decays t to Zq in pp collisions at $$ \sqrt{s}=8 $$ TeV},
   author={S. Chatrchyan and V. Khachatryan and A. M. Sirunyan and others},
   collaboration = {CMS Collaboration},
   journal={\href{http://dx.doi.org/10.1103/PhysRevLett.112.171802}{Phys. Rev. Lett. \textbf{112}, 171802}},
   year={2014},
   }

@article{ATLAS_CMS_conduct_related_studies_5,
    title = "{Search for flavour changing neutral currents in top quark production and decays with three-lepton final state using the data collected at sqrt(s) = 13 TeV}",
    journal={{CMS Collaboration (CMS Collaboration),} \href{https://cms-results.web.cern.ch/cms-results/public-results/preliminary-results/TOP-17-017/index.html}{CMS-PAS-TOP-17-017}},
    year = "2017"
   }

@article{ATLAS_CMS_conduct_related_studies_6,
  title = {Search for flavor-changing neutral-current couplings between the top quark and the $Z$ boson with proton-proton collisions at $\sqrt{s}=13\text{ }\text{ }\mathrm{TeV}$ with the ATLAS detector},
  author = {G. Aad and B. Abbott and D. C. Abbott and others },
  collaboration = {ATLAS Collaboration},
  journal = {\href{https://link.aps.org/doi/10.1103/PhysRevD.108.032019}{Phys. Rev. D \textbf{108}, 032019}},
  year = {2023},
  }

@article{ATLAS_CMS_conduct_related_studies_7,
   titlee = {Search for flavor changing neutral current interactions of the top quark in final states with a photon and additional jets in proton-proton collisions at $\sqrt{s}=13 TeV},
   author={A. Hayrapetyan and A. Tumasyan and W. Adam and others},
   collaboration = {CMS Collaboration},
   journal={\href{http://dx.doi.org/10.1103/PhysRevD.109.072004}{Phys. Rev. D \textbf{109}, 072004}},
   year={2024},
   }

@article{effective_field_theory_1,
   tttitle={Top-charm associated production in high energy $e^{+} e^{-}$ Collisions},
   author={Tao Han and JoAnne L. Hewett},
   journal={\href{http://dx.doi.org/10.1103/PhysRevD.60.074015}{Phys. Rev. D \textbf{60}, 074015}},
   year={1999},
   }

@article{effective_field_theory_2,
   tttitle={Probing new physics from top quark FCNC processes at linear colliders: a mini review},
   author={Jin Min Yang},
   journal={\href{http://dx.doi.org/10.1016/j.aop.2004.11.003}{Ann. Phys. \textbf{316}, 529-539}},
   year={2005},
   }

@article{effective_field_theory_3,
   tttitle={Single top quark production as a probe of anomalous tqγ and tqZ couplings at the FCC-ee},
   author={Khanpour, Hamzeh and Khatibi, Sara and Yanehsari, Morteza Khatiri and Najafabadi, Mojtaba Mohammadi},
   journal={\href{http://dx.doi.org/10.1016/j.physletb.2017.10.047}{Phys. Lett. B \textbf{775}, 25-31}},
   year={2017},
   }

@article{effective_field_theory_5,
   tttitle={Probing FCNC couplings in single top-quark production associated with a neutral gauge boson in future lepton colliders},
   author={Sara Khatibi and Mehrnoosh Moallemi},
   journal={\href{http://dx.doi.org/10.1088/1361-6471/ac09dd}{Nucl. Part. Phys. \textbf{48}, 125004}},
   year={2021},
   }

@article{effective_field_theory_6,
   tttitle={Probing the top quark flavor-changing couplings at CEPC *},
   author={Liaoshan Shi and Cen Zhang},
   journal={\href{http://dx.doi.org/10.1088/1674-1137/43/11/113104}{Chin. Phys. C \textbf{43}, 113104}},
   year={2019},
   }

@article{effective_field_theory_7,
   tttitle={Top-quark physics at the CLIC electron-positron linear collider},
   author={H. Abramowicz and N. Alipour Tehrani and D. Arominski and others},
   collaboration = {CLICdp collaboration},
   journal={\href{http://dx.doi.org/10.1007/JHEP11(2019)003}{JHEP \textbf{1911}, 003}},
   year={2019},
   }

@article{2020_European_Strategy_1,
   tttitle={Muon colliders to expand frontiers of particle physics},
   author={Long, K. R. and Lucchesi, D. and Palmer, M. A. and Pastrone, N. and Schulte, D. and Shiltsev, V.},
   journal={\href{http://dx.doi.org/10.1038/s41567-020-01130-x}{Nature Physics \textbf{17}, 289-292}},
   year={2021},
   }

@misc{2020_European_Strategy_2,
      tttitle={Towards a Muon Collider}, 
      author={Carlotta Accettura and Dean Adams and Rohit Agarwal},
      eprint={\href{https://arxiv.org/abs/2303.08533}{arXiv:2303.08533 [physics.acc-ph]}},
   }

@misc{2020_European_Strategy_3,
      tttitle={A Muon Collider Facility for Physics Discovery}, 
      author={D. Stratakis and N. Mokhov and M. Palmer and others},
      eprint={\href{https://arxiv.org/abs/2203.08033}{arXiv:2203.08033 [physics.acc-ph]}},
   }

@article{Lagrangian_approach,
   tttitle={A minimal set of top anomalous couplings},
   author={Aguilar-Saavedra, J.A.},
   journal={\href{http://dx.doi.org/10.1016/j.nuclphysb.2008.12.012}{Nucl. Phys. B \textbf{812}, 181-204}},
   year={2009},
   }

@article{dif_chan_0,

   tttitle = {Probing tqZ anomalous couplings in the trilepton signal at the HL-LHC, HE-LHC and FCC-hh},
   author = {Liu, Yao-Bei and Moretti, Stefano},
   journal = {\href{https://iopscience.iop.org/article/10.1088/1674-1137/abe0c0}{Chin. Phys. C \textbf{45}, 043110}},
   year = "2021"
   }

@article{t_to_wb,
  tttitle = {QCD corrections to $t\ensuremath{\rightarrow}{W}^{+}+b$},
  author = {Chong Sheng Li and Robert J. Oakes and Tzu Chiang Yuan},
  journal = {\href{https://link.aps.org/doi/10.1103/PhysRevD.43.3759}{Phys. Rev. D \textbf{43}, 3759}},
  year = {1991},
  }

@article{FeynRules_package,
   tttitle={FeynRules  2.0 — A complete toolbox for tree-level phenomenology},
   author={Adam Alloul and Neil D. Christensen and Céline Degrande and Claude Duhr and Benjamin Fuks },
   journal={\href{http://dx.doi.org/10.1016/j.cpc.2014.04.012}{Comput. Phys. Commun. \textbf{185}, 2250-2300}},
   year={2014},
   }

@article{UFO_files,
   tttitle={UFO – The Universal FeynRules Output},
   author={Degrande, Céline and Duhr, Claude and Fuks, Benjamin and Grellscheid, David and Mattelaer, Olivier and Reiter, Thomas},
   journal={\href{http://dx.doi.org/10.1016/j.cpc.2012.01.022}{Comput. Phys. Commun. \textbf{183}, 1201-1214}},
   year={2012},
   }

@article{MadGraph5_NLO,
   tttitle={The automated computation of tree-level and next-to-leading order differential cross sections, and their matching to parton shower simulations},
   journal={\href{http://dx.doi.org/10.1007/JHEP07(2014)079}{JHEP \textbf{1407}, 079}},
   author = {Alwall, J. and Frederix, R. and Frixione, S. and others},
   year={2014},
   }

@article{PDG_2024,
    title = {Review of particle physics},
    author = {Navas, S. and others},
    collaboration = {Particle Data Group},
    journal = {\href{https://journals.aps.org/prd/abstract/10.1103/PhysRevD.110.030001}{Phys. Rev. D \textbf{110}, 030001}},
    year = "2024"
   }

@article{polar_2,
   tttitle={Top flavour-changing neutral coupling signals at a linear collider},
   author={Aguilar-Saavedra, J.A},
   journal={\href{http://dx.doi.org/10.1016/S0370-2693(01)00162-9}{Phys. Lett. B \textbf{502}, 115-124}},
   year={2001},
   }

@misc{polar_1,
      tttitle={Physics Opportunities with Polarized e- and e+ Beams at TESLA}, 
      author={Gudrid Moortgat-Pick and Herbert M. Steiner},
      eprint={\href{https://arxiv.org/abs/hep-ph/0106155}{arXiv:hep-ph/0106155}},
   }

@article{polar_3,
    tttitle = {Lepton flavor violating top quark FCNC processes at the {\ensuremath{\mu}}TRISTAN},
    author = "Sarkar, Abhik",
    journal = {\href{https://journals.aps.org/prd/abstract/10.1103/jnnt-6t8f}{Phys. Rev. D \textbf{113}, 095010}},
    year = {2026}
   }

@article{top_quark_fcnc_at_muc,

    tttitle = {Top Quark Flavor Changing Couplings at a Muon Collider},
    author = {Ake, Daniel and Bouzas, Antonio O. and Larios, F.},
    journal = {\href{https://inspirehep.net/literature/2723222}{Adv. High Energy Phys. \textbf{2024}, 2038180}},
    year = "2024"
   }

@article{pythia8,
   tttitle={An introduction to PYTHIA 8.2},
   author={Torbjörn Sjöstrand and Stefan Ask and Jesper R. Christiansen and others},
   journal={\href{http://dx.doi.org/10.1016/j.cpc.2015.01.024}{Comput. Phys. Commun. \textbf{191}, 159-177}},
   year={2015},
   }

@article{delphes,
    title_e = {DELPHES 3, A modular framework for fast simulation of a generic collider experiment},
    author = {J. de Favereau and C. Delaere and P. Demin and others},
    collaboration = {DELPHES 3 collaboration},
    journal = {\href{https://link.springer.com/article/10.1007/JHEP02(2014)057}{JHEP \textbf{1402}, 057}},
    year = {2014}
    }

@article{FastJet_user_manual,
   tttitle={FastJet user manual: (for version 3.0.2)},
   author={Cacciari, Matteo and Salam, Gavin P. and Soyez, Gregory},
   journal={\href{http://dx.doi.org/10.1140/epjc/s10052-012-1896-2}{Eur. Phys. J. C \textbf{72}, 1896}},
   year={2012},
   }

@article{VLC_algorithm_1,
   tttitle={A robust jet reconstruction algorithm for high-energy lepton colliders},
   author={Boronat, M. and Fuster, J. and García, I. and Ros, E. and Vos, M.},
   journal={\href{http://dx.doi.org/10.1016/j.physletb.2015.08.055}{Phys. Lett. B \textbf{750}, 95-99}},
   year={2015},
   }

@article{VLC_algorithm_2,
   tttitle={Jet reconstruction at high-energy electron–positron colliders},
   author={Boronat, M. and Fuster, J. and Garcia, I. and others},
   journal={\href{http://dx.doi.org/10.1140/epjc/s10052-018-5594-6}{Eur. Phys. J. C \textbf{78}, 144}},
   year={2018},
   }

@article{delta_R_1,
   tttitle={Jet substructure at the Large Hadron Collider: A review of recent advances in theory and machine learning},
   author={Larkoski, Andrew J. and Moult, Ian and Nachman, Benjamin},
   journal={\href{http://dx.doi.org/10.1016/j.physrep.2019.11.001}{Phys. Rep. \textbf{841}, 1-63}},
   year={2020},
   }

@article{delta_R_2,
   tttitle={Jet substructure at the Large Hadron Collider},
   author={Kogler, Roman and Nachman, Benjamin and Schmidt, Alexander and others},
   journal={\href{http://dx.doi.org/10.1103/RevModPhys.91.045003}{Rev. Mod. Phys. \textbf{91}, 045003}},
   year={2019},
   }

@article{delta_R_3,
   tttitle={Identifying boosted objects with N-subjettiness},
   author={Thaler, Jesse and Van Tilburg, Ken},
   journal={\href{http://dx.doi.org/10.1007/JHEP03(2011)015}{JHEP \textbf{1103}, 015}},
   year={2011},
   }

@article{easyscan,
   tttitle={EasyScan_HEP: A tool for connecting programs to scan the parameter space of physics models},
   author={Shang, Liangliang and Zhang, Yang},
   journal={\href{http://dx.doi.org/10.1016/j.cpc.2023.109027}{Comput. Phys. Commun. \textbf{296}, 109027}},
   year={2024},
   }

@article{MadAnalysis5_1,
   tttitle={MadAnalysis 5, a user-friendly framework for collider phenomenology},
   author={Conte, Eric and Fuks, Benjamin and Serret, Guillaume},
   journal={\href{http://dx.doi.org/10.1016/j.cpc.2012.09.009}{Comput. Phys. Commun. \textbf{184}, 222-256}},
   year={2013},
   }

@article{MadAnalysis5_2,
   tttitle={Designing and recasting LHC analyses with MadAnalysis  5},
   author={Conte, Eric and Dumont, Béranger and Fuks, Benjamin and Wymant, Chris},
   journal={\href{http://dx.doi.org/10.1140/epjc/s10052-014-3103-0}{Eur. Phys. J. C \textbf{74}, 3103}},
   year={2014},
   }

@article{z_excl_1,
   tttitle={Asymptotic formulae for likelihood-based tests of new physics},
   author={Cowan, Glen and Cranmer, Kyle and Gross, Eilam and Vitells, Ofer},
   journal={\href{http://dx.doi.org/10.1140/epjc/s10052-011-1554-0}{Eur. Phys. J. C \textbf{71}, 1554}},
   year={2011},
   }

@article{z_excl_2,
   tttitle={Vectorlike leptons at the Large Hadron Collider},
   author={Kumar, Nilanjana and Martin, Stephen P.},
   journal={\href{http://dx.doi.org/10.1103/PhysRevD.92.115018}{Phys. Rev. D \textbf{92}, 115018}},
   year={2015},
   }

@article{z_excl<1.645,
    title = "{Exotic Higgs Decays in Type-II 2HDMs at the LHC and Future 100 TeV Hadron Colliders}",
    author = "Kling, Felix and Li, Honglei and Pyarelal, Adarsh and Song, Huayang and Su, Shufang",
    journal = {\href{https://inspirehep.net/literature/1707050}{JHEP \textbf{1906}, 031}},
    year = "2019"
   }

@article{dif_chan_1,
   tttitle={Discovery potential for T ′ → tZ in the trilepton channel at the LHC},
   author={Basso, Lorenzo and Andrea, Jeremy},
   journal={\href{http://dx.doi.org/10.1007/JHEP02(2015)032}{JHEP \textbf{1502}, 032}},
   year={2015},
   }

@article{dif_chan_2,
   tttitle={Ultraboosted Zt and $$\gamma t$$ γ t production at the HL-LHC and FCC-hh},
   author={J. A. Aguilar-Saavedra },
   journal={\href{http://dx.doi.org/10.1140/epjc/s10052-017-5375-7}{Eur. Phys. J. C \textbf{77},769}},
   year={2017},
   }

@article{dif_chan_3,
   tttitle={Probing top quark FCNC couplings in the triple-top signal at the high energy LHC and future circular collider},
   author={Hamzeh Khanpour},
   journal={\href{http://dx.doi.org/10.1016/j.nuclphysb.2020.115141}{Nucl. Phys. B \textbf{958}, 115141}},
   year={2020},
   }

@article{dif_chan_4,
   tttitle={Fingerprinting the top quark FCNC via anomalous 
   Ztq couplings at the LHeC},
   author={Behera, Subhasish and Islam, Rashidul and Kumar, Mukesh and Poulose, Poulose and Rahaman, Rafiqul},
   journal={\href{http://dx.doi.org/10.1103/PhysRevD.100.015006}{Phys. Rev. D \textbf{100}, 015006}},
   year={2019},
   }

@article{dif_chan_5,
   tttitle={Probing top quark FCNC tqγ and tqZ couplings at future electron-proton colliders},
   author={Cakir, O. and Yilmaz, A. and Turk Cakir, I. and Senol, A. and Denizli, H.},
   journal={\href{http://dx.doi.org/10.1016/j.nuclphysb.2019.114640}{Nucl. Phys. B \textbf{944}, 11640}},
   year={2019},
   }

@article{dif_chan_6,
   tttitle={Single top quark production as a probe of anomalous tqγ and tqZ couplings at the FCC-ee},
   author={Khanpour, Hamzeh and Khatibi, Sara and Yanehsari, Morteza Khatiri and Najafabadi, Mojtaba Mohammadi},
   journal={\href{http://dx.doi.org/10.1016/j.physletb.2017.10.047}{Phys. Lett. B \textbf{775}, 25-31}},
   year={2017},
   }

@misc{dif_chan_7,
      tttitle={Probing top flavour-changing neutral couplings at TESLA}, 
      author={J. A. Aguilar-Saavedra and T. Riemann},
      eprint = {\href{https://arxiv.org/abs/hep-ph/0102197}{arXiv:hep-ph/0102197}},
}

@article{dif_chan_9,
   tttitle={Probing the top quark flavor-changing couplings at CEPC *},
   author={Shi, Liaoshan and Zhang, Cen},
   journal={\href{http://dx.doi.org/10.1088/1674-1137/43/11/113104}{Chin. Phys. C \text{43}, 113104}},
   year={2019},
   }

@article{dif_chan_10,
   tttitle={Top FCNC interactions through dimension-six four-fermion operators at the electron proton collider},
   author={Liu, Wei and Sun, Hao},
   journal={\href{http://dx.doi.org/10.1103/PhysRevD.100.015011}{Phys. Rev. D \textbf{100}, 015011}},
   year={2019},
   }

@misc{dif_chan_11,
      tttitle={Sensitivity to top-quark FCNC interactions at future muon colliders}, 
      author={A. Senol and B. S. Ozaltay and M. Tekin and H. Denizli},
      eprint = {\href{https://arxiv.org/abs/2604.13562}{arXiv:2604.13562 [hep-ph]}},
      }

@misc{dif_chan_12,
      tttitle={Top quark FCNC in Randall-Sundrum models: post-LHC allowed rates and searches at $e^+e^-$ and $\mu^+ \mu^-$ colliders}, 
      author={Sagar Airen and Roberto Franceschini},
      eprint = {\href{https://arxiv.org/abs/2601.14966}{arXiv:2601.14966 [hep-ph]}},
      }

@article{BDT_SAJA,
    title = "{Identification of tqg flavor-changing neutral current interactions using machine learning techniques}",
    author = "Ko, Byeonghak and Heo, Jeewon and Jang, Woojin and Lee, Jason Sang Hun and Roh, Youn Jung and Watson, Ian James and Yang, Seungjin",
    journal = {\href{https://rd.springer.com/article/10.1007/s40042-024-01277-3}{J. Korean Phys. Soc. \textbf{86}, 269-279}},
    year = "2025"
   }

@article{BDT_2,
    tttitle = "{Top quark anomalous FCNC production via $tqg$ couplings at FCC-hh}",
    author = "Oyulmaz, K. Y. and Senol, A. and Denizli, H. and Cakir, O.",
    journal = {\href{https://journals.aps.org/prd/abstract/10.1103/PhysRevD.99.115023}{Phys. Rev. D \textbf{99}, 115023}},
    year = "2019"
    }

\end{document}